\documentclass[12pt]{article}

\pdfoutput=1

\usepackage{ifpdf}
\ifpdf
\usepackage{graphicx}
\usepackage{hyperref}
\usepackage[dvipsnames]{xcolor}
\else
\usepackage[dvipdfmx]{graphicx}
\usepackage[dvipdfmx]{hyperref}
\usepackage[dvipsnames]{xcolor}
\fi
\hypersetup{
  colorlinks,
  citecolor=magenta,
  linkcolor=Maroon,
  menucolor=black,
  urlcolor=Violet}
\usepackage{amssymb,amsfonts,amsmath,cancel,cite,multirow}
\usepackage{tikz}
\usepackage[capitalise]{cleveref}
\usepackage[margin=1in]{geometry}
\usepackage{bbm}
\usepackage{braket}
\numberwithin{equation}{section}
\newcommand{\eqs}[1]{\begin{equation}\begin{split} #1 \end{split}\end{equation}}

\begin{document}

\begin{titlepage}

\begin{flushright}
\end{flushright}

\vskip 1.35cm
\begin{center}

{\large
\textbf{
        Revisiting Q-ball Interactions with Matters
}}
\vskip 1.2cm

\mbox{
	Ayuki Kamada$^{a}$,
	Takumi Kuwahara$^{b,c}$, 
	and Keiichi Watanabe$^{d}$
}

\vskip 0.4cm

\textit{$^a$
    Institute of Theoretical Physics, Faculty of Physics, University of Warsaw, ul. Pasteura 5, PL-02-093 Warsaw, Poland
}

\textit{$^b$
    Center for Theoretical Physics and College of Physics, Jilin University, Changchun, 130012, China
}
\textit{$^c$ 
    Center for High Energy Physics, Peking University, Beijing 100871, China
}

\textit{$^d$
    Institute for Cosmic Ray Research, The University of Tokyo, Kashiwa, Chiba 277-8582, Japan
}

\vskip 1.5cm
\begin{abstract}
	\noindent
	Q-ball dark matter is one of the candidates for the macroscopic dark matter: Q-ball is a non-topological solitonic configuration, whose stability can be ensured by global charge and energy conservation. 
  One of the crucial factors for discovering signatures from the Q-ball dark matter, is the interactions of the Q-ball dark matter with ordinary matter. 
  In particular, the scattering of ordinary matter off the Q-ball dark matter is important for the direct detection searches, such as paleo-detectors. 
  It was conjectured that quarks incident on the Q-ball were reflected as anti-quarks with a probability of order unity, but it costs the energy of the squark in the Q-ball, which cannot be paid in the scattering of ordinary matter off the Q-ball dark matter. 
  In addition, once a proton is reflected as an anti-proton, the Q-ball obtains the electromagnetic charge. 
  In this study, we revisit the scattering process of quarks with the Q-ball with taking into account the energy cost of the scattering and the electromagnetic charge-up of the Q-ball.
\end{abstract}
\end{center}
\end{titlepage}

\section{Introduction}

Much of particle nature of dark matter (DM) remains unrevealed even though the cosmological observations have established the existence of DM.
For several decades, many experiments and observations have focused on the weakly interacting massive particles (WIMP) as one of the plausible candidates of the particle DM. 
Despite the best efforts of the experiments, no crucial evidence for the WIMP paradigm has been found. 
Macroscopic DM is one of the candidates for DM models beyond WIMP paradigm, and it requires alternative methods searching for macroscopic DM. 
Q-ball DM is one of the candidates for the macroscopic DM (with a mass of $\mathcal{O}(1) \, \mathrm{g}$ and a radius of $\mathcal{O}(10) \, \mathrm{fm}$~\cite{Kasuya:2015uka}): Q-ball is a non-topological solitonic (spherically symmetric) configuration~\cite{Rosen:1968mfz,Friedberg:1976me,Coleman:1985ki}, whose stability can be ensured by global charge and energy conservation.
Supersymmetric extensions of the standard model (SM) may have stable Q-ball (in particular, with gauge-mediated supersymmetry breaking)~\cite{Kusenko:1997zq,Kusenko:1997si}. 
One of scalar counterparts of quarks and leptons (squarks and sleptons) gets a field value along a flat direction of the scalar potential~\cite{Gherghetta:1995dv,Dine:1995kz}.

Interactions of Q-ball DM with matter are crucial keys for discovering signals from the Q-ball at experiments and observations.
There are several studies about the Q-ball interaction: decay into ordinary matter~\cite{Cohen:1986ct,Kawasaki:2012gk,Kawasaki:2019ywz,Kasuya:2024ldq} and absorption of baryons~\cite{Kusenko:1997it}.
In particular, scattering with a nucleon is important for direct detection searches for Q-ball DM.
It was conjectured in Ref.~\cite{Kusenko:1997vp} that the Q-ball absorbs the nucleon during the scattering. 
The point is that a scalar field gets its field value inside the Q-ball, and its value gradually decreases with the distance from the Q-ball center. 
The scalar field has a relatively small field value near the boundary of the Q-ball, where the quark mass from the field value can be smaller than the quantum chromodynamics (QCD) dynamical scale, while the color symmetry is broken by the field value. 
Then, the nucleon incident on the Q-ball is dissociated into quarks on the layer, and the energy of nucleons can be released by emitting pions~\cite{Kusenko:1997vp,Super-Kamiokande:2006sdq}. 
Through the annihilation process of the incident quark into the scalars inside the Q-ball, nucleons are absorbed on the Q-ball (leaving one quark left).
Meanwhile, the scattering of a Q-ball with an ordinary matter has been considered from the quantum-mechanical perspective in Refs.~\cite{Cohen:1986ct,Multamaki:1999an,Kusenko:2004yw}.%
\footnote{
  The scattering of Q-ball with its own scalar particles has also been considered in Refs.~\cite{Smolyakov:2017axd,Smolyakov:2019cld,Ciurla:2024ksm,Azatov:2024npx}. 
}
In Ref.~\cite{Kusenko:2004yw}, the Q-ball is considered as the static background, and we find that quarks incident on the Q-ball are reflected as anti-quarks with a probability of order unity.
It leads to quick energy release through annihilation of the reflected anti-quarks with the surrounded materials. 
However, after the nucleon (quark)--Q-ball scattering, the squarks should be added in the Q-ball in accordance with the baryon number conservation, and it costs the energy of squarks in the Q-ball, which we refer to as the chemical potential $\omega$.
When the scattering process cannot pay the energy cost, the nucleon (quark) cannot be reflected as the anti-nucleon (anti-quark).
In addition, once a proton is reflected as an anti-proton, the Q-ball obtains the electromagnetic charge.
In the literature, the energy cost and charge up have not been taken into account. 
In this study, we revisit the scattering process of quarks with the Q-ball with incorporating these effects. 
We focus on the Q-ball with the $udd$ flat direction, which carries only the baryon number. 

Recently, a new avenue has been proposed for direct searches of DM using ancient mineral as target materials, so-called paleo-detector~\cite{Drukier:2018pdy,Baum:2023cct}.
Instead of constructing detectors with a large target mass as conventional direct detection experiments, ancient minerals can record scratches from the DM-nucleon scattering for quite long time scale, about a billion years. 
For the material of 1\,kg recorded for about a billion years, the exposure for this kind of experiments can achieve to the same orders of that for conventional direct detection experiments with the target mass of $10^4\,\mathrm{kg}$ achieved in a year. 
As for the macroscopic DM, it is challenging to explore the recoil signals from the DM-nucleon scattering in the conventional direct detection experiments since the DM flux coming into a detector is limited.
Meanwhile, the paleo-detector is a good candidate for the direct detection experiments searching for macroscopic DM since they have recorded the recoil signals for a billion years. 
In Ref.~\cite{Baum:2023cct}, the electromagnetically-charged Q-ball \cite{Arafune:2000yv,Shoemaker:2008gs,Hong:2016ict,Hong:2017qvx} has been discussed: due to the high energy loss of the charged Q-ball in materials, the expected tracks would be longer than the WIMP-induced tracks. 
Even for the electromagnetically-neutral Q-ball, there exist the constraints on the flux from experiments searching for absorption of nucleons into a Q-ball \cite{Arafune:2000yv,Super-Kamiokande:2006sdq,IceCube:2013ydv,Kasuya:2015uka}.
It is also expected that the Q-ball would leave unique signals in ancient minerals if incident quarks would be reflected as anti-quarks and would lead to energy release through the pair-annihilation as shown in Ref.~\cite{Kusenko:2004yw}. 

This paper is organized as follows:  
In \cref{sec:quark}, we revisit scattering of Q-ball DM with quarks in the presence of the chemical potential.
After briefly reviewing the basics of Q-ball scattering in \cref{sec:basics}, we first discuss the scattering on the infinitely large Q-ball wall in \cref{sec:infinite}, and then the three-dimensional scattering in \cref{sec:spherical}.
We discuss the implication of the quark scattering for the nucleon scattering in \cref{sec:nucleon}.
\cref{sec:conclusion} is devoted to concluding our study.

\section{Scattering of Quarks \label{sec:quark}}

In this section, we revisit quark scattering under a Q-ball background oscillating in time (in contrast to that under the static Q-ball background in Ref.~\cite{Kusenko:2004yw}).
We first review the basics of Q-ball scattering and summarize our notation: we consider a toy model that describes scattering of quark without the color degrees of freedom with Q-ball.
Then, we discuss the scattering of quark with infinitely large Q-ball wall and the three-dimensional scattering with finite-size Q-ball.

\subsection{Basics of Q-ball scattering
\label{sec:basics}}

Before discussing scattering, let us briefly summarize relevant macroscopic properties of a Q-ball (DM), while more discussion about the scalar-field potential and relation to microscopic parameters can be found in \cref{app:Q-potential}.
The chemical potential is inversely proportional to the radius as
\begin{align}
  \omega \simeq R^{-1} \simeq 20 \, \mathrm{MeV} \left( \frac{10 \, \mathrm{fm}}{R} \right) \,.
\end{align}
For stable Q-ball DM, $3 \times \omega$ should be lower than the nucleon mass $m_{N} \simeq 1 \, \mathrm{GeV}$.
The mass (and charge) of the Q-ball is given by the scalar-field value $\varphi_0$ inside the Q-ball as
\begin{align}
  M_{Q} \simeq \omega^{2} \varphi_{0}^{2} R^{3} \simeq \omega Q \simeq 0.2 \, \mathrm{g} \left( \frac{\varphi_{0}}{10^{11} \, \mathrm{GeV}}\right)^{2} \left( \frac{R}{10 \, \mathrm{fm}} \right)
\end{align}
The mass is bounded from below as supersymmetric breaking scale $M_{F}$:
\begin{align}
  M_{Q} \gtrsim  2 \times 10^{-3} \, \mathrm{g} \left( \frac{R}{10 \, \mathrm{fm}} \right)^{3}
\end{align}
for $M_{F} \gtrsim 10 \, {\rm TeV}$.
As far as the energy of an incoming quark is below $\omega Q^{1/3} \sim \omega (M_Q/\omega)^{1/3}$, the scattering process cannot resolve the valence squarks consisting of the Q-ball.
We can treat the Q-ball as a coherent potential in the low-energy scattering.
Moreover, the quarks obtain the mass proportional to $\varphi_0$ inside the Q-ball, and thus do not penetrate into the Q-ball: namely, the Q-ball is approximated by a wall.
In general, the Q-ball radius $R$ does not precisely identical to the radius used in the geometric cross section. 
This is because quarks can feel the scalar-field value as their mass even much below $\varphi_0$, leading to a larger radius than $R$ by a factor of $\mathcal{O}(1)$.
We need to take into account the Q-ball profile (namely, the spatial distribution of the scalar-field value) for the precise treatment of $R$. 
In this work, we approximate the Q-ball profile by the step function, though the scalar-field value varies gradually around $R$. 

Let us consider a toy model in which we disregard the color degrees of freedom and that captures the key features of fermion scattering under a Q-ball background. 
In the next section, we discuss the implication of analysis in this section for a realistic setup. 
Although we do not take into account the color degrees of freedom, we utilize the same names for the fields as in the supersymmetric models. 
The squarks $\phi$ that have a flat direction couples to the quark $\psi$ and the gluino $\lambda$. 
We here use two-component notation for fermions. 
The relevant parts of the Lagrangian are 
\begin{align}
  \mathcal{L} 
    = - \sqrt{2} g_s (\lambda \psi_i \phi_i^\ast) + \mathrm{h.c.} \,,
\end{align}
where $g_s$ denotes a gauge coupling constant and $i = 1, 2$ represents flavor indices. 
$\lambda$ denotes gluino that does not have its own charge, and $\psi_1$ and $\psi_2$ have the opposite charges. 
The Majorana mass term for gluino is 
\begin{align}
  \mathcal{L}_M = - M_\lambda \lambda \lambda \,.
\end{align}
Here, the Majorana mass is assumed to be real-valued.

The squarks obtain a non-zero field value $\phi_i = \varphi_{i,0} / \sqrt{2}\neq 0$ in the Q-ball background, then the quark gets a mass mixing with gluino. 
The mass matrix for fermions in the Q-ball background is parametrized as 
\begin{align}
  M_\psi = 
    \begin{pmatrix}
        0 & 0 & M_1 \\
        0 & 0 & M_2 \\
        M_1 & M_2 & M_\lambda \\
    \end{pmatrix} \,.
\end{align}
Here, $M_{1(2)}$ denotes the mass mixing between quark $\psi_1$ ($\psi_2$) and gluino. 
We assume that the Dirac mass of quarks is negligible compared to the other mass parameters, the gluino mass and the field values of the squark fields. 
We introduce the mixing angles $\alpha \,, \beta$ and the dimensionful parameter $M$, and the mass parameters are written as 
\eqs{
    M_1 & = \overline M \cos \alpha \,, \quad 
    M_2 = \overline M \sin\alpha \,, \\
    \overline M & = g_s \varphi_0 = M \sin 2 \beta \,, \quad \varphi_0 = \sqrt{\varphi_{1,0}^2 + \varphi_{2,0}^2} \,, \\
    M_\lambda & = 2 M \cos 2 \beta \,,
}
Here, $\alpha$ denotes the ratio of the field values of the squark fields: $\tan \alpha = \varphi_{1,0}/\varphi_{2,0}$.
The field rotation by $\alpha$ makes only a linear combination of the fermions, which is denoted by $\psi$, to couple to the gluino.
\begin{align}
  \begin{pmatrix}
    \psi_1 \\
    \psi_2
\end{pmatrix}
= 
\begin{pmatrix}
    \sin \alpha & \cos \alpha \\
    -\cos \alpha & \sin \alpha \\
\end{pmatrix}
\begin{pmatrix}
    \chi_0\\
    \psi
\end{pmatrix} \,.
\end{align}
Here, $\chi_0$ denotes another linear combination, which does not have a mass mixing with the gluino.
The gluino and the massive combination of quarks mix with each other with the mixing angle $\beta$.
Then, the mass eigenstates of fermions are given by the rotation from the basis $(\psi \,, \lambda)$ as follows. 
\begin{align}
  \begin{pmatrix}
        \psi \\
        \lambda
    \end{pmatrix}
    = 
    \begin{pmatrix}
        \cos \beta & \sin \beta \\
        - \sin\beta & \cos \beta 
    \end{pmatrix}
    \begin{pmatrix}
        \chi_1 \\
        \chi_2
    \end{pmatrix} \,.
\end{align}
Here, $\chi_1$ and $\chi_2$ denote the mass eigenstates with the Majorana masses $- M_- = - 2 M \sin^2 \beta$ and $M_+ = 2 M \cos^2 \beta$, respectively.

Now, we consider wave functions for fermions inside and outside the Q-ball.
As far as the Dirac mass terms for quarks are negligible, quark $\psi$ and gluino $\lambda$ outside the Q-ball obey the following equations.
\begin{align}
  \begin{cases}
    i \overline \sigma \cdot \partial \psi = 0 \,, \\
    i \overline \sigma \cdot \partial \lambda - M_\lambda \lambda^\dag = 0 \,, 
  \end{cases}
\end{align}
where $\lambda^\dag = i \sigma^{2} \cdot \lambda^{*}$. 
Meanwhile, quark and gluino have the mass term proportional to the field value of the squark field $\varphi_{0}$ inside Q-ball. 
Thus, wave functions for quark and gluino inside Q-ball obey the following equations.
\begin{align}
  \begin{cases}
    i \overline \sigma^{\mu} \partial_{\mu} \psi - \sqrt{2} g_{s} \phi \lambda^\dag = 0 \,,  \\
    i \overline \sigma^{\mu} \partial_{\mu} \lambda - M_\lambda \lambda^\dag - \sqrt{2} g_{s} \phi \psi^\dag = 0 \,.
  \end{cases}
\end{align}

Let us consider the squark field oscillating in time, $\phi = \varphi_{0} e^{-i \omega t}/ \sqrt{2} $.
A time-dependent phase factor appears in the mass term, but the phase factor will be removed by the field redefinition $\psi = e^{-i \omega t} \psi_0$.
The equations will be reduced as follows.
\begin{align}
  \begin{cases}
    i \overline{\sigma}^{\mu} \partial_{\mu} \psi_0 + \omega \overline{\sigma}^{0} \psi_0 - \overline M \lambda^\dag=0 \,, \\
    i \overline{\sigma}^{\mu} \partial_{\mu} \lambda - M_\lambda \lambda^\dag - \overline M \psi_0^\dag=0\,.
  \end{cases}
\end{align}
There is an extra term proportional to $\omega$ in the equation for $\psi_0$. 
The chemical potential $\omega$ is typically much smaller than the field value $\varphi_{0}$ and the gluino mass $M_\lambda$.
In the following, we obtain the wave functions by solving the equation with ignoring the $\omega$ term, and the oscillating phase factor appears only in the matching conditions for the wave function of the matter fermions.

Finally, we comment on the four-component notations for fermions in the Weyl representation. 
The four-component Majorana fermions for quark are written as 
\begin{align}
  \Psi_W^\mathrm{in} = 
  \begin{pmatrix}
    e^{-i \omega t} \psi_0 \\ 
    e^{i \omega t} \psi_0^\dag
  \end{pmatrix} \,, \qquad 
  \Psi_W^\mathrm{out} = 
  \begin{pmatrix}
    \psi \\ 
    \psi^\dag
  \end{pmatrix} \,, 
\end{align}
where the superscripts, ``in'' and ``out'', indicate the quarks inside and outside the Q-ball, respectively. 
We note that the phase factor appears only for the quarks inside the Q-ball.
The four-component fermions for gluinos are defined in a similar way to $\Psi_W^\mathrm{out}$, namely the four-component fermion without the time-dependent phase factor, both inside and outside the Q-ball. 

\subsection{Scattering on the infinitely large Q-ball wall \label{sec:infinite}}

\begin{figure}[t]
  \centering
    \includegraphics[width=0.45\linewidth]{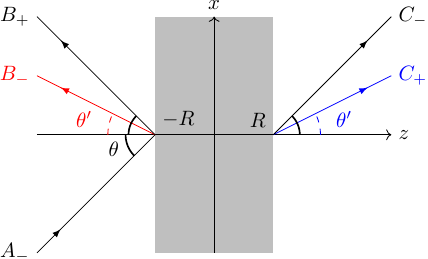}
  \caption{
  Scattering on the infinitely large Q-ball wall with the left-handed quark coming in: the shaded area corresponds to the Q-ball. 
  The black solid lines depict the incoming, reflected, and transmitted quarks with the energy of $E$, while the colored lines depict the reflected and transmitted quarks with the energy of $E-2\omega$.
  Since the energy $E$ is much less than the masses of the fermions inside the Q-ball, the massive fermions are assumed to run on the $z$-axis and the transmitted fermions are emitted from the $z$-axis.
  }
  \label{fig:2dpict}
\end{figure}

We begin to discuss the (colorless) quark wave functions and their matching conditions for the scattering on the infinitely large Q-ball wall. 
We assume that Q-ball boundaries are located at $z = \pm R$ and are infinitely extended in other directions as shown in \cref{fig:2dpict}.
The fermions get massive in $|z| < R$ through the field value of squarks. 
A Majorana fermion $\psi$, collectively denoting quarks and gluinos outside Q-ball and massive fermions inside Q-ball, obeys the following equation:
\begin{align}
  i \overline \sigma \cdot \partial \psi - i M \sigma^2 \psi^\ast = 0 \,,
  \label{eq:majorana_eq}
\end{align}
where $M$ is the Majorana mass. 
A solution to the equation is given by 
\begin{align}
  \begin{cases}
    \psi = \sqrt{p \cdot \sigma} \left( A e^{- i p \cdot x} - i \sigma^2 A^\ast e^{i p \cdot x} \right)\,,  & M > 0 \,, \\
    \psi = \sqrt{p \cdot \sigma} \left( A e^{- i p \cdot x} + i \sigma^2 A^\ast e^{i p \cdot x} \right)\,,  & M < 0 \,. 
  \end{cases}
\end{align}
Here, $A$ denotes an arbitrary two-component spinor, and $p = (E, \mathbf{p})$ with $\mathbf{p}^2 = E^2 - M^2$.
As for the Majorana fermions inside the Q-ball, the injected energy $E$ is less than masses of Majorana fermions, namely $E < M$.
In such a case, we consider the wave functions with an analytically-continued momentum $p$. 
We use the wave function with the analytically-continued momentum given by
\begin{align}
  \begin{cases}
    \psi = \sqrt{p \cdot \sigma} A e^{- i p \cdot x} + \sqrt{p^\ast \cdot \sigma} ( - i \sigma^2) A^\ast e^{i p^\ast \cdot x}\,,   & M > 0 \,, \\ 
    \psi = \sqrt{p \cdot \sigma} A e^{- i p \cdot x} - \sqrt{p^\ast \cdot \sigma} ( - i \sigma^2) A^\ast e^{i p^\ast \cdot x}\,,   & M < 0 \,.
  \end{cases}
  \label{eq:complex_sol}
\end{align}
We summarize several formulae regarding the analytically-continued wave functions in \cref{app:formulae}.

There are five kinds of fermions in the matching conditions for the fermions at the Q-ball boundaries.
The incoming quark (whose wave function is denoted by $\psi_A$) enters the Q-ball from $z < - R$, and there are the reflected quark ($\psi_B$) in $z < - R$ and the transmitted quark ($\psi_C$) in $z > R$.
When the energy of the incoming quark is large enough, there are the reflected gluino ($\lambda_B$) and the transmitted gluino ($\lambda_C$). 
The matching conditions of wave functions at the Q-ball boundary $z = - R$ are 
\begin{align}
  \psi_A + \psi_{B} = \sum_{\pm} \left( \sin \beta \chi_{D}^\pm + \cos \beta \chi^\pm_{F} \right) \,, \quad 
  \lambda_{B} = \sum_{\pm} \left( \cos \beta \chi_{D}^\pm - \sin \beta \chi^\pm_{F} \right) \,.
\end{align}
Here, $\chi^\pm_{D\,,F}$ denote the wave functions for the Majorana fermions inside the Q-ball: $\chi_D^\pm$ denotes the fermion with the mass of $M_+$, while $\chi_F^\pm$ denotes the fermion with the mass of $M_-$.
The superscript of $\chi^{+(-)}_{D\,,F}$ indicates the growing (decaying) mode.
The matching conditions at the other Q-ball boundary $z = R$ are:
\begin{align}
  \psi_{C} = \sum_{\pm} \left( \sin \beta \chi_{D}^\pm + \cos \beta \chi_{F}^\pm \right) \,, \quad 
  \lambda_C = \sum_{\pm} \left( \cos \beta \chi_{D}^\pm - \sin \beta \chi_{F}^\pm \right) \,.
\end{align}
Now, we obtain the spinor relations among the fermions for given spinor of the incoming fermion, $A$. 
We focus only on positive and negative helicities for $A$ and find the relations for each helicity, but we obtain the generic relations by combining the relations for each helicity. 

In this subsection, we consider that the incoming quark comes into the Q-ball wall with an oblique angle $\theta$: $\theta$ denotes the angle between the $z$ axis and the momentum of the incoming quark.
We assume that the momenta lie in the $x$--$z$ plane as illustrated in \cref{fig:2dpict}. 
Assuming the helicity of the incoming quark, we construct the other spinors by solving the matching conditions. 
We consider the case that the left-handed quark comes in. 
The four-momenta for massless quarks are assigned as follows: 
\eqs{
  p_{A-} & = (E, E \sin \theta, 0, E \cos\theta) \,, \\
  p_{B+} & = (E, E \sin \theta, 0, - E \cos\theta) \,, \qquad 
  p_{B-} = (E-2\omega, (E-2\omega) \sin \theta', 0, - (E-2\omega) \cos \theta') \,, \\
  p_{C+} & = (E-2\omega, (E-2\omega)\sin\theta', 0, (E-2\omega)\cos\theta') \,, \qquad 
  p_{C-} = (E, E\sin\theta, 0, E\cos\theta) \,.
  \label{eq:p_assign1}
}
Here, we decompose the massless fermions into components with definite spin directions (denoted by the label $\pm$), so that their spinors projected onto the $z$-axis satisfy the proper projection operation, discussed later. 
We assume the energy to be $E$ for the left-handed quarks, $B_+$ and $C_-$, while the energy to be $E - 2 \omega > 0$ for the right-handed anti-quarks, $B_-$ and $C_+$ due to the energy cost for adding the left-handed squarks in the Q-ball.
Meanwhile, we do not decompose the massive fermions into each spinor part since both spinors with different spins are assumed to have the same energy for them.
\eqs{
  \kappa_{B} & = (E-\omega, E \sin\theta, 0, - i M_\lambda) \,, \qquad 
  \kappa_{C} = (E-\omega, E \sin\theta, 0, i M_\lambda) \,, \\
  \kappa_{D}^\pm & = (E-\omega, E \sin\theta, 0, \mp i M_+) \,, \qquad 
  \kappa_{F}^\pm = (E-\omega, E \sin\theta, 0, \mp i M_-) \,. 
  \label{eq:p_assign2}
}
Hereafter, we take the limit that the mass parameters $M_\lambda$ and $M_\pm$ are much larger than the injected energy $E$ and the chemical potential $\omega$.

The massless fermions with the different energy can be transmitted and reflected with a different angle, denoted by $\theta'$.
The momentum conservation in the $x$-direction gives the following relation between the angles:
\begin{align}
  \frac{\sin \theta'}{\sin \theta} = \frac{E}{E-2\omega} \,.
\end{align}
We note that the scattering is available even when $E \sin \theta > E- 2 \omega$ (namely, $\sin \theta' > 1$) once we consider the analytic continuation of $\theta'$ as we will see later. 
Since the spins of massless quarks are assumed to be aligned into the momentum direction, we impose the projection operation on the spinors projected onto the $z$-axis as follows. 
\eqs{
  \frac12 (\mathbbm{1}+\sigma^3) e^{i\theta \frac{\sigma^2}{2}} A_- & = 0 \,, \\
  \frac12 (\mathbbm{1}+\sigma^3) e^{- i\theta' \frac{\sigma^2}{2}} B_- & = 0 \,, \qquad 
  \frac12 (\mathbbm{1}-\sigma^3) e^{-i\theta \frac{\sigma^2}{2}} B_+ = 0 \,, \\
  \frac12 (\mathbbm{1}+\sigma^3) e^{i\theta \frac{\sigma^2}{2}} C_- & = 0 \,, \qquad 
  \frac12 (\mathbbm{1}-\sigma^3) e^{i\theta' \frac{\sigma^2}{2}} C_+ = 0 \,.
  \label{eq:spinor_proj1}
}
Here, we note that the rotation angle of $B_-$ and $C_+$ is different from others due to the different energy of $B_-$ and $C_+$.
The matching conditions for the wave functions give the constant spinors in terms of the spinor of the incoming quark $A_-$. 
The detail of the calculation is given in \cref{app:mc_wf}.
In the limit of large $M_{(\pm)} R$, we find the constant spinors for the reflected quark and the transmitted quark as follows. 
\begin{align}
  \sqrt{E} e^{-i\theta \frac{\sigma^2}{2}} B_+ e^{i E R  \cos\theta} 
  & = \frac{\sin\left( \frac{\theta-\theta'}{2} \right)}{\cos\left( \frac{\theta+\theta'}{2} \right)}(i \sigma^2) \sqrt{E} e^{i\theta \frac{\sigma^2}{2}} A_- e^{- i E R  \cos\theta} \,, \\
  \sqrt{E-2\omega} e^{- i\theta' \frac{\sigma^2}{2}} B_- e^{i (E-2\omega) R  \cos\theta'} 
  & = i \frac{\cos \theta}{\cos\left( \frac{\theta+\theta'}{2} \right)} \sqrt{E} e^{i\theta \frac{\sigma^2}{2}} A_- e^{- i E R  \cos\theta} \,, \\
  \sqrt{E-2\omega} e^{i\theta' \frac{\sigma^2}{2}} C_+ e^{i (E-2\omega) R  \cos\theta} 
  & = 2 e^{- M_- R} (1+\tan^2 \beta) \frac{\cos\theta \sin\left( \frac{\theta-\theta'}{2} \right)}{\cos^2\left( \frac{\theta+\theta'}{2} \right)} \sigma^2 \sqrt{E} e^{i\theta \frac{\sigma^2}{2}} A_- e^{- i E R  \cos\theta} \,, \\
  \sqrt{E} e^{i\theta \frac{\sigma^2}{2}} C_- e^{i E R  \cos\theta} 
  & = 2 e^{- M_- R} (1+\tan^2 \beta) \frac{\cos\theta \cos \theta'}{\cos^2\left( \frac{\theta+\theta'}{2} \right)} \sqrt{E} e^{i\theta \frac{\sigma^2}{2}} A_- e^{- i E R  \cos\theta} \,.
\end{align}

We comment on the case with $E \sin \theta > E- 2 \omega$, where the quark injects into the Q-ball relatively parallel to the Q-ball surface. 
Since the incoming quark is just passing through on the Q-ball surface when $\theta = \pi/2$, we naively expect that the incoming quark is elastically reflected in this case.
When $E \sin \theta > E- 2 \omega$, we can analytically continue the sine function such that $\sin \theta' > 1$, and the analytically-continued cosine function is the pure imaginary. 
We can easily realize this analytic continuation by choosing $\theta' = \pi/2 - i \eta$ with a positive-real parameter $\eta$ (namely, $\sin\theta' = \cosh \eta$ and $\cos \theta' = i \sinh \eta$).
The prefactor of $B_+$ is computed as follows. 
\begin{align}
  \frac{\sin^2\left( \frac{\theta-\theta'}{2} \right)}{\cos^2\left( \frac{\theta+\theta'}{2} \right)}
  = \frac{1 - \sin\theta \cosh\eta + i \cos\theta \sinh\eta}{1 - \sin\theta \cosh\eta - i \cos\theta \sinh\eta} \,.
\end{align}
Hence, the absolute value of the prefactor of $B_+$ is unity, while the absolute value of the prefactor of $B_-$ is $e^{- i (E-2\omega) R  \cos\theta'} = e^{(E-2\omega) R \sinh \eta}$ at the Q-ball boundary.
The analytically-continued momentum for $B_-$ is given by 
\begin{align}
  p_{B-} = (E-2\omega, (E-2\omega) \cosh \eta, 0, - i (E-2\omega) \sinh \eta) \,, 
\end{align}
and hence, the right-handed anti-quark $B_-$ is no longer the propagating wave. 

The probability flux is defined in terms of the direction normal to the Q-ball surface. 
There are two non-zero constant spinors for the reflected quarks, $B_\pm$, while the constant spinors for the transmitted quarks are quite suppressed in the large $M_{(\pm)} R$ limit and we ignore them. 
Therefore, the probability flux after the scattering consists only of the reflected quarks in this limit, and we find 
\begin{align}
  (E-2\omega) \cos \theta' B_-^\dag B_- 
  + E \cos \theta B_+^\dag B_+
  = E \cos \theta A_-^\dag A_- \,.
\end{align}

Once we choose the oblique angle to be $\theta = \theta' = 0$, it corresponds to the scattering with normal incident.
In the limit of large Q-ball radius $R$, we find $B_+ = C_+ = 0$ and 
\begin{align}
  \sqrt{E-2\omega} B_- e^{i (E-2\omega)R} 
  & = i \sqrt{E} A_- e^{-i E R} \,, 
  \label{eq:B-1dscatteringA-}\\
  \sqrt{E} C_- e^{i E R} 
  & = 2(1 + \tan^2 \beta) e^{- 2 M_- R} \sqrt{E} A_- e^{-i E R} \,.
  \label{eq:C-1dscatteringA-}
\end{align}
The spin directions of both the transmitted and reflected quarks are the same as that of the incoming quarks. 
The momentum of the reflected quark is reversed compared to that of the incoming quark. 
This implies that the left-handed quark is reflected as the right-handed anti-quark as far as $E > 2 \omega$.
The spinor $C_-$ is suppressed by $e^{- 2 M_- R}$, and hence the transmission rate is also suppressed in the large $R$ limit.
Our results are consistent with Ref.~\cite{Kusenko:2004yw} where the authors have treated the Q-ball as the semi-infinite large wall located at $z \geq 0$.
Our result implies that baryons coming to the Q-ball at normal incident are reflected as anti-baryons with the probability of order unity.%
\footnote{
  In \cref{app:NRpot}, we discuss the non-relativistic scattering of nucleon--anti-nucleon system under the Q-ball background. 
}
The Q-ball does not get the angular momentum as far as we treat the Q-ball as a background and impose the spin conservation in the matter side.
One may consider the slowly rotating Q-ball~\cite{Almumin:2023wwi} as the excited state, and then we may be able to incorporate the spin flip of the incoming quark.%
\footnote{
  We cannot deal with such excitation of Q-ball in quantum-mechanical scattering, and thus approaches beyond ours are required.
  We can, however, find that the event rate for the process not changing helicity is suppressed in a naive Feynman-diagrammatic estimate of the processes.
  An incoming quark is reflected as a quark when one squark excitation (the squark field is expanded as its field value and excitation inside the Q-ball as $\varphi = \varphi_0 e^{-i \omega t} + \delta \varphi$) is absorbed and one squark excitation with the angular momentum of $\ell =1$ is emitted. 
  Therefore, the amplitude for the process is proportional to the energy of incoming quark, which arise from the quark propagator (dressed by the field value $\varphi_0$). 
  On the other hand, an incoming quark is reflected as an anti-quark when two squark excitations are emitted with the insertion of Majorana mass of gluino due to the chirality flip. 
}

We may consider the case where the left-handed incoming quark with the energy of $E - 2 \omega < 0$.
Since the energy of the reflected quark and the transmitted quark should be positive, the momentum assignment to the quarks, \cref{eq:p_assign1}, in the case of normal incident should be changed as follows. 
\begin{align}
  p_{B-} = (-E+2\omega, 0, 0, E-2\omega) \,, \qquad 
  p_{C+} = (-E+2\omega, 0, 0, -E+2\omega) \,,
\end{align}
and same as \cref{eq:p_assign1} with $\theta = \theta' = 0$ for other quarks. 
In this case, the positive-frequency solution for $B_-$ matches to the negative-frequency solution of the incoming quark $A_-$ to satisfy the matching conditions at $x$ for any time. 
We get the constant spinor $B_-$ in the large $R$ limit as follows.
\begin{align}
  \sqrt{2\omega-E} B_- e^{i (2\omega-E)R} 
  & = i \sqrt{E} (- i \sigma^2) A_-^\ast e^{-i E R} \,.
\end{align}
This relation seems to be naively regarded as the elastic scattering of the incoming quark.
However, this relation cannot be interpreted as the quantum mechanical scattering process. 
The amplitude for the process is given by the transition of the spinor $A_-$ into its conjugate $A_-^\ast$, but the unitary evolution of the state does not give this relation.
In addition, the energy of the system changes from $E$ to $2 \omega - E$.
We cannot deal with this process within the scattering problem in the relativistic quantum mechanics, and we must interpret this process in the quantum field theory (e.g., $A_-$ into its conjugate $A_-^\ast$ as in-state annihilation operator into out-state creation operator). 

We can consider the opposite helicity for the incoming quark, namely its spinor satisfying $(\mathbbm{1}-\sigma^3) e^{i\theta \frac{\sigma^2}{2}} A_+ = 0$. 
We obtain the similar result even in this case: the right-handed quark is completely reflected as the left-handed anti-quark when the incoming quark enters the Q-ball at normal incidence. 
We discuss this case in detail in \cref{app:mc_wf}.

We have considered the scattering of quarks on the Q-ball in the mass basis so far: for the scattering with $\omega = 0$ and normal incident ($\theta = 0$), the quark corresponding to the massive Majorana fermion inside the Q-ball is reflected as the anti-quark, while the quark corresponding to the massless fermion inside the Q-ball just passes through the Q-ball.
The $S$-matrix for the scattering is block-diagonalized with respect to the mass eigenstates: 
\begin{align}
  S' = 
  \begin{pmatrix}
    \mathbf{1}_4 & 0 \\
    0 & S_4
  \end{pmatrix} \,, \qquad 
  S_4 = i e^{-2 i E R}
  \begin{pmatrix}
    0 & 1 & 0 & 0 \\
    1 & 0 & 0 & 0 \\
    0 & 0 & 0 & 1 \\
    0 & 0 & 1 & 0 \\
  \end{pmatrix} \,.
\end{align}
Here, the in-state for each block is labelled by the helicity and the moving direction, so there is a $4 \times 4$ sub-matrix for each block.
The upper-left block corresponds to the massless fermion $\chi_0$ inside the Q-ball, while the lower-right block corresponds to the massive Majorana fermion $\psi$.
The detail of the non-trivial matrix $S_4$ is given in \cref{app:QM_Qball}. 
We obtain the $S$-matrix element in terms of the quarks in the original (interaction) basis by the field rotation. 
\begin{align}
  S = 
  \begin{pmatrix}
    \sin^2\alpha \mathbf{1}_4 + \cos^2 \alpha S_4 & \cos \alpha \sin\alpha (\mathbf{1}_4 - S_4) \\
    \cos \alpha \sin\alpha (\mathbf{1}_4 - S_4) & \cos^2\alpha \mathbf{1}_4 + \sin^2 \alpha S_4
  \end{pmatrix} \,.
\end{align}
This indicates that the quark $\psi_1$ ($\psi_2$) in the interaction basis can be reflected as not only its anti-quark $\overline \psi_1$ ($\overline \psi_2$) with the probability of $\cos^2 \alpha$ ($\sin^2 \alpha$) but also the anti-quark of the other species $\overline \psi_2$ ($\overline \psi_1$) with the probability of $\cos \alpha \sin\alpha$.

\subsection{Scattering in Spherical Wave \label{sec:spherical}}

Finally, we generalize the previous discussion to the three-dimensional scattering of quarks with the finite-size Q-ball.
The Q-ball is assumed to be spherically symmetric, and hence we can use the spherical wave states, which are eigenstates of energy and total angular momentum. 
For spin-zero states, we expand the wave function in terms of the spherical harmonics $Y_{\ell,m}(\theta,\phi)$ with angular coordinates $(\theta,\phi)$ and the eigenvalues of the orbital angular momentum operators $L^2$ and $L_z$ (denoted by $\ell$ and $m$).
We use the spin-$1/2$ analog of the spherical harmonics called the spinor spherical harmonics in this study. 
We discuss the spinor spherical harmonics in \cref{app:SSH} in detail.
The spinor spherical harmonics are simultaneous eigenfunctions of the total angular momentum $J^2$ and $J_z$, the orbital angular momentum $L^2$, and the spin $S^2$. 
The spinor spherical harmonics are given in a two-component spinor form by
\begin{align}
\Omega_{\ell\pm\frac12,\ell, m}(\theta,\phi) & 
  =  \frac{1}{\sqrt{2\ell+1}}
  \begin{pmatrix}
    \pm \sqrt{\ell \pm m+1/2} Y_{\ell,m-\frac12}(\theta,\phi)\\
    \sqrt{\ell \mp m+1/2} Y_{\ell,m+\frac12}(\theta,\phi)
  \end{pmatrix} \,.
\end{align}
Here, the first subscript corresponds to the quantum number of $J^2$, $j = \ell \pm 1/2$ . 
$\ell$ denotes the quantum number of $L^2$, while $m$ denotes the quantum number of $J_z$.
The spinor for given quantum numbers $j$ and $m$ is decomposed into the radial part and the angular part.
\begin{align}
  \psi_{j,m} (\mathbf{r}) = 
  \frac{1}{r} 
  \begin{pmatrix}
    i P_{\kappa}(r) \Omega_{j, j \pm 1/2, m}(\theta,\phi) \\
    Q_{\kappa}(r) \Omega_{j, j \mp 1/2, m}(\theta,\phi) 
  \end{pmatrix} 
  \equiv 
  \frac{1}{r}
  \begin{pmatrix}
    i p_{\kappa,m}(\mathbf{r}) \\
    q_{\kappa,m}(\mathbf{r}) 
  \end{pmatrix} \,.
\end{align}
Here, we introduce a label $\kappa = \mp (j+1/2)$ (for $j = \ell \pm 1/2$) for the eigenvalues of an operator $K = - 1 - 2 \mathbf{L} \cdot \mathbf{S}$ associated with the spin-orbit coupling.
For a free Dirac fermion, the radial functions $P_\kappa(r)$ and $Q_\kappa(r)$ satisfy the coupled differential equations:
\eqs{
  M P_\kappa + \left(\frac{d}{dr} - \frac{\kappa}{r}\right) Q_\kappa & = E P_\kappa \,, \\
  -\left(\frac{d}{dr} + \frac{\kappa}{r}\right) P_\kappa - M Q_\kappa & = E Q_\kappa \,.
}
Here, $E$ is the energy of the Dirac fermion, and $M$ denotes the mass of the Dirac fermion. 
When the scalar background oscillates, $\varphi = \varphi_0 e^{- i \omega t}$, the wave function inside the Q-ball absorbs the phase factor by the field redefinition. 
The time-dependent wave function for the four-component Majorana fermions in the Dirac representation (with energy $E$) are written as follows.
\begin{align}
  \Psi^\mathrm{in}_{j,m} (\mathbf{r},t) = 
  \frac{e^{-iEt}}{r} e^{i \omega t \gamma_5}
  \begin{pmatrix}
    i p_{\kappa,m}(\mathbf{r}) \\
    q_{\kappa,m}(\mathbf{r}) 
  \end{pmatrix} \,, \quad
  \Psi^\mathrm{out}_{j,m} (\mathbf{r},t) = 
  \frac{e^{-iEt}}{r} 
  \begin{pmatrix}
    i p_{\kappa,m}(\mathbf{r}) \\
    q_{\kappa,m}(\mathbf{r}) 
  \end{pmatrix} \,.
  \label{eq:wf_oscbkg}
\end{align}
We also have the charge-conjugated part inside and outside Q-ball.
Once we obtain the solution to the matching conditions among $\Psi^\mathrm{in \,, out}_{j,m}$, it is trivial that the charge-conjugated part satisfy the matching condition. 
Hence, we consider the matching conditions for the Dirac spinors $\Psi^\mathrm{in \,, out}_{j,m}$ in the following. 

We assign different momenta to different massless quarks with the different helicity similarly to the previous subsection. 
Summing a specific linear combination over $\kappa$ gives a definite chirality of the plane-wave solution for a massless fermion (shown in \cref{app:SSH}). 
We define the wave functions leading to definite chirality after the summation as follows. 
\begin{align}
  \Psi_{+,j,m} & = 
  \frac{e^{-iEt}}{r} 
  \begin{pmatrix}
    i p_{\varkappa,m}(\mathbf{r}) + i p_{- \varkappa,m}(\mathbf{r}) \\
    q_{\varkappa,m}(\mathbf{r}) + q_{- \varkappa,m}(\mathbf{r}) 
  \end{pmatrix} \,, & \quad 
  \Psi_{-,j,m} & = 
  \frac{e^{-iEt}}{r} 
  \begin{pmatrix}
    - i p_{\varkappa,m}(\mathbf{r}) + i p_{- \varkappa,m}(\mathbf{r}) \\
    - q_{\varkappa,m}(\mathbf{r}) + q_{- \varkappa,m}(\mathbf{r}) 
  \end{pmatrix} \,, 
  \label{eq:wf_chirality}
\end{align}
Here, $\varkappa$ is positive integer, and $\pm$ indicates the chirality. 

The incoming quark is assumed to have a definite chirality as before, and we construct other fermions by solving the matching conditions for the wave functions at the Q-ball boundary $(r = R)$.
The wave functions inside Q-ball should be regular at the origin, while the wave functions outside Q-ball should not grow at large $r$. 
In contrast to the previous subsection, we have only to take into account the one wave function inside Q-ball for each massive fermion with a specific chirality and $m$.
Thus, the matching conditions for the wave functions at $r = R$ are 
\begin{align}
  \Psi_{A} + \sum_{m,\pm} \Psi_{B_{\pm,m}} & = \sum_{m, \pm}\left( \sin \beta X_{D_{\pm,m}} + \cos \beta X_{F_{\pm,m}} \right) \,, 
  \label{eq:mc_3d_quark} \\
  \sum_{m, \pm} \Lambda_{C_{\pm,m}} & = \sum_{m, \pm} \left( \cos \beta X_{D_{\pm,m}} - \sin \beta X_{F_{\pm,m}} \right) \,.
  \label{eq:mc_3d_gluino}
\end{align}
Here, the index for $j$ is implicit. 
$\Psi_A$ denotes the wave function for the incoming quark with a definite chirality and thus $m$.
$\Psi_{B_{\pm,m}}$ and $\Lambda_{C_{\pm,m}}$ are the wave functions for the reflected quark and gluino, respectively.
$X_{{D,F}_{\pm,m}}$ denote the wave functions for the massive fermions inside Q-ball.

Now, we consider a left-handed chirality for the incoming quark (with $m = -1/2$ for all fermions).
We assume the four-momentum of each fermions as follows. 
\eqs{
  p_{A_-} & = (E, - E \hat{\mathbf{r}}) \,, \qquad 
  p_{B_{+}} = (E-2\omega, (E-2\omega) \hat{\mathbf{r}}) \,, \qquad 
  p_{B_{-}} = (E, E \hat{\mathbf{r}}) \,, \\
  \kappa_{C_{\pm}} & = (E-\omega, i M_\lambda \hat{\mathbf{r}}) \,, \qquad 
  \kappa_{D_{\pm}} = (E-\omega, i M_+ \hat{\mathbf{r}}) \,, \qquad 
  \kappa_{F_{\pm}} = (E-\omega, i M_- \hat{\mathbf{r}}) \,.
  \label{eq:E_assign}
}

The outgoing and incoming Dirac fermions are described by the spherical Hankel functions $h_\kappa^{(1)}(x)$ and $h_\kappa^{(2)}(x)$, which behave as $e^{i x}/x$ and $e^{- i x}/x$ at large $x$, respectively.
We choose the overall coefficients of $P_\kappa$ and $Q_\kappa$ to be convenient for constructing the plain-wave solution in terms of the spinor spherical harmonics. 
Therefore, the radial functions for the massless quarks for $\varkappa > 0$ ($j = \ell + 1$) are 
\begin{align}
  P_\varkappa^{A_{-}} & = - \sqrt{E} r h_\varkappa^{(2)} (E r) A_{-} \,, & 
  Q_\varkappa^{A_{-}} & = \sqrt{E} r h_{\varkappa-1}^{(2)} (E r) A_{-} \,, 
  \label{eq:PQ_A} \\
  P_\varkappa^{B_{+}} & = - \sqrt{E-2\omega} r h_\varkappa^{(1)} [(E-2\omega) r] B_{+} \,, & 
  Q_\varkappa^{B_{+}} & = \sqrt{E-2\omega} r h_{\varkappa-1}^{(1)} [(E-2\omega) r] B_{+} \,, \\
  P_\varkappa^{B_{-}} & = - \sqrt{E} r h_\varkappa^{(1)} (E r) B_{-} \,, & 
  Q_\varkappa^{B_{-}} & = \sqrt{E} r h_{\varkappa-1}^{(1)} (E r) B_{-} \,, 
\end{align}
while these for $ - \varkappa < 0$ ($j = \ell - 1$) are 
\begin{align}
  P_{-\varkappa}^{A_{-}} & = - i \sqrt{E} r h_{\varkappa-1}^{(2)} (E r) A_{-} \,, & 
  Q_{-\varkappa}^{A_{-}} & = - i \sqrt{E} r h_{\varkappa}^{(2)} (E r) A_{-} \,, \\
  P_{-\varkappa}^{B_{+}} & = - i \sqrt{E-2\omega} r h_{\varkappa-1}^{(1)} [(E-2\omega) r] B_{+} \,, & 
  Q_{-\varkappa}^{B_{+}} & =  - i \sqrt{E-2\omega} r h_{\varkappa}^{(1)} [(E-2\omega) r] B_{+} \,, \\
  P_{-\varkappa}^{B_{-}} & = - i \sqrt{E} r h_{\varkappa-1}^{(1)} (E r) B_{-} \,, & 
  Q_{-\varkappa}^{B_{-}} & = - i \sqrt{E} r h_{\varkappa}^{(1)} (E r) B_{-} \,.
\end{align}
Here, $A_-$ and $B_{\pm}$ denote the coefficients for the quark wave functions, not spinors unlike the previous subsection. 
We use the same coefficients for both $\varkappa$ and $-\varkappa$ for the wave functions with the definite chirality as shown in \cref{eq:wf_chirality} (see also \cref{app:SSH} for the construction of the wave functions with a definite chirality).

Since we assume that the injected energy $E$ is less than the gluino mass and the field value of the scalar field, the wave function for the heavy fermions is given by a linear combination of the modified spherical Bessel functions. 
Gluinos propagate outside the Q-ball and are written by the use of the spherical Bessel functions if the energy of incoming quark is sufficiently large.
Otherwise (like our case), the wave function of gluinos should not grow at $r \to \infty$ and be described by the second kind of the modified spherical Bessel function $k_n(x)$.
Meanwhile, the wave functions for the fermions in the mass basis inside the Q-ball should be regular at $r \to 0$, and hence the wave functions are described by the first kind of the modified spherical Bessel function $i_n(x)$.
The explicit form of the radial functions for the massive fermions is given by
\begin{align}
  P^{C_{\pm}}_\varkappa(r) & = - \sqrt{M_\lambda} r k_{\varkappa}(M_\lambda r) C_{\pm} \,, \quad &
  Q^{C_{\pm}}_\varkappa(r) & = - \sqrt{M_\lambda} r k_{\varkappa-1}(M_\lambda r) C_{\pm} \,, \\
  P^{D_{\pm}}_\varkappa(r) & = \sqrt{M_+} r i_{\varkappa}(M_+ r) D_{\pm} \,, \quad &
  Q^{D_{\pm}}_\varkappa(r) &= - \sqrt{M_+} r i_{\varkappa-1}(M_+ r) D_{\pm} \,, \\    
  P^{F_{\pm}}_\varkappa(r) & = \sqrt{M_-} r i_{\varkappa}(M_- r) F_{\pm} \,, \quad &
  Q^{F_{\pm}}_\varkappa(r) & = \sqrt{M_-} r i_{\varkappa-1}(M_- r) F_{\pm} \,,
\end{align}
for positive integer $\varkappa$, and 
\begin{align}
  P^{C_{\pm}}_{-\varkappa}(r) & = - \sqrt{M_\lambda} r k_{\varkappa-1}(M_\lambda r) C_{\pm} \,, \quad &
  Q^{C_{\pm}}_{-\varkappa}(r) &= - \sqrt{M_\lambda} r k_{\varkappa}(M_\lambda r) C_{\pm} \,, \\
  P^{D_{\pm}}_{-\varkappa}(r) & = \sqrt{M_+} r i_{\varkappa-1}(M_+ r) D_{\pm} \,, \quad &
  Q^{D_{\pm}}_{-\varkappa}(r) & = - \sqrt{M_+} r i_{\varkappa}(M_+ r) D_{\pm} \,, \\    
  P^{F_{\pm}}_{-\varkappa}(r) & = \sqrt{M_-} r i_{\varkappa-1}(M_- r) F_{\pm} \,, \quad &
  Q^{F_{\pm}}_{-\varkappa}(r) & = \sqrt{M_-} r i_{\varkappa}(M_- r) F_{\pm} \,,
  \label{eq:PQ_F}
\end{align}
for negative integer $-\varkappa$.

\begin{figure}[t]
  \centering
    \includegraphics[width=0.45\linewidth]{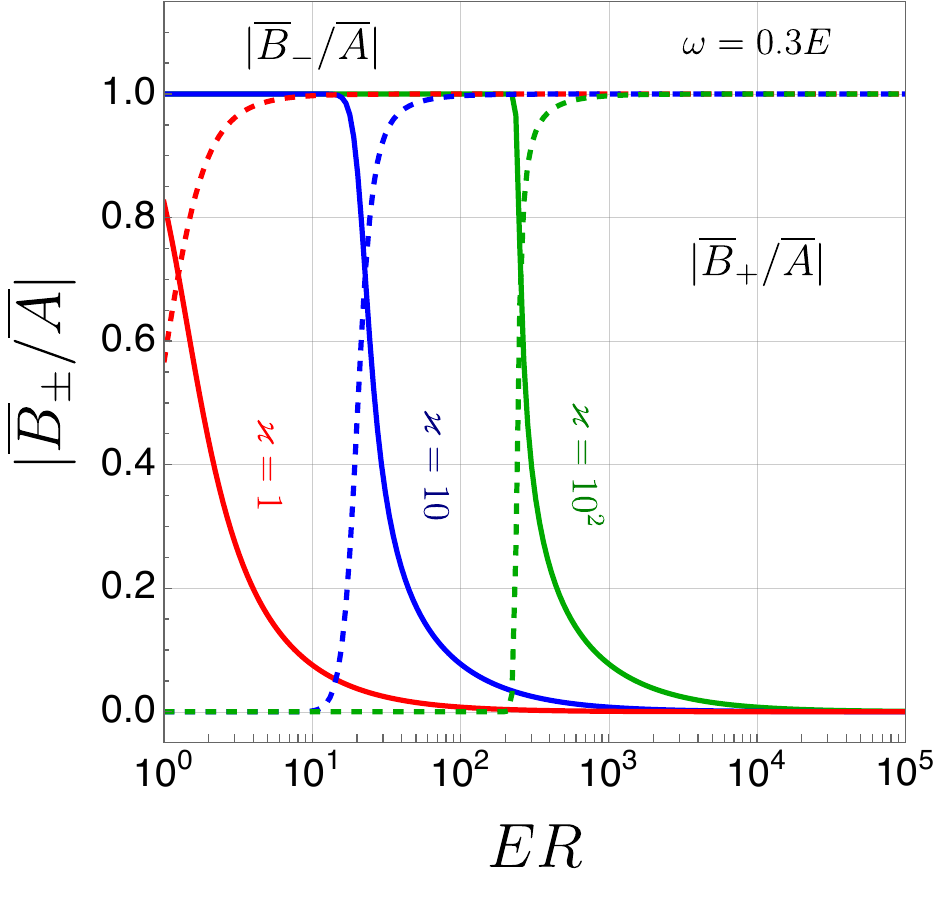}
    \includegraphics[width=0.45\linewidth]{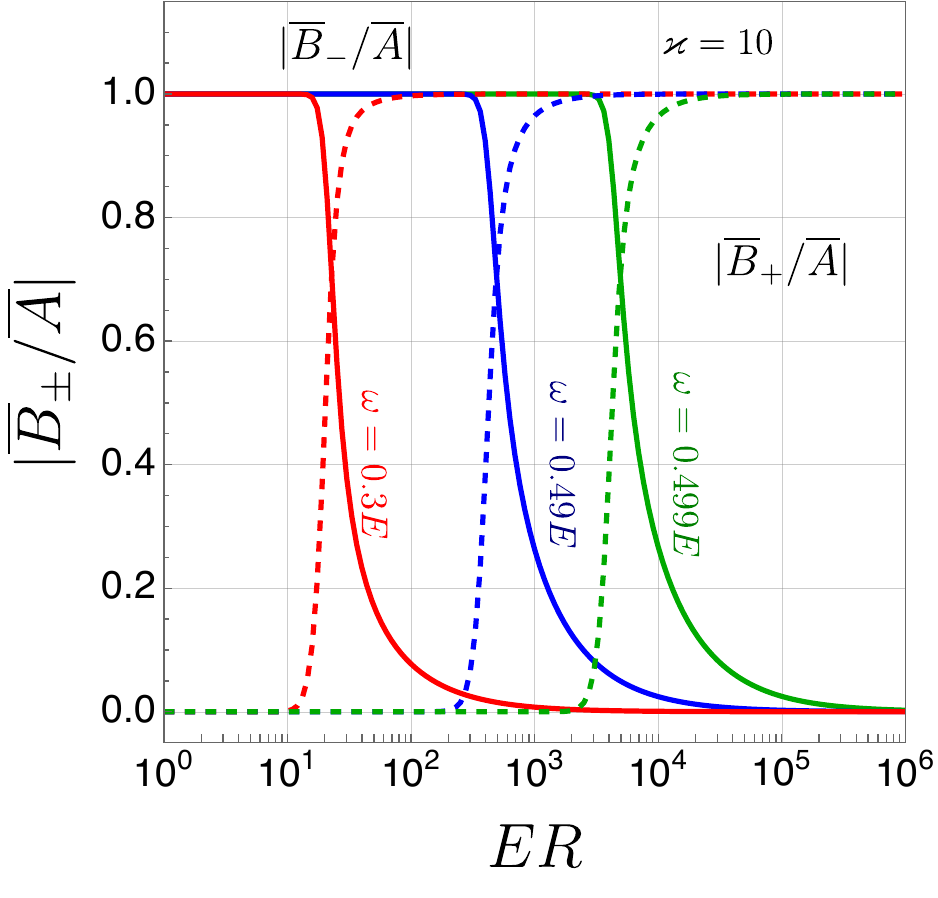}
  \caption{
    The $ER$-dependence of the absolute values of the coefficients $\overline B_{\pm}$ for fixed parameters: fixed $\varkappa = 10$ in the right panel, while fixed $\omega/E = 0.3$ in the left panel.
    $\overline B_{-}$ is depicted as the solid lines, while $\overline B_{+}$ is depicted as the dashed lines.
  }
  \label{fig:reflectedquark}
\end{figure} 

We find the coefficients after matching the wave functions at the Q-ball boundary. 
We assume $M_{(\pm)} R \gg \varkappa$ and use the asymptotic form of the modified spherical Bessel functions at large $M_{(\pm)} R$.
Then, we find the solution for the reflected quark as follows:
\begin{align}
  B_{+} & = \sqrt{\frac{E}{E-2\omega}} \frac{h_{\varkappa-1}^{(1)} (E R) h_\varkappa^{(2)} (E R) - h_\varkappa^{(1)} (E R) h_{\varkappa-1}^{(2)} (ER)}{h_{\varkappa-1}^{(1)} (E R) h_{\varkappa-1}^{(1)} [(E-2\omega)R] - h_\varkappa^{(1)} (E R) h_\varkappa^{(1)} [(E-2\omega)R]} A_{-} \,, \\
  B_{-} & = - \frac{h_{\varkappa-1}^{(2)} (E R) h_{\varkappa-1}^{(1)} [(E-2\omega)R] - h_\varkappa^{(2)} (E R) h_\varkappa^{(1)} [(E-2\omega)R]}{h_{\varkappa-1}^{(1)} (E R)h_{\varkappa-1}^{(1)} [(E-2\omega)R] - h_\varkappa^{(1)} (E R) h_\varkappa^{(1)} [(E-2\omega)R]} A_{-} \,.
  \label{eq:B_3d}
\end{align}
We consider the analytic solution in several limits: $E R \,, (E - 2 \omega) R \ll \varkappa$, $(E - 2 \omega) R \ll \varkappa \ll E R$, and $\varkappa \ll E R \,,(E - 2 \omega) R$.
The asymptotic forms of these coefficients at the leading order of expansion parameters for each limit are as follows. 
\begin{align}
  B_{+} & = \frac{2 i}{[(2\varkappa-1)!!]^2} \sqrt{\frac{E-2\omega}{E}} (E R)^{\varkappa} [(E - 2\omega)R]^{\varkappa} A_{-} \,, \nonumber \\
  B_{-} & = A_{-} \,, & E R \,,(E - 2 \omega) R \ll \varkappa  \,, 
  \label{eq:Blimit_1} \\
  B_{+} & = - i^\varkappa \sqrt{\frac{E-2\omega}{E}} \frac{[(E-2\omega)R]^\varkappa}{(2\varkappa-1)!!} \frac{e^{- i E R} A_{-}}{i (2\varkappa-1)} \,, \nonumber \\
  B_{-} & = - (- 1)^{\varkappa+1} e^{- 2 i E R} A_{-}\,, & (E - 2 \omega) R \ll \varkappa \ll E R \,, 
  \label{eq:Blimit_2}\\
  B_{+} & = i (-1)^\varkappa \sqrt{\frac{E-2\omega}{E}} e^{- 2 i (E - \omega) R} A_{-} \,, \nonumber \\
  B_{-} & = i (-1)^\varkappa \frac{2 \varkappa \omega R }{E R (E - 2 \omega) R} e^{- 2 i E R} A_{-} \,, & \varkappa \ll E R \,, (E - 2 \omega) R \,.
  \label{eq:Blimit_3}
\end{align}
An orbital angular momentum for classical scattering is given by the product of the injected energy $E$ and the impact parameter $b$.
When the orbital angular momentum satisfies $E R \ll \ell \simeq \varkappa$ (i.e., $R \ll b$), the incoming quark does not collide with the Q-ball. 
As shown in the asymptotic forms, $B_{+}$ is quite suppressed by $(E R)^{\varkappa}$ and $[(E - 2\omega)R]^{\varkappa}$ while $B_{-}$ is not.
Therefore, the final-state quark has a left-handed chirality in the $E R \,,(E - 2 \omega) R \ll \varkappa$ limit. 
In the region of $(E - 2 \omega) R \ll \varkappa \ll E R$  (i.e., $b \ll R$), the incoming quark collides with the Q-ball, but this is scattering with a sizable oblique angle. 
The asymptotic form of $B_{+}$ in the region is only suppressed by $[(E - 2\omega)R]^{\varkappa}$, and hence the scattering is interpreted as the elastic scattering. 
Finally, as $\ell \simeq \varkappa \ll E R \,,(E - 2 \omega) R$, the incoming quark enters almost perpendicular to the Q-ball surface.
Contrary to the previous two cases, the asymptotic form of $B_{-}$ is only suppressed by $\varkappa/ER$.
Therefore, as we saw in \cref{sec:infinite}, the reflected quark has a right-handed chirality, namely the incoming quark is reflected as the right-handed anti-quark. 

\cref{fig:reflectedquark} shows the $E R$ dependence of the absolute value of the coefficients $B_{\pm}$.
All of coefficients are normalized as $\overline A \equiv A/\sqrt{E}$, $\overline B_{+} \equiv B_{+}/\sqrt{E-2\omega}$, and $\overline B_{-} \equiv B_{-}/\sqrt{E}$.
In this figure, we plot $\overline B_{+}$ as dashed lines and $\overline B_{-}$ as solid lines. 
The chemical potential is fixed to be $\omega = 0.3 E$ and take $\varkappa = 1$ (red), $\varkappa = 10$ (blue), and $\varkappa = 10^2$ (green) in the left panel. 
As expected, the constant $\overline B_{-}$ begins to decay at $ER \simeq \varkappa$ and goes to zero at large $E R$, while the constant $\overline B_{+}$ begins to develop at $ER \simeq \varkappa$ and goes to unity at large $E R$.
The total angular momentum $\varkappa$ is fixed $\varkappa = 10$, and we take $\omega = 0.3 E$ (red), $\omega = 0.49 E$ (blue), and $\omega = 0.499 E$ (green) in the right panel.
Even if $E R \gtrsim \varkappa$, the reflected quark has a left chirality unless the injected energy is sufficiently larger than the chemical potential, namely $(E - 2 \omega) R \gg \varkappa$.

\begin{figure}[t]
  \centering
    \includegraphics[width=0.45\linewidth]{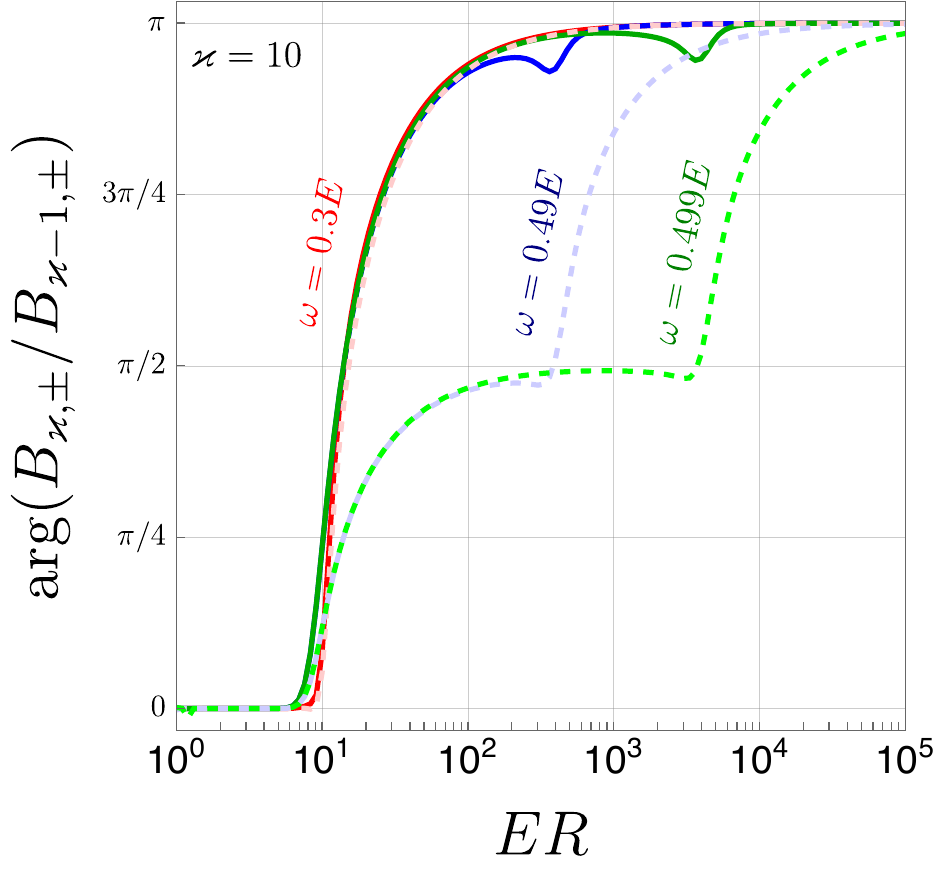}
  \caption{
    The relative phase of the coefficients $B_{\pm}$ with fixed $\varkappa = 10$.
    $B_{-}$ is depicted as the solid lines with the same color code as \cref{fig:reflectedquark}, while $B_{+}$ is depicted as the dashed lines with light colors.
  }
  \label{fig:arg}
\end{figure}

We show the phase of the coefficients $B_{\pm}$ with fixed $\varkappa$ in \cref{fig:arg}.
There, we take the ratio of the coefficients with $\varkappa$ and $\varkappa-1$ to remove the common phases, which approach to $e^{- i (E - \omega) R}$ and $e^{- i E R}$ in the large $E R \,, (E - 2 \omega) R$ limits. 
We note that we add the subscript $\varkappa$ in $B_{\varkappa, \pm}$ to show $\varkappa$-dependence explicitly. 
As shown in the asymptotic forms, the ratio of the coefficients $B_{\varkappa, -}/B_{\varkappa-1, -}$ approximately goes to $(-1)$ as $E R \gg \varkappa$ even for small $(E-2\omega)R$ as expected from the hard sphere scattering~(see Ref.~\cite{Sakurai:2011zz}).
Meanwhile, the ratio $B_{\varkappa, +}/B_{\varkappa-1, +}$ approaches to $i$ in the region $E R \gg \varkappa \gg (E - 2 \omega) R$, and then approximately goes to $(-1)$ as $E R  (E - 2 \omega) R \gg \varkappa$. 
Therefore, as shown in the figure, the relative phase goes to $\pi/2$ as $E R > \varkappa$, and then it approaches to $\pi$ for $(E-2\omega) R > \varkappa$ again as expected from the hard sphere scattering.

As with the previous subsection, we consider the $S$-matrix for the scattering with $\omega = 0$. 
For given $\varkappa$ (in other words, the total angular momentum $j$), the $S$-matrix for the Q-ball scattering with quark is block-diagonalized with respect to the mass eigenstates:
\begin{align}
  S'^\varkappa = 
  \begin{pmatrix}
    \mathbf{1}_2 & 0 \\
    0 & S^\varkappa_2
  \end{pmatrix} \,.
\end{align}
Here, the upper block corresponds to the scattering of the massless fermion inside the Q-ball, while the lower block corresponds to that of the massive fermion. 
The in-state for each block is labelled by the helicity, so each is a $2 \times 2$ sub-matrix.
The sub-matrix for $\varkappa$ is given by
\begin{align}
  S^\varkappa_2 = (-1)^{\varkappa} i e^{-2iER}
  \begin{pmatrix}
    0 & 1 \\
    1 & 0 \\
  \end{pmatrix} \,. 
\end{align}

Similarly to the previous case, we can write the $S$-matrix in terms of the interaction basis of quarks outside the Q-ball by the field rotation. 
\begin{align}
  S^\varkappa = 
  \begin{pmatrix}
    \sin^2\alpha \mathbf{1}_2 + \cos^2 \alpha S^\varkappa_2 & \cos \alpha \sin\alpha (\mathbf{1}_2 - S^\varkappa_2) \\
    \cos \alpha \sin\alpha (\mathbf{1}_2 - S^\varkappa_2) & \cos^2\alpha \mathbf{1}_2 + \sin^2 \alpha S^\varkappa_2
  \end{pmatrix} \,.
\end{align}
Again, this indicates that the incident quark can be reflected as not only its anti-quark but also the anti-quark of the other species. 

\section{Implication for Nucleon Scattering with Q-ball \label{sec:nucleon}}

In this section, we discuss the implication for the nucleon scattering with the Q-ball from the observation in the last section. 
We naively expect that the incoming nucleon is reflected as nucleon, pions, or anti-nucleon. 
However, due to the energy cost by the chemical potential, it is challenging to be reflected as an anti-nucleon in gravitationally bounded halos such as our Milky Way.
The Q-ball DM scatters off the nucleon with a typical velocity of $v \sim 10^{-3}$, and hence the kinetic energy of the nucleon in the Q-ball rest frame is 
\begin{align}
  E_\mathrm{kin} = \frac12 m_N v^2 \simeq 0.47 \,\mathrm{keV} \left( \frac{v}{10^{-3}} \right)^2 \,. 
\end{align}
Meanwhile, the energy cost is $\omega \simeq 20 \, \mathrm{MeV} \left( 10 \, \mathrm{fm} / R \right)$.
As shown in the previous section, the nucleon can be reflected as the anti-nucleon only when the scattering process can pay the energy cost $3 \times 2 \omega$. 
It is impossible to emit the anti-nucleon due to the energy cost.
Meanwhile, it would be possible that a constituent quark inside a nucleon is reflected as an anti-quark since it has a momentum of $\Lambda_\mathrm{QCD} \simeq 200 \,\mathrm{MeV}$, which can be larger than the energy cost $2 \omega$.
Due to the linear potential by the strong interaction, the single anti-quark emission is prohibited. 
This process can be interpreted as excitation of the Q-ball via the nucleon absorption.
The excited (or intermediate) Q-ball can decay into the more stable configuration after emitting a nucleon or pions (note that the relaxed Q-ball configuration is fermionic rather than bosonic in the latter case). 
Since our treatment does not incorporate such a back-reaction process, studying it is beyond the scope of this work. 

The nucleon scattering, with conversion into pions or a nucleon, would charge up and down the Q-ball.
If the Q-ball is negatively charged, the Coulomb interaction between proton and Q-ball is attractive, and the charged Q-ball scatters with protons until the Q-ball is positively charged, which is an attractor for the following reason. 
Let assume that the Q-ball has the radius $R$ and the electromagnetic (positive) charge $Z_Q$. 
The maximal charge of the Q-ball that can scatter with proton is classically determined when the Coulomb potential equals to the kinetic energy of proton. 
Then, we find the maximal charge of the Q-ball: 
\begin{align}
  Z_Q \simeq \frac{m_N R}{2 \alpha} v^2 \simeq 0.3 \left( \frac{R}{10^3 \, \mathrm{fm}} \right) \left( \frac{v}{10^{-3}} \right)^2 \,.
\end{align}
Here, $\alpha^{-1} \simeq 137$ is the fine structure constant. 
Even if the charge is larger than the maximal value, the Q-ball scatters with the proton via the tunneling, and the scattering cross section between the proton and the Q-ball is suppressed by the Coulomb barrier. 
The scattering wave function is estimated by the WKB approximation in this case. 
In the limit of the small DM velocity limit, the Coulomb barrier factor for the wave function is approximately given by (see Ref.~\cite{2015lqm..book.....W})
\begin{align}
  \exp \left[ - \int^{R_2}_R dr \sqrt{2 m_N \left( \frac{Z_Q \alpha}{r} - \frac{Z_Q \alpha}{R_2} \right)} \right] \simeq \exp \left[ - \frac{\pi Z_Q \alpha}{v} \left( 1 - \frac{4}{\pi} \sqrt{\frac{R}{R_2}} \right) \right] \,.
\end{align}
Here, $R_2 > R$ denotes the point where the kinetic energy equals to the potential energy, 
\begin{align}
  R_2 = \frac{2 Z_Q \alpha}{m_N v^2} \simeq 3.07 Z_Q \, \mathrm{pm} \left( \frac{10^{-3}}{v} \right)^2 \,.
\end{align}
The Coulomb barrier suppresses the scattering cross section as far as the Q-ball radius satisfies $R \lesssim R_2$. 
The scattering cross section consists of the geometric one and the Coulomb barrier factor, and we find
\begin{align}
  \sigma & \simeq \pi R^2 \exp\left[ - \frac{2 \pi Z_Q \alpha}{v} \left( 1 - \frac{4}{\pi} \sqrt{\frac{R}{R_2}} \right) \right] \nonumber \\
  & \simeq 3 \times 10^{-24} \, \mathrm{cm}^2 \left( \frac{R}{10 \, \mathrm{fm}} \right)^2 \left( 1.2 \times 10^{-20} \right)^{Z_Q} \,.
\end{align}
We assume the $R/R_2$ correction is negligible in order to obtain the last numerical expression. 

\begin{figure}[t]
  \centering
    \includegraphics[width=0.45\linewidth]{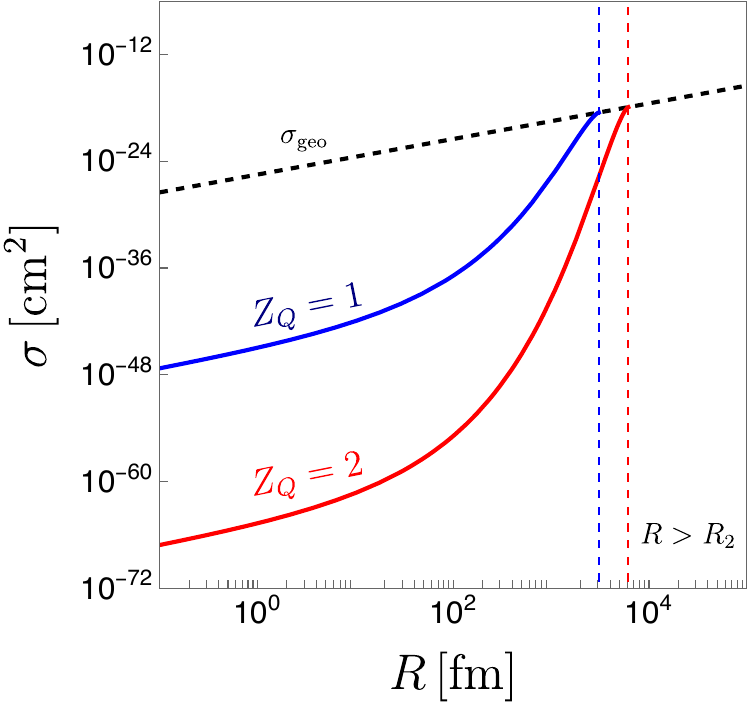}
  \caption{
    The proton--charged Q-ball scattering cross section. 
    The relative velocity of the Q-ball and the proton is assumed to be $v \simeq 10^{-3}$.
    The electromagnetic charge of the Q-ball $Z_Q$ is fixed for each solid lines: $Z_Q = 1$ (blue) and $Z_Q = 2$ (red).
    The thin-dashed vertical lines show the point where $R = R_2$ for each $Z_Q$ with the same color code as the solid lines.
    The geometric cross section $\sigma_\mathrm{geo} = \pi R^2$ is depicted as the black dashed line. 
    The radius $R$ of our interest is $10\,\mathrm{fm} \lesssim R \lesssim \mathcal{O}(1) \, \mathrm{nm}$~\cite{Kasuya:2015uka}.
  }
  \label{fig:C_Barrier}
\end{figure}

\cref{fig:C_Barrier} depicts the radius $R$ dependence of the scattering cross section for fixed the electromagnetic charge of the Q-ball $Z_Q$.
We assume the relative velocity of the Q-ball and the proton to be $v \simeq 10^{-3}$. 
The scattering cross section is quite suppressed with the Coulomb barrier factor of $\sim 10^{-20 Z_Q}$ at small $R$. 
The Coulomb barrier starts to change by an order of magnitude around $R \simeq 10^{-4} R_2$, and then the cross section approaches to the geometric one (shown as the black-dashed line) as $R$ gets larger and the barrier disappears at $R = R_2$.

The discussion of the electromagnetic charge of Q-ball also depends on the corresponding flat direction: we focus on the $udd$ flat direction so far. 
One may consider the Q-balls associated with other flat directions, such as with $u u d e$ and $q q q \ell$ which carry not only the baryon number but also the lepton number~\cite{Gherghetta:1995dv,Dine:1995kz}.
In such a case, the Q-ball configuration is stable only when the chemical potential is less than lepton mass ($m_e$ for the $u u d e$ Q-ball, while neutrino mass for the $q q q \ell$ Q-ball).
When we consider the $uude$ ($qqq\ell$) flat direction, the Q-ball scattering with the electron would make the positively-charged Q-ball discharged by converting $e^- \to e^+$.
Thus, if the injected energy satisfies $E > 2 \omega$ for the electron scattering, the charged Q-ball is easily discharged. 
It would also be possible that the injected proton is reflected as the positron.

The charged Q-ball can also be discharged by forming the bound state with the free electrons. 
The bound state formation cross section at large $\eta = Z_Q\alpha/v$ limit is approximately given by (see Ref.~\cite{kotelnikov2019electron} for example)
\begin{align}
  \sigma_\mathrm{BSF} \simeq \frac{32 \pi}{3 \sqrt3} \frac{\eta^4 \alpha^3 a_B^2}{n (\eta^2 + n^2)} 
  & \simeq 1.1 \times 10^4 \, \mathrm{barn} \left( \frac{10^{-3}}{v} \right)^2 \,, \qquad (\eta \gg n) \,, \\
  & \simeq 6.0 \times 10^4 \, \mathrm{barn} \left( \frac{10^{-3}}{v} \right)^4 \,, \qquad (\eta \ll n) \,.
\end{align}
Here, $a_B = Z_Q/m_e \alpha$ is the Bohr radius and $n$ is the principal quantum number.
The typical radius for this system is given by 
\begin{align}
  r_c(n) =
  \frac{n^2}{m_e \alpha Z_Q} \simeq 52.9 \frac{n^2}{Z_Q} \mathrm{pm} \,,
\end{align}
and hence the typical radius can be smaller than $R_2$ when 
\begin{align}
  Z_Q > 4.15 n \left( \frac{v}{10^{-3}} \right) \,.
\end{align}
In such a case, the electromagnetic force is screened for $r > r_c(n) $, and the Coulomb barrier factor is modified as follows. 
\begin{align}
  \exp \left[ - \int^{r_c(n) }_R dr \sqrt{2 m_N \left( \frac{Z_Q \alpha}{r} - \frac{Z_Q \alpha}{R_2} \right)} \right] \,.
\end{align}
Similarly to \cref{fig:C_Barrier}, it leads to a huge suppression of the scattering cross section by the Coulomb barrier unless the Q-ball radius $R$ is close to $r_c(n)$.

\section{Conclusion and Discussion \label{sec:conclusion}}

It is important to understand the interactions between the Q-ball DM and matters, in order to investigate the experimental signals from the Q-ball DM~\cite{Arafune:2000yv}, e.g., in paleo-detectors~\cite{Baum:2023cct}. 
We have studied the scattering of the Q-ball DM and (colorless) quarks with taking into account the chemical potential $\omega$, which is overlooked in the literature. 
The incoming quark is reflected as the anti-quark when the injected energy of the incoming quark exceeds the energy cost for increasing the baryon number of the Q-ball, $2 \omega$.
As for the scattering between the Q-ball DM with the radius of $\mathcal{O}(1) \, \mathrm{fm}$ and the nucleons in our Milky Way (namely, with relative speed of $v \sim 10^{-3}$), the nucleon is not reflected as the anti-nucleon due to the energy cost.

The Q-ball DM can be electromagnetically charged via absorption of nucleons and subsequent emission of a nucleon or pions in our Milky Way.
Once the Q-ball is positively charged, the scattering cross section to proton is quite suppressed by the Coulomb barrier and the experimental signals are different from the neutral Q-ball~\cite{Arafune:2000yv}.
It is worthwhile to clarify what a fraction of Q-balls is electromagnetically charged in our Milky Way by taking into account the dependence on the flat directions. 

\section*{Acknowledgements}

A. K. thanks Kohta Murase and Shigenobu Hirose for encouraging us to work on this study, and also Alexander Kusenko for sharing comments on the draft.
A. K. acknowledges partial support from Norwegian Financial Mechanism for years 2014-2021, grant nr 2019/34/H/ST2/00707; and from National Science Centre, Poland, grant DEC-2018/31/B/ST2/02283.
The work of T. K. is supported in part by the National Science Foundation of China under Grant Nos. 11675002, 11635001, 11725520, 12235001, and the NSFC Research Fund for International Scientists Grant No.~12250410248.

\appendix
\section{Q-ball Properties \label{app:Q-potential}}

In this appendix, we discuss the scalar potential for Q-ball, which gives the relations among macroscopic parameters of Q-ball, such as the radius $R$, the mass $M_Q$, the total charge $Q$, and the chemical potential $\omega$.

The scalar potential is composed of two contributions: one is from the gauge mediation contribution, and another is from the gravity mediation contribution~\cite{deGouvea:1997afu}. 
\begin{align}
  V = M_F^4 \left[ \ln \left( 1 + \frac{g |\phi|}{M_\mathrm{mess}} \right)\right]^2 + m_{3/2}^2 \left( 1 + K \ln \frac{|\phi|^2}{\mu^2} \right) |\phi|^2 \,.
\end{align}
Here, $M_\mathrm{mess}$ is the messenger mass, $M_F$ is related to the $F$-term supersymmetry breaking, and $m_{3/2}$ is the gravitino mass. 
$g$ denotes the gauge coupling constant, and $K$ is negative and arises one-loop corrections (therefore, $|K| \simeq O(0.01 \mathrm{-} 0.1)$).
These dimensionful parameters are associated with the vacuum expectation values of the supersymmetry breaking chiral supermultiplet $S$: $S =  \langle S \rangle +  \langle F_S \rangle \theta^2$.
\begin{align}
  M_F^2 \simeq \frac{g}{(4 \pi)^2} \langle F_S \rangle \,, \quad 
  m_{3/2} \simeq \frac{\langle F_S \rangle}{\sqrt 3 M_\mathrm{Pl}} \,.
\end{align}
Here, $M_\mathrm{Pl}$ is the reduced Planck mass. 
The Q-ball is a localized condensate of a complex scalar field $\phi$. 
Assuming the spherically symmetric configuration, the scalar field is written as 
\begin{align}
  \phi(x) = \frac{1}{\sqrt{2}}\varphi(r) e^{-i \omega t} \,.
\end{align}
The gauge-mediation contribution dominates the potential as $\varphi \gtrsim \varphi_\mathrm{eq} \equiv \sqrt2 M_F^2/m_{3/2}$, while the gravity-mediation contribution dominates the potential as $\varphi \lesssim \varphi_\mathrm{eq}$.
We obtain the equation of the scalar configuration by minimizing the energy with the Q-ball charge $Q$ fixed. 
\begin{align}
  E_\omega & = E + \omega \left[ Q + \frac{1}{i} \int d^3 x \left(\phi^\ast \partial_0 \phi - \partial_0 \phi^\ast \phi \right) \right] \\
  & = \int d^3 x \frac{1}{2}\left| \partial_0 \varphi \right|^2 + \int d^3 x \left[ \frac{1}{2}|\nabla \varphi|^2 + V - \frac{1}{2}\omega^2 |\varphi|^2 \right] + \omega Q \,.
\end{align}
The equation for the radial scalar field 
\begin{align}
  \frac{\partial^2}{\partial r^2} \varphi + \frac{2}{r} \frac{\partial}{\partial r} \varphi + \omega^2 \varphi - \frac{\partial}{\partial \varphi} V = 0 \,,
\end{align} 
leads to the approximate solution for the field in the flat potential limit, and we also obtain the charge $Q$ by the approximate solution as follows (see \cite{Hisano:2001dr}).
\begin{align}
  \varphi(r) \simeq \varphi_0 \frac{\sin \omega r}{\omega r}\,, \qquad 
  Q = \int d^3 x \, \omega \varphi^2 = \frac{2 \pi^2 \varphi_0^2}{\omega^2} \,.
\end{align}
Here, $\varphi_0$ is the field value of the scalar field at the center of Q-ball. 
The radius of the Q-ball is set by $R = \pi/\omega$, beyond which $\varphi(r) = 0$ is assumed.

As for the gauge-mediation domination, the minimization of the energy, $\omega Q + 4 \pi^4 V / (3 \omega^3)$, with respect to $\omega$ once the logarithmic dependence is ignored.
Therefore, $\omega$ and $\varphi_0$ are written in terms of $M_F$ and $Q$ as 
\begin{align}
  \omega = \sqrt{2} \pi M_F Q^{- \frac{1}{4}} \,, \quad 
  \varphi_0 = M_F Q^{\frac{1}{4}} 
\end{align}
and the radius and the total energy, in other words the total mass, of the Q-ball are
\begin{align}
  R = \frac{1}{\sqrt{2}} M_F^{-1} Q^{\frac{1}{4}} \,, \quad 
  E_\omega = \frac43 \sqrt{2} \pi M_F Q^{\frac34} \,.
\end{align}

Regarding the potential dominated by the gravity-mediation contribution, we assume the Gaussian profile of the approximate solution following \cite{Enqvist:1997si,Enqvist:1998en,Kasuya:2000sc}.
The charge of the Q-ball is computed under the Gaussian ansatz as follows.
\begin{align}
  \varphi(r) \simeq \varphi_0 e^{-r^2/(2 R^2)}\,, \quad 
  Q = \pi^{3/2} \omega \varphi_0^2 R^3 \,,
\end{align}
From the equation of the radial scalar field, one can find 
\begin{align}
  R^2 = \frac{1}{|K| m_{3/2}^2}\,, \quad 
  \omega^2 = m_{3/2}^2(1+2|K|) \simeq m_{3/2}^2\,.
\end{align}
and the field value $\varphi_0$ is written as 
\begin{align}
  \varphi_0 \simeq \frac{1}{\pi^{3/4}} |K|^{3/4} m_{3/2} Q^{1/2} \,.
\end{align}

\section{Useful Formulae \label{app:formulae}}
In this appendix, we summarize the useful formulae for our study. 

\subsection{Formulae for Analytically-continued momentum}

The square root of the matrices $\sigma \cdot p$ and $\overline \sigma \cdot p$ with the four-momentum $p = (E, \mathbf{p})$ (the mass $M^2 = E^2 - \mathbf{p}^2$) is expanded as follows. 
\begin{align}
  \sqrt{\sigma \cdot p}
  & = \frac{1}{\sqrt2} \left( \sqrt{E+M} \mathbf{1}_2 - \frac{1}{\sqrt{E+M}} \mathbf{p} \cdot \boldsymbol{\sigma} \right) \,, \nonumber \\
  \sqrt{\overline \sigma \cdot p}
  & = \frac{1}{\sqrt2} \left( \sqrt{E+M} \mathbf{1}_2 + \frac{1}{\sqrt{E+M}} \mathbf{p} \cdot \boldsymbol{\sigma} \right) \,.
\end{align}
It is straightforward to extend this formula for the analytically continued momentum in this form.  
In particular, we just replace the three-momentum $\mathbf{p} \to i \boldsymbol{\kappa}$ with a real vector $\boldsymbol{\kappa}$ as $E < M$.
The complex conjugate of the matrix is 
\begin{align}
  \sigma^2 \left(\sqrt{\sigma \cdot p^\ast}\right)^{*} \sigma^2 = \sqrt{\overline \sigma \cdot p} \,.
\end{align}
Using this equality, the complex conjugate of the negative-frequency part of the wave function in \cref{eq:complex_sol} is calculated as follows. 
\begin{align}
  \left[\sqrt{\sigma \cdot p^\ast}\left(-i \sigma_{2}\right) A^{*} e^{i p^\ast x}\right]^{*}=\left(-i \sigma_{2}\right) \sqrt{\overline \sigma \cdot p} A e^{-i p x}\,, 
  \label{eq:cc_negative}
\end{align}

\subsection{Spinor Spherical Harmonics \label{app:SSH}}

We construct the spherical-wave solutions to the Dirac equation, which are important for describing the scattering of the Dirac fermion (see \cite{Landau:1991wop,Berestetskii:1982uol,johnson2007atomic}). 
Now, we consider a state of the total angular momentum $\mathbf{J} = \mathbf{L} + \mathbf{S}$ with the orbital angular momentum operator $\mathbf{L}$ and the spin operator $\mathbf{S}$.
The state of the total angular momentum is a simultaneous eigenstate of $J^2$, $J_z$, $L^2$, and $S^2$ (with the eigenvalues of $j(j+1)$, $m$, $\ell(\ell+1)$, and $s(s+1)$, respectively), and we denote the eigenstate as $|j,m;\ell,s\rangle$. 
The state of the total angular momentum is given by the linear combination of the direct product of the states of $\mathbf{L}$ and $\mathbf{S}$.
\begin{align}
  |j,m;\ell,s\rangle = \sum_{m_s} C(\ell,s,j;m_\ell,m_s,m) |\ell,m _\ell\rangle |s,m_s\rangle \,.
\end{align}
Here, $C(\ell,s,j;m_\ell,m_s,m)$ is the Clebsch-Gordan coefficient (CGC), and $m = m_\ell + m_s$.

Now, we consider the state with the spin-1/2 and the orbital angular momentum $\ell$. 
The total angular momentum is given by $j = \ell \pm 1/2$. 
For fixed $j$ and $m$, we have two independent CGCs. 
\begin{align}
  x = C(\ell,1/2,j;m-1/2,1/2,m) \,, \qquad
  y = C(\ell,1/2,j;m+1/2,-1/2,m) \,.
\end{align}
Using these coefficients, we write the state as 
\begin{align}
  \left|j,m;\ell,\frac12\right\rangle
  = x \left|\ell,m-\frac12\right\rangle \left|\frac12,\frac12\right\rangle
  + y \left|\ell,m+\frac12\right\rangle \left|\frac12,-\frac12\right\rangle \,.
  \label{eq:spinhalf_expansion}
\end{align}
The orthogonality relation $\langle j',m';\ell',1/2|j,m;\ell,1/2\rangle = \delta_{j',j} \delta_{m',m}$ gives $x^2 + y^2 = 1$.
The state $|j,m;\ell,1/2\rangle$ is the eigenstate of the spin-orbit coupling $\Lambda = 2 \mathbf{L} \cdot \mathbf{S} = \mathbf{J}^2 - \mathbf{L}^2 - \mathbf{S}^2$.
\begin{align}
  \Lambda \left|j,m;\ell,\frac12\right\rangle
  = \lambda \left|j,m;\ell,\frac12\right\rangle\,, 
\end{align}
with the eigenvalue $\lambda = \ell$ ($j = \ell + 1/2$) or $\lambda = - \ell - 1$ ($j = \ell - 1/2$).
Meanwhile, acting $\Lambda = 2 \mathbf{L} \cdot \mathbf{S} = 2 L_z S_z + L_+ S_- + L_- S_+$ on the right-hand side of \cref{eq:spinhalf_expansion}, we have two independent equations that are obtained from coefficients of each state, $|\ell,m-1/2\rangle |1/2,1/2\rangle$ and $|\ell,m+1/2\rangle |1/2,-1/2\rangle$.
\begin{align}
  \lambda x & = \left( m - \frac12 \right) x + \sqrt{\left(\ell - m + \frac12\right)\left(\ell + m + \frac12\right)} y \,, \\
  \lambda y & = \sqrt{\left(\ell - m + \frac12\right)\left(\ell + m + \frac12\right)} x - \left( m + \frac12 \right) y \,,
\end{align}
and we find the ratio $y/x$ to be
\begin{align}
  \frac{y}{x} = \frac{\lambda - m + \frac12 }{\sqrt{\left(\ell - m + \frac12\right)\left(\ell + m + \frac12\right)}} \,.
\end{align}
We take a sign convention, $y > 0$.
Combining them, we obtain the CGCs for the spin and the orbital angular momentum as shown in \cref{tab:CDCsList}.

\begin{table}
  \centering
  \caption{List of the Clebsch-Gordan coefficients for $\mathbf{L}+\mathbf{S}$.}
  \begin{tabular}{|c|c|c|}
     & $j = \ell + 1/2$ & $j = \ell - 1/2$ \\ \hline
    $m_s = 1/2$ 
    & $\displaystyle \sqrt{\frac{\ell + m + \frac12}{2 \ell + 1}}$ 
    & $\displaystyle - \sqrt{\frac{\ell - m + \frac12}{2 \ell + 1}}$ \\
    $m_s = - 1/2$ 
    & $\displaystyle \sqrt{\frac{\ell - m + \frac12}{2 \ell + 1}}$
    & $\displaystyle \sqrt{\frac{\ell + m + \frac12}{2 \ell + 1}}$ \\
  \end{tabular}
  \label{tab:CDCsList}
\end{table}

The spherical spinor $\Omega_{j\ell m}$ is an eigenfunction of the total angular momentum $\mathbf{J}^2$ and $J_z$.
$\Omega_{j,\ell, m}$ is given by a combination of the spherical harmonics $Y_{\ell,m}$, the eigenfunction of the orbital angular momentum, and the two-component spinor $\chi_m$.
In other words, the spherical spinor is given by the projection of a state onto the angle coordinate and the spinor space. 
\eqs{
  \Omega_{j,\ell, m}(\theta,\phi) & = \langle\theta,\phi; \chi|j,m;\ell,1/2\rangle \\
  & = \sum_{m_s} C(\ell,1/2,j;m-m_s,m_s,m) \langle\theta,\phi|\ell,m - m_s\rangle \langle \chi |s,m_s\rangle \\
  & = \sum_{m_s} C(\ell,1/2,j;m-m_s,m_s,m) Y_{\ell,m-m_s}(\theta,\phi) \chi_{m_s} \,.
}
Here, the two-component spinor is given by 
\begin{align}
  \chi_{\frac12} = 
  \begin{pmatrix}
    1 \\
    0
  \end{pmatrix} \,, \quad 
  \chi_{-\frac12} = 
  \begin{pmatrix}
    0 \\
    1
  \end{pmatrix} \,.
\end{align}
The CGCs for the spin-$1/2$ and the orbital angular momentum $\ell$ are given by \cref{tab:CDCsList}, and hence we have explicit forms of the spherical spinors as follows. 
\eqs{
  \Omega_{\ell+\frac12,\ell, m}(\theta,\phi) & = 
  \begin{pmatrix}
    \sqrt{\frac{\ell+m+1/2}{2\ell+1}} Y_{\ell,m-\frac12}(\theta,\phi)\\
    \sqrt{\frac{\ell-m+1/2}{2\ell+1}} Y_{\ell,m+\frac12}(\theta,\phi)
  \end{pmatrix} \,, \\
  \Omega_{\ell-\frac12,\ell, m}(\theta,\phi) & 
  = 
  \begin{pmatrix}
    - \sqrt{\frac{\ell-m+1/2}{2\ell+1}} Y_{\ell,m-\frac12}(\theta,\phi)\\
    \sqrt{\frac{\ell+m+1/2}{2\ell+1}} Y_{\ell,m+\frac12}(\theta,\phi)
  \end{pmatrix} \,.
}

Now, we recall properties of the spherical harmonics $Y_{\ell, m}$.
The spherical harmonics are written with the associated Legendre functions $P^{m}_\ell$ as follows. 
\eqs{
  Y_{\ell,m} (\theta, \phi) & = (-1)^{(m+|m|)/2}\sqrt{\frac{2 \ell + 1}{4 \pi} \frac{(\ell-|m|)!}{(\ell+|m|)!}} e^{i m \phi} P^{|m|}_\ell(\cos\theta) \,, \\
  P^{|m|}_\ell(x) & = \frac{1}{2^\ell \ell!} (1-x^2)^{|m|/2} \frac{d^{\ell+|m|}}{d x^{\ell+|m|}} (x^2 - 1)^\ell \,.
}
We choose the phase convention as in the above equation.
The normalization of the spherical harmonics is chosen so that the orthogonal relation is given by 
\begin{align}
  \int d \cos\theta  d \phi \ Y_{\ell' m'}^\ast (\theta,\phi) Y_{\ell m} (\theta,\phi)
 = \delta_{\ell',\ell} \delta_{m',m} \,.
\end{align}
Under the parity transformation, $\mathbf{r} \to - \mathbf{r}$ (in terms of angles, $\{ \theta\,, \phi \} \to \{ \pi-\theta, \phi+\pi\} $), the spherical harmonics change only by a sign factor. 
\begin{align}
  P Y_{\ell,m} (\theta, \phi) = Y_{\ell,m} (\pi-\theta, \phi+\pi) = (-1)^\ell Y_{\ell,m} (\theta, \phi) \,.
\end{align}

Since the state $|j,m;\ell,s\rangle$ is an eigenstate of the spin-orbit coupling, the spherical spinors are also the eigenfunctions of the coupling. 
Conventionally, the spherical spinors are labelled by the eigenvalues $\kappa$ of an operator $K = - 1 - 2 \mathbf{L} \cdot \mathbf{S}$.
The operation of $K$ on the spherical spinors is given by 
\begin{align}
  K \Omega_{j,\ell, m}(\theta,\phi) = \kappa \Omega_{j,\ell, m}(\theta,\phi)\,.
\end{align}
Here, $\kappa = - 1 - \ell$ for $j = \ell +1/2$, and $\kappa = \ell$ for $j = \ell -1/2$.
In other words, $\kappa = \mp (j+1/2)$ for $j = \ell \pm 1/2$.
The absolute value of $\kappa$ determines $j$, and the sign of $\kappa$ determines the relative sign of $j - \ell$.
Hence, a compact notation for the spherical spinors, labelled by $\kappa$, is often used. 
\begin{align}
  \Omega_{\kappa, m}(\theta,\phi) = \Omega_{j,\ell, m}(\theta,\phi) \,.
\end{align}

Let us now discuss the properties of the spherical spinors. 
The orthogonal relation of the spherical spinors follows from that of the spherical harmonics $Y_{\ell,m}$ and is given by 
\begin{align}
  \int d \cos\theta d \phi \ \Omega_{\kappa', m'}^\dag \Omega_{\kappa, m}
 = \delta_{\kappa',\kappa} \delta_{m',m} \,.
\end{align}
The parity transformation of the spherical spinors also follows from that of the spherical harmonics as follows.
\begin{align}
  P \Omega_{\kappa, m}(\theta,\phi) 
  = (-1)^\ell \Omega_{\kappa, m}(\theta,\phi) \,.
\end{align}
Since $\ell = j \pm 1/2$, the sign factor differs only by $(-1)$ for fixed $j$ with different sign of $\kappa$.
Therefore, the spherical spinors have opposite parities for $\kappa$ and $-\kappa$. 

Let us consider operations $\boldsymbol{\sigma} \cdot \hat{\mathbf{r}}$ with $\hat{\mathbf{r}} = \mathbf{r}/r$ and $\boldsymbol{\sigma} \cdot \mathbf{p}$ on the spherical spinors for constructing the solution to the Dirac equation.
The parity operator changes the sign of $\boldsymbol{\sigma} \cdot \hat{\mathbf{r}}$.
Therefore, the operation of $\boldsymbol{\sigma} \cdot \hat{\mathbf{r}}$ on $\Omega_{\kappa, m}$ is proportional to $\Omega_{-\kappa, m}$,
\begin{align}
  \boldsymbol{\sigma} \cdot \hat{\mathbf{r}} \Omega_{\kappa, m}(\theta,\phi)
  = a \Omega_{-\kappa, m}(\theta,\phi) \,.
\end{align}
Here, $a$ is a constant satisfying $a^2 = 1$ because of an identity $(\boldsymbol{\sigma} \cdot \hat{\mathbf{r}}) (\boldsymbol{\sigma} \cdot \hat{\mathbf{r}}) = 1$.
We evaluate both sides of the above equation with $\theta = 0$, and one can establish $a = -1$ independent of the sign of $\kappa$.
Using the identity $\sigma_{i} \sigma_{j} = \delta_{ij} + i \epsilon_{ijk} \sigma_{k}$, one can rewrite $\boldsymbol{\sigma} \cdot \mathbf{p}$ as follows. 
\begin{align}
  \boldsymbol{\sigma} \cdot \mathbf{p} 
  = (\boldsymbol{\sigma} \cdot \hat{\mathbf{r}}) \left( \hat{\mathbf{r}} \cdot \mathbf{p} + i \frac{\boldsymbol{\sigma} \cdot \mathbf{L}}{r} \right) \,.
\end{align}
We can use the identity $\boldsymbol{\sigma} \cdot \mathbf{L} = - 1 - K$ for simplifying the second term in the bracket. 
Assuming the function $f(r)$ depending only on the radius coordinate $r$, the operation on a function $f(r) \Omega_{\kappa m}(\theta,\phi)$ is given by
\begin{align}
  \boldsymbol{\sigma} \cdot \mathbf{p} 
  \left[ f(r) \Omega_{\kappa m}(\theta,\phi) \right]
  = i \left[ \frac{df}{dr} + \frac{1 + \kappa}{r} f(r) \right] \Omega_{-\kappa m}(\theta,\phi)\,.
  \label{eq:sigmap_op}
\end{align}

We are now ready to discuss the solution to the Dirac equation with the spherically symmetric (central force) potential $V(r)$.
The four-component Dirac spinor $\Psi$ satisfies the time-independent Dirac equation: 
\begin{align}
  H_D \Psi = E \Psi\,, \qquad
  H_D = \boldsymbol{\alpha} \cdot \mathbf{p} + \beta M + V(r) \,, 
  \label{eq:Dirac_eq}
\end{align}
where $H_D$ is the Dirac Hamiltonian, $M$ denotes the mass of the Dirac fermion, and $E$ denotes the energy. 
$\boldsymbol{\alpha}$ and $\beta$ correspond to the $4 \times 4$ Dirac matrices in the Dirac basis%
\footnote{
  The notation for $\gamma$ in the Dirac basis is given by 
  \begin{align}
    \gamma_{D}^{0} = \beta = 
    \begin{pmatrix}
      \mathbf{1}_2 & 0 \\
      0 & -\mathbf{1}_2 \\
    \end{pmatrix} \,, \quad
    \boldsymbol{\gamma}_{D} = \beta \boldsymbol{\alpha} = 
    \begin{pmatrix}
      0 & \boldsymbol{\sigma} \\
      -\boldsymbol{\sigma} & 0 \\
    \end{pmatrix}\,.
  \end{align}
  They are related with those in the Weyl basis as
  \begin{align}
    \gamma_{W}^{0} = U \gamma_{D}^{0} U^{\dagger}= 
    \begin{pmatrix}
      0 & \mathbf{1}_2 \\
      \mathbf{1}_2 & 0 \\
    \end{pmatrix} \,, \quad
    \boldsymbol{\gamma}_{W} = U \boldsymbol{\gamma}_{D}^{0} U^{\dagger}= 
    \begin{pmatrix}
      0 & \boldsymbol{\sigma} \\
      -\boldsymbol{\sigma} & 0 \\
    \end{pmatrix}\,, \quad
    \gamma_{5W} = U \gamma_{5D} U^{\dagger}= 
    \begin{pmatrix}
      -\mathbf{1}_2 & 0 \\
      0 & \mathbf{1}_2 \\
    \end{pmatrix}\,,
  \end{align}
  with the rotation matrix 
  \begin{align}
    U = \frac{1}{\sqrt{2}}
    \begin{pmatrix}
      \mathbf{1}_2 & -\mathbf{1}_2 \\
      \mathbf{1}_2 & \mathbf{1}_2 \\
    \end{pmatrix}\,.
  \end{align}
  The spinors in the Weyl basis are 
  \eqs{
    u_{W}(\mathbf{p},\chi) & = U u_{D}(\mathbf{p},\chi)
    =  
    \begin{pmatrix}
      \sqrt{\sigma \cdot k} \chi \\
      \displaystyle
      \sqrt{\overline \sigma \cdot k} \chi
    \end{pmatrix} \,, \\
    v_{W}(\mathbf{p},\chi) & = U v_{D}(\mathbf{p},\chi)
    = 
    \begin{pmatrix}
      -\sqrt{\sigma \cdot k} \chi \\
      \displaystyle
      \sqrt{\overline \sigma \cdot k} \chi
    \end{pmatrix} \,.
  }
}: 
\eqs{
  \boldsymbol{\alpha} = 
  \begin{pmatrix}
    0 & \boldsymbol{\sigma} \\
    \boldsymbol{\sigma} & 0 \\
  \end{pmatrix} \,, \quad 
  \beta = 
  \begin{pmatrix}
    \mathbf{1}_2 & 0 \\
    0 & -\mathbf{1}_2 \\
  \end{pmatrix}\,, 
}
and we note that the $\gamma_5$-matrix in the Dirac basis is 
\eqs{
  \gamma_5 = 
  \begin{pmatrix}
    0 & \mathbf{1}_2 \\
    \mathbf{1}_2 & 0 \\
  \end{pmatrix} \,.
}

When the potential is spherically symmetric, the total angular momentum $\mathbf{J} = \mathbf{L}+\mathbf{S}$ commutes with the Dirac Hamiltonian. 
The solution is the simultaneous eigenstate of the energy and the total angular momentum $J^2$ and $J_z$.
As mentioned above, the eigenfunction for the total angular momentum $J^2$ and $J_z$ is given by the spherical spinors. 
We have two spherical spinors $\Omega_{\kappa, m}$ and $\Omega_{-\kappa, m}$ for given $j$, but they have opposite parity.
The wave function is decomposed into the radial part and two spherical spinors:
\eqs{
  \Psi_{\kappa,m} (\mathbf{r}) = 
  \frac{1}{r} 
  \begin{pmatrix}
    i P_\kappa(r) \Omega_{\kappa, m}(\theta,\phi) \\
    Q_\kappa(r) \Omega_{-\kappa, m}(\theta,\phi) 
  \end{pmatrix} \,.
}
We label the wave function the quantum numbers $(\kappa \,, m)$ instead of $(j \,, \ell \,, m)$. 
We note again that the absolute value of $\kappa$ determines the size of $j$ and the sign of $\kappa$ determines $j-\ell$.
Using \cref{eq:sigmap_op}, one obtains the coupled first-order differential equations for the radial functions $P_\kappa(r)$ and $Q_\kappa(r)$.
\eqs{
  [V(r)+M] P_\kappa + \left(\frac{d}{dr} - \frac{\kappa}{r}\right) Q_\kappa & = E P_\kappa \,, \\
  -\left(\frac{d}{dr} + \frac{\kappa}{r}\right) P_\kappa + [V(r)-M] Q_\kappa & = E Q_\kappa \,.
  \label{eq:spherical_Dirac}
}

In order to consider the scattering of fermions under the spherical potential, we first expand the plane-wave Dirac wave functions in terms of the spherical spinors.
The plane-wave solutions to \cref{eq:Dirac_eq} (with $V = 0$) are 
\eqs{
  \Psi(\mathbf{r},t,\chi) = u(\mathbf{p},\chi) e^{- i E t + i \mathbf{p} \cdot \mathbf{r}} \,, \qquad 
  u(\mathbf{p},\chi)
  = \sqrt{E+M} 
  \begin{pmatrix}
    \chi \\
    \displaystyle
    \frac{\boldsymbol{\sigma} \cdot \mathbf{p}}{E + M} \chi
  \end{pmatrix} \,, \\
  \Psi(\mathbf{r},t,\chi) = v(\mathbf{p},\chi) e^{i E t - i \mathbf{p} \cdot \mathbf{r}} \,, \qquad 
  v(\mathbf{p},\chi)
  = \sqrt{E+M} 
  \begin{pmatrix}
    \displaystyle
    \frac{\boldsymbol{\sigma} \cdot \mathbf{p}}{E + M} \chi \\
    \chi
  \end{pmatrix} \,.
  \label{eq:planewave_sol}
}
Here, we take the normalization of the Dirac spinors as 
\begin{align}
  \overline u(\mathbf{p},\chi') u(\mathbf{p},\chi)
  = 2 M \chi'^\dag \chi \,, \qquad 
  \overline v(\mathbf{p},\chi') v(\mathbf{p},\chi)
  = - 2 M \chi'^\dag \chi \,.
\end{align}

The Dirac equation for the radial functions (\ref{eq:spherical_Dirac}) is rewritten as the second-order differential equations under the assumption of $V = 0$.
For given $p^2 = E^2 - M^2$, one easily finds the solutions that are regular at $pr \to 0$.
The solution is given by 
\eqs{
  P_\varkappa(r) & = A_\varkappa r j_{\varkappa}(pr) \,, \quad
  Q_\varkappa(r) = B_\varkappa r j_{\varkappa-1}(pr) \,, \\
  P_{-\varkappa}(r) & = A_{-\varkappa} r j_{-\varkappa-1}(pr) \,, \quad
  Q_{-\varkappa}(r) = B_{-\varkappa} r j_{-\varkappa}(pr) \,.
 }
Here, $A$'s and $B$'s are constants, and $\varkappa >0$. 
Inserting these solutions to the coupled Dirac equations, \cref{eq:spherical_Dirac}, for the positive energy solution, one finds 
\begin{align}
  B_\kappa = \mp \frac{p}{E+M} A_\kappa \,, \qquad (\kappa \gtrless 0) \,.
\end{align}
For the negative energy solution, we can replace $E \to - E$. 
Instead of $B$ in terms of $A$, to match the plane-wave solution with negative energy, we use the following $A$ in terms of $B$:
\begin{align}
  A_\kappa = \pm \frac{p}{E+M} B_\kappa \,, \qquad (\kappa \gtrless 0) \,.
\end{align} 
The spherical-wave solutions for the positive energy and for the negative energy are 
\begin{align}
  \Psi(\mathbf{r},t,\chi) = u_{\kappa,m}(\mathbf{r},p) e^{- i E t} \,, \quad 
  \Psi(\mathbf{r},t,\chi) = v_{\kappa,m}(\mathbf{r},p) e^{i E t} \,.
\end{align}
with the Dirac spinors for the spherical-wave solutions:
\begin{align}
  u_{\varkappa, m} (\mathbf{r}, p) & = 
  \sqrt{4 \pi (E+M)}
  \begin{pmatrix}
    - i j_{j +\frac12}(pr) \Omega_{j, j +\frac12, m}(\theta,\phi) \\
    \frac{p}{E+M} j_{j -\frac12}(pr) \Omega_{j, j -\frac12, m}(\theta,\phi) 
  \end{pmatrix} \,, \\
  u_{- \varkappa, m} (\mathbf{r}, p) & = 
  \sqrt{4 \pi (E+M)}
  \begin{pmatrix}
    j_{j -\frac12}(pr) \Omega_{j,j -\frac12, m}(\theta,\phi) \\
    - i\frac{p}{E+M} j_{j +\frac12}(pr) \Omega_{j,j +\frac12, m}(\theta,\phi) 
  \end{pmatrix} \,, 
\end{align}
for the positive energy (with positive $\varkappa$), and with 
\begin{align}
  v_{\varkappa, m} (\mathbf{r}, p) & = 
  \sqrt{4 \pi (E+M)}
  \begin{pmatrix}
    \frac{i p}{E+M} j_{j +\frac12}(pr) \Omega_{j, j +\frac12, m}(\theta,\phi) \\
    j_{j -\frac12}(pr) \Omega_{j,j -\frac12, m}(\theta,\phi) 
  \end{pmatrix} \,, \\
  v_{-\varkappa, m} (\mathbf{r}, p) & = 
  \sqrt{4 \pi (E+M)}
  \begin{pmatrix}
    \frac{p}{E+M} j_{j -\frac12}(pr) \Omega_{j,j -\frac12, m}(\theta,\phi) \\
    ij_{j +\frac12}(pr) \Omega_{j, j +\frac12, m}(\theta,\phi) 
  \end{pmatrix} \,, 
\end{align}
for the negative energy. 
Here, we use $\Omega_{j,\ell,m}$ instead of $\Omega_{\kappa,m}$ since we specify the sign of $\kappa$.
Following the parity transformation of the spherical spinors, $P\Omega_{j,\ell,m} = (-1)^\ell \Omega_{j,\ell,m}$, the spherical-wave solutions change under the parity transformation as follows.
\begin{align}
  P u_{\varkappa, m} (\mathbf{r}, p) 
  & = (-1)^{j +\frac12} \beta u_{\varkappa, m} (\mathbf{r}, p) \,, & 
  P v_{\varkappa, m} (\mathbf{r}, p) 
  & = (-1)^{j +\frac12} \beta v_{\varkappa, m} (\mathbf{r}, p) \,, \\
  P u_{- \varkappa, m} (\mathbf{r}, p) 
  & = (-1)^{j -\frac12} \beta u_{- \varkappa, m} (\mathbf{r}, p) \,, & 
  P v_{- \varkappa, m} (\mathbf{r}, p) 
  & = (-1)^{j -\frac12} \beta v_{- \varkappa, m} (\mathbf{r}, p) \,.
\end{align}
Here, $\beta$ is one of the Dirac matrices.

Now, we consider the partial-wave expansion for the Dirac spinors. 
We choose the coordinate for the spherical harmonics so that the incoming momentum $\mathbf{p}$ is aligned with the (positive) $z$-axis.
We recall that, in this coordinate, the partial-wave expansion of the plane-wave solution to the Schr\"odinger equation is given by 
\begin{align}
  e^{i p r \cos\theta} = \sum_{\ell=0}^\infty i^\ell (2 \ell+1) j_\ell(pr) P_\ell (\cos\theta) \,, \quad 
  e^{- i p r \cos\theta} = \sum_{\ell=0}^\infty i^\ell (2 \ell+1) j_\ell(pr) P_\ell (- \cos\theta)\,. 
\end{align}
This expansion may be rewritten in terms of the spherical harmonics by using a relation:
\begin{align}
  P_\ell(\cos\theta) 
  = \sqrt{\frac{4 \pi}{2 \ell + 1}} Y_{\ell,0}(\theta,\phi) \,.
\end{align}
The partial-wave expansion of the plane-wave Dirac solutions must have the following form: 
\eqs{
  u(\mathbf{p},\chi_{m_s}) e^{i p r \cos\theta} 
  = \sum_{\varkappa = 1}^\infty \left[ a_{\varkappa, m_s} u_{\varkappa, m_s} (\mathbf{r}, p) + b_{\varkappa, m_s} u_{-\varkappa, m_s} (\mathbf{r}, p) \right] \,, \\
  v(\mathbf{p},\chi_{m_s}) e^{- i p r \cos\theta} 
  = \sum_{\varkappa = 1}^\infty \left[\overline a_{\varkappa,m_s} v_{\varkappa, m_s} (\mathbf{r}, p) + \overline b_{\varkappa,m_s} v_{-\varkappa, m_s} (\mathbf{r}, p) \right] \,,
}
where $a_{\varkappa,m_s}\,, \overline a_{\varkappa,m_s} \,, b_{\varkappa,m_s}$ and $\overline b_{\varkappa,m_s}$ are the expansion coefficients. 
Since the left-hand side depends only on $\cos \theta$, the right-hand side should contain only the terms proportional to $Y_{\ell,0}$.
In other words, we choose the expansion coefficients to satisfy this condition. 

We consider the positive-energy solution (\ref{eq:planewave_sol}) with up-spin $m_{s} = 1/2$ as an example.
In this case, the upper component of the two-component spinor should be proportional to $Y_{\ell,0}$, namely $m = 1/2$ in the right-hand side. 
Both upper and lower two-component spinors of a linear combination $u_{\varkappa, \frac12} (\mathrm{r}, p) + u_{-\varkappa, \frac12} (\mathrm{r}, p)$ (with a positive $\varkappa$) are proportional to a linear combination of spherical spinors:
\eqs{
  - i j_{j+\frac12} (pr) \Omega_{j, j+\frac12, \frac12}  + j_{j-\frac12} (pr) \Omega_{j, j-\frac12, \frac12} \,.
}
Choosing $a_{\varkappa,m_s} = b_{\varkappa,m_s} = i^{\varkappa-1} \sqrt{\varkappa}$ and summing over $\varkappa$, the upper component of the two-component spinor remains while the lower component vanishes. 
As a result, one finds the summation is proportional to $u(\mathbf{p},\chi_{1/2})$.
In a similar way, we obtain the partial-wave expansion of the other plane-wave solutions: 
\eqs{
  u\left(\mathbf{p},\chi_{\frac12} \right) e^{i p r \cos\theta} & = \sum_{\varkappa = 1}^\infty i^{\varkappa-1} \sqrt{\varkappa} \left[ u_{\varkappa, \frac12} (\mathbf{r}, p) + u_{-\varkappa, \frac12} (\mathbf{r}, p) \right]  \\
  u\left(\mathbf{p},\chi_{-\frac12} \right) e^{i p r \cos\theta} & = \sum_{\varkappa = 1}^\infty i^{\varkappa-1} \sqrt{\varkappa} \left[ - u_{\varkappa, -\frac12} (\mathbf{r}, p) + u_{-\varkappa, -\frac12} (\mathbf{r}, p) \right] \,,  \\
  v\left(\mathbf{p},\chi_{\frac12} \right) e^{-i p r \cos\theta} & = \sum_{\varkappa = 1}^\infty (-i)^{\varkappa-1} \sqrt{\varkappa} \left[ v_{\varkappa, \frac12} (\mathbf{r}, p) + v_{-\varkappa, \frac12} (\mathbf{r}, p) \right]\,,  \\
  v\left(\mathbf{p},\chi_{-\frac12} \right) e^{-i p r \cos\theta} & = \sum_{\varkappa = 1}^\infty (-i)^{\varkappa-1} \sqrt{\varkappa} \left[ - v_{\varkappa, -\frac12} (\mathbf{r}, p) + v_{-\varkappa, -\frac12} (\mathbf{r}, p) \right]\,. 
  \label{eq:plain_expansion}
}
One may consider other combinations that are orthogonal to the above combinations. 
One finds that the sum of these combinations gives the parity-transformed plane-wave solutions.
\eqs{
  \beta u\left(\mathbf{p},\chi_{\frac12}\right) e^{- i p r \cos\theta}
  & =
  \sum_{\varkappa = 1}^\infty (-i)^{\varkappa-1} \sqrt{\varkappa} \left[ - u_{\varkappa, \frac12} (\mathbf{r}, p) + u_{-\varkappa, \frac12} (\mathbf{r}, p) \right] \,, \\
  \beta u\left(\mathbf{p},\chi_{-\frac12}\right) e^{- i p r \cos\theta} 
  & =
  \sum_{\varkappa = 1}^\infty (-i)^{\varkappa-1} \sqrt{\varkappa} \left[ u_{\varkappa, -\frac12} (\mathbf{r}, p) + u_{-\varkappa, -\frac12} (\mathbf{r}, p) \right] \,, \\
  - \beta v\left(\mathbf{p},\chi_{\frac12}\right) e^{i p r \cos\theta} 
  & =
  \sum_{\varkappa = 1}^\infty i^{\varkappa-1} \sqrt{\varkappa} \left[ - v_{\varkappa, \frac12} (\mathbf{r}, p) + v_{-\varkappa, \frac12} (\mathbf{r}, p) \right] \,, \\
  - \beta v\left(\mathbf{p},\chi_{-\frac12}\right) e^{i p r \cos\theta} 
  & =
  \sum_{\varkappa = 1}^\infty i^{\varkappa-1} \sqrt{\varkappa} \left[ v_{\varkappa, -\frac12} (\mathbf{r}, p) + v_{-\varkappa, -\frac12} (\mathbf{r}, p) \right] \,.
}

\section{Matching Conditions of Wave-functions \label{app:mc_wf}}

In this appendix, we give calculations of the matching conditions for the wave functions discussed in \cref{sec:quark} in detail. 

\subsection{Scattering on the infinitely large Q-ball wall}

Now, we give the detail of the calculation of the matching conditions for the scattering of the colorless quarks on the infinitely large Q-ball wall in \cref{sec:infinite}.
First, we consider the case that the incoming quark is left-handed. 
The momentum assignments to the massless fermions and the massive fermions are given in \cref{eq:p_assign1,eq:p_assign2}, and the scattering process is illustrated in \cref{fig:2dpict}.
The projection operations which the constant spinors have to satisfy is shown in \cref{eq:spinor_proj1}.

We solve the matching conditions for Majorana fermions with positive and negative frequency modes. 
At the left boundary $z = - R$, the matching conditions for the positive frequency solutions give
\begin{align}
  & \sqrt{\sigma \cdot p_{A-}} A_- e^{-i p_{A-} \cdot x} + \sum_{s = \pm} \sqrt{\sigma \cdot p_{Bs}} B_s e^{-i p_{Bs} \cdot x} \nonumber \\
  & =
  \sum_{m = \pm} \left[
    \sin \beta \left( \sqrt{\sigma \cdot \kappa_{D}^{m}} e^{- i \kappa_D^m \cdot x} D^{m} \right)
    + \cos \beta \left( \sqrt{\sigma \cdot \kappa_{F}^{m}} e^{- i \kappa_D^m \cdot x} F^{m} \right) 
  \right] e^{- i \omega t} \,, \\
  & \sqrt{\sigma \cdot \kappa_{B}} G_B e^{- i \kappa_B \cdot x} \nonumber \\
  & =
  \sum_{m = \pm} \left[
    \cos \beta \left( \sqrt{\sigma \cdot \kappa_{D}^{m}} e^{- i \kappa_D^m \cdot x} D^{m} \right)
    - \sin \beta \left( \sqrt{\sigma \cdot \kappa_{F}^{m}} e^{- i \kappa_F^m \cdot x} F^{m}  \right)
  \right] \,.
\end{align}
Here, ${A_-}$ and $B_\pm$ denote the spinors for the incoming and reflected quarks, $G_B$ is the spinor for the gluino, and $D^\pm$ and $F^\pm$ denote the spinors for the fermions inside the Q-ball. 
$x$ denotes the four-vector of any position on the left boundary.
We note that the wave function of the quark inside the Q-ball is multiplied by the time-dependent phase factor $e^{-i\omega t}$.
We recall that the subscript of $B_\pm$ indicates its spin direction while the superscripts of $D^\pm$ and $F^\pm$ indicate the decaying mode or the growing mode. 
On the other hand, the matching conditions for the negative frequency solutions give 
\begin{align}
  & \sqrt{\overline \sigma \cdot p_{A-}} A_- e^{-i p_{A-} \cdot x} + \sum_{s = \pm} \sqrt{\overline \sigma \cdot p_{Bs}} B_s e^{-i p_{Bs} \cdot x} \nonumber \\
  & = \sum_{m = \pm} \left[
    \sin \beta \left( \sqrt{\overline{\sigma} \cdot\kappa_{D}^{ m}} e^{- i \kappa_D^m \cdot x} D^{m} \right)
    + \cos \beta \left( - \sqrt{\bar{\sigma} \cdot\kappa_{F}^{ m}} e^{- i \kappa_F^m \cdot x} F^{m} \right)
  \right] e^{i \omega t} \,, \\
  & \sqrt{\overline \sigma \cdot \kappa_{B}} G_B e^{- i \kappa_B \cdot x} \nonumber \\
  & = \sum_{m = \pm} \left[
    \cos \beta \left( \sqrt{\overline{\sigma} \cdot\kappa_{D}^{ m}} e^{- i \kappa_D^m \cdot x} D^{m} \right)
    - \sin \beta \left( - \sqrt{\bar{\sigma} \cdot\kappa_{F}^{ m}} e^{- i \kappa_F^m \cdot x} F^{m} \right)
  \right] \,.
\end{align}
Here, we take the complex conjugate of the negative frequency solutions using \cref{eq:cc_negative}.
The negative sign in front of $F^m$ originates from the relative sign of the Majorana wave function for negative mass in \cref{eq:complex_sol}.
At the right boundary $z = R$, the matching conditions for the positive and negative frequency solutions give
\begin{align}
  \sum_{s}\sqrt{\sigma \cdot p_{Cs}} C_s e^{-i p_{Cs} \cdot x'} 
  & = \sum_{m} \left[
    \sin \beta \left( \sqrt{\sigma \cdot \kappa_{D}^{m}} e^{- i \kappa_D^m \cdot x'} D^{m} \right)
    + \cos \beta \left( \sqrt{\sigma \cdot \kappa_{F}^{m}} e^{- i \kappa_D^m \cdot x'} F^{m} \right)
  \right] e^{- i \omega t} \,, \\
  \sqrt{\sigma \cdot \kappa_{C}} G_C e^{- i \kappa_C \cdot x'} 
  & = \sum_{m} \left[
    \cos \beta \left( \sqrt{\sigma \cdot \kappa_{D}^{m}} e^{- i \kappa_D^m \cdot x'} D^{m} \right)
    - \sin \beta \left( \sqrt{\sigma \cdot \kappa_{F}^{m}} e^{- i \kappa_F^m \cdot x'} F^{m} \right)
  \right] \,, \\
  \sum_{s}\sqrt{\overline \sigma \cdot p_{Cs}} C_s e^{-i p_{Cs} \cdot x'}
  & = \sum_{m} \left[
    \sin \beta \left( \sqrt{\overline{\sigma} \cdot\kappa_{D}^{m}} e^{- i \kappa_D^m \cdot x'} D^{m} \right)
    + \cos \beta \left( - \sqrt{\bar{\sigma} \cdot\kappa_{F}^{m}} e^{- i \kappa_F^m \cdot x'} F^{m} \right)
  \right] e^{i \omega t} \,, \\
  \sqrt{\overline \sigma \cdot \kappa_{C}} G_C e^{- i \kappa_C \cdot x'} 
  & = \sum_{m} \left[
    \cos \beta \left( \sqrt{\overline{\sigma} \cdot\kappa_{D}^{m}} e^{- i \kappa_D^m \cdot x'} D^{m} \right)
    - \sin \beta \left( - \sqrt{\bar{\sigma} \cdot\kappa_{F}^{m}} e^{- i \kappa_F^m \cdot x'} F^{m} \right)
  \right] \,.
\end{align}
Here, $x'$ denotes the four-vector of any position on the right boundary.

In the following, we take the limit that the mass parameters $M_\lambda$ and $M_\pm$ are much larger than the injected energy $E$ and the chemical potential $\omega$.
Thanks to our momentum assignment and helicity [see \cref{eq:p_assign1,eq:p_assign2}], the matching conditions for the gluino wave function at the right boundary are rewritten as the time and $x$-independent form as follows.
\begin{align}
  2 \sqrt{M_\lambda} G_{B} e^{- M_\lambda R} & =
  2 \sqrt{M_+} \cos \beta e^{- M_+ R} D^+
  - 2 \sqrt{M_-} \sin \beta e^{- M_- R} F^+ \nonumber\\
  & - 2 i \sigma^3\sqrt{M_+} \cos \beta e^{M_+ R} D^-
  + 2 i \sigma^3 \sqrt{M_-} \sin \beta e^{M_- R} F^- \,, \\
  2 \sqrt{M_\lambda} G_{B} e^{- M_\lambda R} & =
  2 \sqrt{M_+} \cos \beta e^{- M_+ R} D^+
  + 2 \sqrt{M_-}\sin \beta e^{- M_- R} F^+ \nonumber \\
  & + 2 i \sigma^3 \sqrt{M_+} \cos \beta e^{M_+ R} D^-
  + 2 i \sigma^3 \sqrt{M_-} \sin \beta e^{M_- R} F^- \,.
\end{align}
In a similar way, the matching conditions for the gluino wave function at left boundary are rewritten in the time-independent form as follows.
\begin{align}
  2 \sqrt{M_\lambda} G_{C} e^{- M_\lambda R} & =
  2 i \sigma^3 \sqrt{M_+} \cos \beta e^{M_+ R} D^+
  - 2 i \sigma^3 \sqrt{M_-} \sin \beta e^{M_- R} F^+ \nonumber \\
  & + 2\sqrt{M_+} \cos \beta e^{- M_+ R} D^-
  - 2 \sqrt{M_-} \sin \beta e^{- M_- R} F^- \,, \\
  2 \sqrt{M_\lambda} G_{C} e^{- M_\lambda R} & =
  - 2 i \sigma^3 \sqrt{M_+} \cos \beta e^{M_+ R} D^+
  - 2 i \sigma^3 \sqrt{M_-}\sin \beta e^{M_- R} F^+ \nonumber \\
  & + 2 \sqrt{M_+} \cos \beta e^{- M_+ R} D^-
  + 2 \sqrt{M_-} \sin \beta e^{- M_- R} F^- \,.
\end{align}
By removing the spinors of gluino, $G_B$ and $G_C$, we find the relations between the spinors $D^\pm$ and $F^\mp$ as follows. 
\begin{align}
  D^- & = i \sigma^3 \sqrt{\frac{M_-}{M_+}} \tan \beta e^{- (M_++M_-)R} F^+ \,, \label{eq:D-F+rel} \\
  D^+ & = - i \sigma^3 \sqrt{\frac{M_-}{M_+}} \tan \beta e^{- (M_+ + M_-) R} F^- \,. \label{eq:D+F-rel}
\end{align}

We replace $D^\pm$ in the matching conditions for quark wave functions with $F^\pm$.
Considering the matching conditions for $C_\pm$, we give the constant spinors $F^\pm$ in terms of $C_\pm$. 
All the time and $x$-dependent factor are the same thanks to our momentum assignment and helicity [see \cref{eq:p_assign1,eq:p_assign2}], and hence the system of equations is closed. 
\begin{align}
  2 \sqrt{E} e^{- i\theta \frac{\sigma^2}{2}} \overline C_- e^{i E \cos\theta R} 
  & = \sqrt{M_-} \cos \beta (\mathbbm{1} + i \sigma^3) \left( 1 + e^{- 2(M_++M_-)R} \tan^2\beta \right) e^{M_- R} F^+ \nonumber \\
  & \qquad + \sqrt{M_-} \cos \beta (\mathbbm{1} - i \sigma^3) \left( 1 + \tan^2 \beta \right) e^{- M_- R} F^-  \,, \\
  2 \sqrt{E-2\omega}  e^{- i\theta' \frac{\sigma^2}{2}} \overline C_+ e^{i (E-2\omega) \cos\theta' R} 
  & = - \sqrt{M_-} \cos \beta
  (\mathbbm{1} - i \sigma^3) \left( 1 + e^{- 2(M_++M_-)R} \tan^2\beta \right) e^{M_- R} F^+ \nonumber \\
  & \qquad - \sqrt{M_-} \cos \beta (\mathbbm{1} + i \sigma^3) \left( 1 + \tan^2 \beta  \right) e^{- M_- R} F^-
  \,.
\end{align}
We define the spinors for the transmitted quarks projected onto the $z$-axis (spinors with bars) as follows.
\begin{align}
  \overline C_- \equiv e^{i\theta \frac{\sigma^2}{2}} C_- \,, \qquad 
  \overline C_+ \equiv e^{i\theta' \frac{\sigma^2}{2}} C_+\,.
\end{align}
Due to the projections, $\overline C_\pm$ satisfy $(\mathbbm{1}\pm\sigma^3) \overline C_\mp = 0$ and $(\mathbbm{1}\pm\sigma^3) \overline C_\pm = 2 \overline C_\pm$. 
We multiply $(\mathbbm{1} \pm i \sigma^3)$ to make the $F^+$ terms proportional to $\mathbbm{1}$.
$F^\pm$ are given as a function of $\overline C_\pm$ as follows. 
\begin{align}
   & 2 (1 + \tan^2 \beta) \sqrt{M_-} \cos \beta (- i \sigma^3) e^{- M_- R} F^{-} \nonumber \\
   & \quad = 
     \sqrt{E} (1-i\sigma^3) e^{- i\theta \frac{\sigma^2}{2}} \overline C_- e^{i E R}
     + \sqrt{E-2\omega} (1+i\sigma^3) e^{- i\theta' \frac{\sigma^2}{2}} \overline C_+ e^{i (E-2\omega) R}\,, \\
   & 2 (1 + e^{-2(M_++M_-)R} \tan^2 \beta) \sqrt{M_-} \cos \beta e^{M_- R} F^{+} \nonumber \\
   & \quad =
    \sqrt{E} (1-i\sigma^3) e^{- i\theta \frac{\sigma^2}{2}} \overline C_- e^{i E R}
     - \sqrt{E-2\omega} (1+i\sigma^3) e^{- i\theta' \frac{\sigma^2}{2}} \overline C_+ e^{i (E-2\omega) R}\,. 
\end{align}

Similarly to $\overline C_\pm$, we define the spinors for the massless quarks projected onto the $z$-axis (spinors with bars) as follows.
\begin{align}
  \overline A_- \equiv e^{i\theta \frac{\sigma^2}{2}} A_-\,, \qquad 
  \overline B_- \equiv e^{- i\theta' \frac{\sigma^2}{2}} B_- \,, \qquad 
  \overline B_+ \equiv e^{-i\theta \frac{\sigma^2}{2}} B_+ \,.
\end{align}
We note that $(\mathbbm{1}+\sigma^3) \overline A_- = 0$ and $(\mathbbm{1}\mp\sigma^3) \overline B_\pm = 0$, and $(\mathbbm{1}-\sigma^3) \overline A_- = 2 \overline A_-$ and $(\mathbbm{1}\pm\sigma^3) \overline B_\pm = 2 \overline B_\pm$.
The matching conditions for the massless quarks at $z = -R$ are simplified as follows. 
\begin{align}
  & 2 \sqrt{E} e^{- i\theta \frac{\sigma^2}{2}} \overline A_- e^{-i E \cos\theta R} 
  + 2 \sqrt{E} e^{i\theta \frac{\sigma^2}{2}} \overline B_+ e^{i E \cos\theta R} \nonumber \\\
  & \quad = \sqrt{M_-} \cos \beta 
  (\mathbbm{1} + i \sigma^3) \left( 1 + \tan^2\beta \right) e^{- M_- R} F^{+} \nonumber \\
  & \qquad + \sqrt{M_-} \cos \beta 
  (\mathbbm{1} - i \sigma^3) \left( 1 + e^{- 2(M_++M_-)R} \tan^2 \beta \right) e^{M_- R} F^{-} \,, \\
  & 2 \sqrt{E-2\omega} e^{i\theta' \frac{\sigma^2}{2}} \overline B_- e^{i (E-2\omega) \cos\theta' R} \nonumber \\
  & \quad = - \sqrt{M_-} \cos \beta 
  (\mathbbm{1} - i \sigma^3) \left( 1 + \tan^2\beta\right) e^{- M_- R} F^{+} \nonumber \\
  & \qquad - \sqrt{M_-} \cos \beta 
  (\mathbbm{1} + i \sigma^3) \left( 1 + e^{- 2(M_++M_-)R} \tan^2 \beta \right)e^{M_- R} F^{-} \,. 
\end{align}
The spinor $F^\pm$ is given in terms of a function of $\overline C_\pm$.
We obtain the spinor relations among quarks by inserting $F^\pm$ as follows: 
\begin{align}
  & \sqrt{E} e^{- i\theta \frac{\sigma^2}{2}} \overline A_- e^{-i E R \cos\theta} 
  + \sqrt{E} e^{i\theta \frac{\sigma^2}{2}} \overline B_+ e^{i E R \cos\theta} \nonumber \\
  & = \frac{a_- }{4} \sqrt{E-2\omega} (i\sigma^3) e^{-i\theta' \frac{\sigma^2}{2}} \overline C_+ e^{i (E-2\omega) R\cos\theta'}
  + \frac{a_+}{4} =\sqrt{E} e^{-i\theta \frac{\sigma^2}{2}} \overline C_- e^{i E R \cos\theta}\,, \\
  & \sqrt{E-2\omega} (i\sigma^3) e^{i\theta' \frac{\sigma^2}{2}} \overline B_- e^{i (E-2\omega)R} \nonumber \\
  & = \frac{a_+}{4} \sqrt{E-2\omega} (i\sigma^3) e^{-i\theta' \frac{\sigma^2}{2}} \overline C_+ e^{i (E-2\omega) R\cos\theta'}
  + \frac{a_-}{4} \sqrt{E} e^{-i\theta \frac{\sigma^2}{2}} \overline C_- e^{i E R \cos\theta}\,.
\end{align} 
Here, we define the parameters $a_\pm$ as follows.
\begin{align}
  a_\pm = \frac{1 + e^{- 2(M_++M_-)R} \tan^2 \beta}{1 + \tan^2 \beta} e^{2 M_- R}
  \pm \frac{1 + \tan^2\beta}{1 + e^{-2(M_++M_-)R} \tan^2 \beta} e^{- 2 M_- R} \,. 
  \label{eq:apm}
\end{align}

We take the large $R$ limit in the following and keep the dominant terms in order to simplify the solutions. 
The matching conditions for quarks at the right-side boundary of the Q-ball are simplified as follows. 
\begin{align}
  & 2 \sqrt{M_-} \cos \beta (1 + \tan^2 \beta) (- i \sigma^3) e^{- M_- R} F^{-} \nonumber \\
  & \quad = 
    \sqrt{E} (1-i\sigma^3) e^{- i\theta \frac{\sigma^2}{2}} \overline C_- e^{i E R \cos\theta}
    + \sqrt{E-2\omega} (1+i\sigma^3) e^{- i\theta' \frac{\sigma^2}{2}} \overline C_+ e^{i (E-2\omega) R \cos\theta'} \,, \\
  & 2 \sqrt{M_-} \cos \beta e^{M_- R} F^{+} \nonumber \\
  & \quad = 
   \sqrt{E} (1-i\sigma^3) e^{- i\theta \frac{\sigma^2}{2}} \overline C_- e^{i E R \cos\theta}
    - \sqrt{E-2\omega} (1+i\sigma^3) e^{- i\theta' \frac{\sigma^2}{2}} \overline C_+ e^{i (E-2\omega) R \cos\theta'} \,. 
\end{align}
Meanwhile, the matching conditions for quarks at the left-side boundary of the Q-ball are simplified as follows. 
\begin{align}
  & \sqrt{E} (\mathbbm{1} - i \sigma^3) e^{- i\theta \frac{\sigma^2}{2}} \overline A_- e^{-i E R \cos\theta} 
  + \sqrt{E} (\mathbbm{1} - i \sigma^3) e^{i\theta \frac{\sigma^2}{2}} \overline B_+ e^{i E R \cos\theta} \nonumber \\\
  & \quad = \sqrt{M_-} \cos \beta \left( 1 + \tan^2\beta \right) e^{- M_- R} F^{+} 
  + \sqrt{M_-} \cos \beta (- i \sigma^3) e^{M_- R} F^{-} \,, \\
  & \sqrt{E-2\omega} (\mathbbm{1} + i \sigma^3) e^{i\theta' \frac{\sigma^2}{2}} \overline B_- e^{i (E-2\omega) R \cos\theta'} \nonumber \\
  & \quad = - \sqrt{M_-} \cos \beta \left( 1 + \tan^2\beta\right) e^{- M_- R} F^{+} 
  - \sqrt{M_-} \cos \beta (i \sigma^3) e^{M_- R} F^{-} \,. 
\end{align}
Then, we obtain the spinor relations among the massless quarks by removing the spinor $F^\pm$ as follows.
We also simplify the relations by multiplying $1 \pm i \sigma^3$.
\begin{align}
  & \sqrt{E} e^{- i\theta \frac{\sigma^2}{2}} \overline A_- e^{-i E R  \cos\theta} 
  + \sqrt{E} e^{i\theta \frac{\sigma^2}{2}} \overline B_+ e^{i E R  \cos\theta} \nonumber \\
  & \quad =  
  \frac{e^{2 M_- R}}{2 (1 + \tan^2 \beta)} \left[ 
    \sqrt{E} e^{-i\theta \frac{\sigma^2}{2}} \overline C_- e^{i E R \cos\theta}
    + \sqrt{E-2\omega} (i\sigma^3) e^{-i\theta' \frac{\sigma^2}{2}} \overline C_+ e^{i (E-2\omega) R \cos\theta'} \right] \,, \\
  & \sqrt{E-2\omega} e^{i\theta' \frac{\sigma^2}{2}} \overline B_- e^{i (E-2\omega)R \cos\theta'} \nonumber \\
  & \quad =  
  \frac{e^{2 M_- R}}{2(1 + \tan^2 \beta)} \left[ 
    \sqrt{E} (-i\sigma^3) e^{-i\theta \frac{\sigma^2}{2}} \overline C_- e^{i E R \cos\theta}
    + \sqrt{E-2\omega} e^{-i\theta' \frac{\sigma^2}{2}} \overline C_+ e^{i (E-2\omega) R \cos\theta'}
   \right] \,.
\end{align}
Since the spinors with bars are the eigenstate of $\sigma^3$, each spinor has only single component, either upper or lower component: only lower component for $\overline B_-$ and $\overline C_-$, while only upper component for $\overline C_+$.
We multiply $e^{-i\theta' \frac{\sigma^2}{2}}$ both relations and expand the rotation factors $e^{-i(\theta-\theta')\frac{\sigma^2}{2}}$ and $e^{-i\theta' \sigma^2}$ in $\sigma^2$.
We obtain two spinor relations from the relation for $\overline B_-$ by comparing the upper and lower components as follow.
\begin{align}
  \sqrt{E-2\omega} \overline B_- e^{i (E-2\omega)R \cos\theta'} 
  & =  
  \frac{e^{2 M_- R}}{2(1 + \tan^2 \beta)} \sqrt{E} (-i\sigma^3) \cos \frac{\theta-\theta'}{2} \overline C_- e^{i E R \cos\theta} \nonumber \\
  & \quad + \frac{e^{2 M_- R}}{2(1 + \tan^2 \beta)} \sqrt{E-2\omega} \left( - i \sin \theta' \right) \sigma^2 \overline C_+ e^{i (E-2\omega) R \cos\theta'} \,, \label{eq:Bbar-} \\
  0 & = \frac{e^{2 M_- R}}{2(1 + \tan^2 \beta)} \sqrt{E} (-i\sigma^3) \left( - i \sin \frac{\theta-\theta'}{2} \right) \sigma^2 \overline C_- e^{i E R \cos\theta} \nonumber \\
  & \quad + \frac{e^{2 M_- R}}{2(1 + \tan^2 \beta)} \sqrt{E-2\omega} \cos \theta' \overline C_+ e^{i (E-2\omega) R \cos\theta'} \,.
\end{align}
We also obtain two spinor relations from the first equation by the similar procedure. 
\begin{align}
  & \sqrt{E} \overline A_- e^{-i E R  \cos\theta} 
  + \sqrt{E} (i \sin \theta) \sigma^2 \overline B_+ e^{i E R  \cos\theta} \nonumber \\
  & \quad =  
  \frac{e^{2 M_- R}}{2(1 + \tan^2 \beta)} 
  \left[ 
    \sqrt{E} \overline C_- e^{i E R \cos\theta} 
    + \sqrt{E-2\omega} \left( - i \sin \frac{\theta+\theta'}{2} \right) (i\sigma^3 \sigma^2) \overline C_+ e^{i (E-2\omega) R \cos\theta'}
  \right]
   \,, \\
  & \sqrt{E} \cos \theta \overline B_+ e^{i E R  \cos\theta} 
  = \frac{e^{2 M_- R}}{2(1 + \tan^2 \beta)} 
  \left[ 
    \sqrt{E-2\omega} \left( \cos \frac{\theta+\theta'}{2} \right) (i\sigma^3) \overline C_+ e^{i (E-2\omega) R \cos\theta'}
  \right]
   \,,
\end{align}
These four equations give the spinors $\overline B_\pm$ and $\overline C_\pm$ in terms of the spinor $\overline A_-$.
\begin{align}
  \sqrt{E} \overline B_+ e^{i E R \cos\theta} 
  & = \frac{\sin\left( \frac{\theta-\theta'}{2} \right)}{\cos\left( \frac{\theta+\theta'}{2} \right)}(i \sigma^2) \sqrt{E} \overline A_- e^{- i E R  \cos\theta} \,, \\
  \sqrt{E-2\omega} \overline B_- e^{i (E-2\omega) R  \cos\theta} 
  & = i \frac{\cos \theta}{\cos\left( \frac{\theta+\theta'}{2} \right)} \sqrt{E} \overline A_- e^{- i E R  \cos\theta} \,, \\
  \sqrt{E-2\omega} \overline C_+ e^{i (E-2\omega) R  \cos\theta} 
  & = 2 e^{- M_- R} (1+\tan^2 \beta) \frac{\cos\theta \sin\left( \frac{\theta-\theta'}{2} \right)}{\cos^2\left( \frac{\theta+\theta'}{2} \right)} \sigma^2 \sqrt{E} \overline A_- e^{- i E R  \cos\theta} \,, \\
  \sqrt{E} \overline C_- e^{i E R  \cos\theta} 
  & = 2 e^{- M_- R} (1+\tan^2 \beta) \frac{\cos\theta \cos \theta'}{\cos^2\left( \frac{\theta+\theta'}{2} \right)} \sqrt{E} \overline A_- e^{- i E R  \cos\theta} \,,
\end{align}
As discussed in the text, we can consider the scattering with normal incident by taking $\theta = \theta' = 0$ and find $B_+ = C_+ = 0$ and \cref{eq:B-1dscatteringA-,eq:C-1dscatteringA-}. 

We also obtain the spinors for the massive fermions in terms of the incoming spinor $A_-$. 
We show the spinors in the large $R$ limit as confirmation that the matching conditions are solved in a self-consistent way.
As for the Majorana fermions inside the Q-ball, we get
\begin{align}
  \sqrt{M_+} \sin\beta D^+
  & = (1 + i) \frac{\cos \theta}{\cos\left( \frac{\theta+\theta'}{2} \right)} \tan^2\beta e^{- (2 M_-+M_+) R} \left( \cos \frac{\theta'}{2} - \sin \frac{\theta'}{2} \sigma^2 \right) \sqrt{E} \overline A_- e^{- i E R \cos \theta} \,, \\
  \sqrt{M_+} \sin\beta D^-
  & = - (1 - i) \frac{\cos \theta}{\cos^2\left( \frac{\theta+\theta'}{2} \right)} \tan^2\beta (1+\tan^2\beta) e^{- (4 M_-+M_+) R} U \sqrt{E} \overline A_- e^{- i E R \cos \theta} \,, \\
  \sqrt{M_-} \sin\beta F^+
  & = - (1 + i) \frac{\cos \theta}{\cos^2\left( \frac{\theta+\theta'}{2} \right)} (1+\tan^2\beta) e^{- 3 M_- R} U \sqrt{E} \overline A_- e^{- i E R \cos \theta} \,, \\
  \sqrt{M_-} \sin\beta F^-
  & = (1 - i) \frac{\cos \theta}{\cos\left( \frac{\theta+\theta'}{2} \right)} e^{- M_- R} \left( \cos \frac{\theta'}{2} + \sin \frac{\theta'}{2} \sigma^2 \right) \sqrt{E} \overline A_- e^{- i E R \cos \theta} \,.
\end{align}
Here, we define the rotation matrix $U$ as follows. 
\begin{align}
  U = \frac12 \left( \cos \frac{\theta}{2} - 2 \cos \frac{\theta-2\theta'}{2} - \cos \frac{\theta+2\theta'}{2} \right) - \frac12 \left( \sin \frac{\theta}{2} + 2 \sin \frac{\theta-2\theta'}{2} + \sin \frac{\theta+2\theta'}{2} \right) \sigma^2 \,. 
\end{align}
As for the reflected gluino $G_B$ and the transmitted gluino $G_C$, we get 
\begin{align}
  & \sqrt{M_\lambda} G_B
  = -(1 + i) \frac{\cos \theta}{\cos\left( \frac{\theta+\theta'}{2} \right)} \tan\beta \left( \cos \frac{\theta'}{2} - \sin \frac{\theta'}{2} \sigma^2 \right) \sqrt{E} \overline A_- e^{- i E R \cos \theta} \,, \\
  &\sqrt{M_\lambda} G_C
  = (1 - i) \frac{\cos \theta}{\cos^2\left( \frac{\theta+\theta'}{2} \right)} \tan\beta (1+\tan^2\beta) e^{- 2 M_- R} U \sqrt{E} \overline A_- e^{- i E R \cos \theta} \,.
\end{align}

Next, we consider the opposite helicity for the incoming quark, namely right-handed quark, and assign the four-momenta to each fermions as follows. 
\eqs{
  p_{A+} & = (E, E \sin \theta, 0, E \cos\theta) \,, \\
  p_{B+} & = (E + 2 \omega, (E + 2 \omega) \sin \theta', 0, - (E + 2 \omega) \cos\theta') \,, \qquad 
  p_{B-} = (E, E \sin \theta, 0, - E \cos \theta) \,, \\
  p_{C+} & = (E, E\sin\theta, 0, E\cos\theta) \,, \qquad 
  p_{C-} = (E+ 2 \omega, (E + 2 \omega)\sin\theta', 0, (E + 2 \omega)\cos\theta') \,, \\
  \kappa_{B} & = (E+\omega, E \sin\theta, 0, - i M_\lambda) \,, \qquad 
  \kappa_{C} = (E+\omega, E \sin\theta, 0, i M_\lambda) \,, \\
  \kappa_{D}^\pm & = (E+\omega, E \sin\theta, 0, \mp i M_+) \,, \qquad 
  \kappa_{F}^\pm = (E+\omega, E \sin\theta, 0, \mp i M_-) \,. 
  \label{eq:p_assign3}
}
The helicity projections operations that the massless spinors (with the projection onto $z$-axis) are
\eqs{
  \frac12 (\mathbbm{1}-\sigma^3)e^{i\theta \frac{\sigma^2}{2}} A_+ & = 0 \,, \\
  \frac12 (\mathbbm{1}+\sigma^3)e^{- i\theta \frac{\sigma^2}{2}} B_- & = 0 \,, \qquad 
  \frac12 (\mathbbm{1}-\sigma^3)e^{-i\theta' \frac{\sigma^2}{2}} B_+ = 0 \,, \\
  \frac12 (\mathbbm{1}+\sigma^3)e^{i\theta' \frac{\sigma^2}{2}} C_- & = 0 \,, \qquad 
  \frac12 (\mathbbm{1}-\sigma^3)e^{i\theta \frac{\sigma^2}{2}} C_+ = 0 \,,
  \label{eq:spinor_proj2}
}
and the relation between the angles $\theta$ and $\theta'$ is given by 
\begin{align}
  \frac{\sin \theta'}{\sin \theta} = \frac{E}{E+2\omega} \,.
\end{align} 
When the injected momentum is sufficiently smaller than the masses of the fermions, we have the same matching conditions as before. 
Therefore, we have the same relations between $D^\pm$ and $F^\mp$ as given in \cref{eq:D-F+rel,eq:D+F-rel}

We define the spinors for the transmitted quarks projected onto the $z$-axis (spinors with bars) as follows.
\begin{align}
  \overline C_- \equiv e^{i\theta' \frac{\sigma^2}{2}} C_- \,, \qquad 
  \overline C_+ \equiv e^{i\theta \frac{\sigma^2}{2}} C_+ \,.
\end{align}
We note that the spinors with bars are different from those in the previous case since the energy for each helicity is also different from those in the previous case.  
Due to the projections, $\overline C_\pm$ satisfy $(\mathbbm{1}\pm\sigma^3) \overline C_\mp = 0$ and $(\mathbbm{1}\pm\sigma^3) \overline C_\pm = 2 \overline C_\pm$. 
Therefore, the matching condition for quarks at the right-side boundary of the Q-ball is written as follows.  
\begin{align}
  2\sqrt{E+2\omega} e^{- i\theta' \frac{\sigma^2}{2}} \overline C_- e^{i (E+2\omega) \cos\theta' R} 
  & = \sqrt{M_-} \cos \beta (\mathbbm{1} + i \sigma^3) \left( 1 + e^{- 2(M_++M_-)R} \tan^2\beta \right) e^{M_- R} F^+ \nonumber \\
  & \qquad + \sqrt{M_-} \cos \beta (\mathbbm{1} - i \sigma^3) \left( 1 + \tan^2 \beta \right) e^{- M_- R} F^-  \,, \\
  2 \sqrt{E}  e^{- i\theta \frac{\sigma^2}{2}} \overline C_+ e^{i E \cos\theta R} 
  & = - \sqrt{M_-} \cos \beta
  (\mathbbm{1} - i \sigma^3) \left( 1 + e^{- 2(M_++M_-)R} \tan^2\beta \right) e^{M_- R} F^+ \nonumber \\
  & \qquad - \sqrt{M_-} \cos \beta (\mathbbm{1} + i \sigma^3) \left( 1 + \tan^2 \beta  \right) e^{- M_- R} F^-
  \,.
\end{align}
Similarly to the previous case, by multiplying $(\mathbbm{1} \pm i \sigma^3)$ to make the $F^+$ terms proportional to $\mathbbm{1}$, $F^\pm$ are given as a function of $\overline C_\pm$ as follows. 
\begin{align}
   & 2 (1 + \tan^2 \beta) \sqrt{M_-} \cos \beta (- i \sigma^3) e^{- M_- R} F^{-} \nonumber \\
  & \quad =
     \sqrt{E} (1-i\sigma^3) e^{- i\theta' \frac{\sigma^2}{2}} \overline C_- e^{i (E+2\omega) R \cos\theta'}
     + \sqrt{E-2\omega} (1+i\sigma^3) e^{- i\theta \frac{\sigma^2}{2}} \overline C_+ e^{i E R \cos\theta} \,, \\
   & 2 (1 + e^{-2(M_++M_-)R} \tan^2 \beta) \sqrt{M_-} \cos \beta e^{M_- R} F^{+} \nonumber \\
  & \quad = 
    \sqrt{E} (1-i\sigma^3) e^{- i\theta' \frac{\sigma^2}{2}} \overline C_- e^{i (E+2\omega) R \cos\theta}
     - \sqrt{E-2\omega} (1+i\sigma^3) e^{- i\theta \frac{\sigma^2}{2}}\overline C_+ e^{i E R \cos\theta} \,. 
\end{align}

Similarly to $\overline C_\pm$, we define the spinors for the massless quarks at the left-side of the Q-ball projected onto the $z$-axis (spinors with bars) as follows.
\begin{align}
  \overline A_+ \equiv e^{i\theta \frac{\sigma^2}{2}} A_+\,, \qquad 
  \overline B_- \equiv e^{-i\theta \frac{\sigma^2}{2}} B_- \,, \qquad 
  \overline B_+ \equiv e^{-i\theta' \frac{\sigma^2}{2}} B_+ \,.
\end{align}
We note that the bar spinors satisfy $(\mathbbm{1}-\sigma^3) \overline A_+ = 0$, $(\mathbbm{1}+\sigma^3) \overline A_+ = \overline A_+$ and $(\mathbbm{1}\mp\sigma^3) \overline B_\pm = 0$, $(\mathbbm{1}\pm\sigma^3) \overline B_\pm = 2 \overline B_\pm$.
The matching conditions for the massless quarks at $z = - R$ are simplified as follows. 
\begin{align}
  & 
  2 \sqrt{E+2\omega} e^{i\theta' \frac{\sigma^2}{2}} \overline B_+ e^{i (E+2\omega) \cos\theta' R} \nonumber \\\
  & \quad = \sqrt{M_-} \cos \beta 
  (\mathbbm{1} + i \sigma^3) \left( 1 + \tan^2\beta \right) e^{- M_- R} F^{+} \nonumber \\
  & \qquad + \sqrt{M_-} \cos \beta 
  (\mathbbm{1} - i \sigma^3) \left( 1 + e^{- 2(M_++M_-)R} \tan^2 \beta \right) e^{M_- R} F^{-} \,, \\
  & 2 \sqrt{E} e^{- i\theta \frac{\sigma^2}{2}} \overline A_+ e^{-i E \cos\theta R} 
  +2  \sqrt{E} e^{i\theta \frac{\sigma^2}{2}} \overline B_- e^{i E \cos\theta R} \nonumber \\
  & \quad = - \sqrt{M_-} \cos \beta 
  (\mathbbm{1} - i \sigma^3) \left( 1 + \tan^2\beta\right) e^{- M_- R} F^{+} \nonumber \\
  & \qquad - \sqrt{M_-} \cos \beta 
  (\mathbbm{1} + i \sigma^3) \left( 1 + e^{- 2(M_++M_-)R} \tan^2 \beta \right)e^{M_- R} F^{-} \,. 
\end{align}
The spinor $F^\pm$ is given in terms of a function of $\overline C_\pm$.
We obtain the spinor relations among quarks by inserting $F^\pm$ as follows: 
\begin{align}
  & \sqrt{E+2\omega} e^{i\theta' \frac{\sigma^2}{2}} \overline B_+ e^{i (E-2\omega) R \cos\theta'} \nonumber \\
   & =  
   \frac{a_+}{4} \sqrt{E+2\omega} e^{-i\theta' \frac{\sigma^2}{2}} \overline C_- e^{i (E+2\omega) R\cos\theta'} 
  + \frac{a_-}{4} \sqrt{E} (i\sigma^3) e^{-i\theta \frac{\sigma^2}{2}} \overline C_+ e^{i E R \cos\theta} \,, \\
  & \sqrt{E} e^{- i\theta \frac{\sigma^2}{2}} \overline A_+ e^{-i E R \cos\theta} 
  + \sqrt{E} e^{i\theta \frac{\sigma^2}{2}} \overline B_- e^{i E R \cos\theta} \nonumber \\
   & =  
   \frac{a_-}{4} \sqrt{E+2\omega} (-i\sigma^3) e^{-i\theta' \frac{\sigma^2}{2}} \overline C_- e^{i (E+2\omega) R \cos\theta'}
  + \frac{a_+}{4} \sqrt{E} e^{-i\theta \frac{\sigma^2}{2}} \overline C_+ e^{i E R \cos\theta} \,.
\end{align}

As with the previous discussion, we take the large $R$ limit and get the spinors for the massless quarks in terms of $A_+$. 
The spinors $F^\pm$ are given by 
\begin{align}
  & 2 \sqrt{M_-} \cos \beta (1 + \tan^2 \beta) (- i \sigma^3) e^{- M_- R} F^{-} \nonumber \\
  & \quad = 
  \sqrt{E} (1-i\sigma^3) e^{- i\theta' \frac{\sigma^2}{2}} \overline C_- e^{i (E+2\omega) R\cos\theta}
  + \sqrt{E-2\omega} (1+i\sigma^3) e^{- i\theta' \frac{\sigma^2}{2}}\overline C_+ e^{i E R\cos\theta} \,, \\
  & 2 \sqrt{M_-} \cos \beta e^{M_- R} F^{+} \nonumber \\
  & \quad = 
  \sqrt{E+2\omega} (1-i\sigma^3) e^{- i\theta \frac{\sigma^2}{2}} \overline C_- e^{i (E+2\omega) R}
  - \sqrt{E} (1+i\sigma^3) e^{- i\theta' \frac{\sigma^2}{2}} \overline C_+ e^{i E R \cos\theta} \,. 
\end{align}
The matching conditions for quarks at the left-side boundary of the Q-ball are simplified as follows. 
\begin{align}
  & \sqrt{E+2\omega} (\mathbbm{1} - i \sigma^3) e^{i\theta' \frac{\sigma^2}{2}} \overline B_+ e^{i (E+2\omega) R \cos\theta'} \nonumber \\\
  & \quad = \sqrt{M_-} \cos \beta \left( 1 + \tan^2\beta \right) e^{- M_- R} F^{+} 
  + \sqrt{M_-} \cos \beta (- i \sigma^3) e^{M_- R} F^{-} \,, \\
  & \sqrt{E} (\mathbbm{1} + i \sigma^3) e^{- i\theta \frac{\sigma^2}{2}} \overline A_- e^{-i E R \cos\theta} 
  + \sqrt{E} (\mathbbm{1} + i \sigma^3) e^{i\theta \frac{\sigma^2}{2}} \overline B_- e^{i E R \cos\theta} \nonumber \\
  & \quad = - \sqrt{M_-} \cos \beta \left( 1 + \tan^2\beta\right) e^{- M_- R} F^{+} 
  - \sqrt{M_-} \cos \beta (i \sigma^3) e^{M_- R} F^{-} \,. 
\end{align}
The spinor relations among quarks are simplified as follows. 
\begin{align}
  & \sqrt{E+2\omega} e^{i\theta' \frac{\sigma^2}{2}} \overline B_+ e^{i (E+2\omega) R \cos\theta'} \nonumber \\
   & \quad =  
  \frac{e^{2 M_- R}}{2(1 + \tan^2 \beta)} \left[ 
    \sqrt{E+2\omega} e^{-i\theta' \frac{\sigma^2}{2}}  \overline C_- e^{i (E+2\omega) R \cos\theta'}
    + \sqrt{E} (i\sigma^3) e^{-i\theta \frac{\sigma^2}{2}} \overline C_+ e^{i E R \cos\theta}
  \right] \,, \\
  & \sqrt{E} e^{- i\theta \frac{\sigma^2}{2}} \overline A_- e^{-i E R \cos\theta} 
  + \sqrt{E} e^{i\theta \frac{\sigma^2}{2}} \overline B_- e^{i E R \cos\theta} \nonumber \\
   & \quad =  
   \frac{e^{2 M_- R}}{2(1 + \tan^2\beta)} \left[ 
    \sqrt{E+2\omega} (-i\sigma^3) e^{-i\theta' \frac{\sigma^2}{2}} \overline C_- e^{i (E+2\omega) R \cos\theta'}
    + \sqrt{E} e^{-i\theta \frac{\sigma^2}{2}} \overline C_+ e^{i E R \cos\theta}
  \right] \,.
\end{align}

We find the solutions for the massless quarks as follows. 
\begin{align}
  \sqrt{E+2\omega} \overline B_+ e^{i (E+2\omega) R \cos\theta'} 
  & = i \frac{\cos \theta}{\cos\left( \frac{\theta+\theta'}{2} \right)} \sqrt{E} \overline A_+ e^{- i E R \cos\theta} \,, \\
  \sqrt{E} \overline B_- e^{i E R \cos\theta} 
  & = \frac{\sin\left( \frac{\theta-\theta'}{2}\right)}{\cos\left( \frac{\theta+\theta'}{2} \right)} \sqrt{E} (- i \sigma^2) \overline A_+ e^{- i E R  \cos\theta} \,,
\end{align}
\begin{align}
  \sqrt{E} \overline C_+ e^{i E R  \cos\theta} 
  & = 2 e^{- M_- R} (1+\tan^2 \beta) \frac{\cos\theta \cos \theta'}{\cos^2\left( \frac{\theta+\theta'}{2} \right)} \sqrt{E} \overline A_+ e^{- i E R \cos\theta} \,, \\
  \sqrt{E+2\omega} \overline C_- e^{i (E+2\omega) R \cos\theta'} 
  & = 2 e^{- M_- R} (1+\tan^2 \beta) \frac{\cos\theta \sin\left( \frac{\theta-\theta'}{2} \right)}{\cos^2\left( \frac{\theta+\theta'}{2} \right)} \sigma^2 \sqrt{E} \overline A_+ e^{- i E R  \cos\theta} \,, 
\end{align}

\subsection{Spherical Wave Scattering}

Now, we give the detail of the calculation of the matching conditions for the scattering of the colorless quarks with the finite-size Q-ball in \cref{sec:spherical}.
Since we use the spherical wave state, it is convenient to use the four-component fermions. 
In contrast to the previous appendix, a matching condition for the Dirac spinors give two equations corresponding to the upper and the lower two-components of the four-component fermions. 
Hence, once we have the solution for the Dirac spinor, it is trivial that the charge-conjugated solutions (required for Majorana fermions) satisfy the matching conditions.

The massless solution is given by a linear combination of the spherical Bessel functions. 
Therefore, the outgoing and incoming Dirac fermions are described by the spherical Hankel functions $h_\kappa^{(1)}(x)$ and $h_\kappa^{(2)}(x)$, which behave as $e^{i x}/x$ and $e^{- i x}/x$ at large $x$, respectively.
We choose the overall coefficients of $P_\kappa$ and $Q_\kappa$ to be convenient for constructing the plain-wave solution in terms of the spinor spherical harmonics.
Meanwhile, the massive solutions for gluinos and the Majorana fermions inside the Q-ball are described by the modified Bessel function $k_n(x)$ and $i_n(x)$, respectively.
See Eqs.~\eqref{eq:PQ_A} to \eqref{eq:PQ_F} for the explicit expressions of $P_\kappa$ and $Q_\kappa$.

We match the wave functions at the boundary of the Q-ball located at $r = R$.
We assume that the quarks are massless and that the mass of the other fermions is larger than the energy of the incoming quark.
We consider the case where the incoming quark has a left-handed chirality (with $m=-1/2$). 
The wave functions with the definite chirality are given by \cref{eq:wf_chirality}.
The matching conditions for upper and lower two-components of the gluino wave function (\ref{eq:mc_3d_gluino}) read, respectively,
\begin{align}
  & \sum_m \left\{ 
    e^{-i(E-\omega)t} i \left[ p^{C_{+}}_{\varkappa,m} + p^{C_{+}}_{-\varkappa,m} \right] 
    + e^{-i(E-\omega)t} i \left[ - p^{C_{-}}_{\varkappa,m} + p^{C_{-}}_{-\varkappa,m} \right] 
  \right\} \nonumber \\
  & = \sum_m \left\{ 
    \cos\beta e^{-i(E-\omega)t} i \left[ p^{D_{+}}_{\varkappa,m} + p^{D_{+}}_{-\varkappa,m} \right] 
    + \cos\beta e^{-i(E-\omega)t} i \left[ - p^{D_{-}}_{\varkappa,m} + p^{D_{-}}_{-\varkappa,m} \right] 
  \right. \nonumber \\ 
  & \left. 
    - \sin \beta e^{-i(E-\omega)t} i \left[ p^{F_{+}}_{\varkappa,m} + p^{F_{+}}_{-\varkappa,m} \right] 
    - \sin \beta e^{-i(E-\omega)t} i \left[ - p^{F_{-}}_{\varkappa,m} + p^{F_{-}}_{-\varkappa,m} \right] 
  \right\} \,, \\
  & \sum_m \left\{ 
    e^{-i(E-\omega)t} \left[ q^{C_{+}}_{\varkappa,m} + q^{C_{+}}_{-\varkappa,m} \right] 
    + e^{-i(E-\omega)t} \left[ - q^{C_{-}}_{\varkappa,m} + q^{C_{-}}_{-\varkappa,m} \right] 
  \right\} \nonumber \\
  & = \sum_m \left\{ 
    \cos\beta e^{-i(E-\omega)t} \left[ q^{D_{+}}_{\varkappa,m} + q^{D_{+}}_{-\varkappa,m} \right] 
    + \cos\beta e^{-i(E-\omega)t} \left[ - q^{D_{-}}_{\varkappa,m} + q^{D_{-}}_{-\varkappa,m} \right] 
  \right. \nonumber \\ 
  & \quad \left. 
    - \sin \beta e^{-i(E-\omega)t} \left[ q^{F_{+}}_{\varkappa,m} + q^{F_{+}}_{-\varkappa,m} \right] 
    - \sin \beta e^{-i(E-\omega)t} \left[ - q^{F_{-}}_{\varkappa,m} + q^{F_{-}}_{-\varkappa,m} \right] \right\} \,.
\end{align}
Here, we suppress the arguments of the functions $p$ and $q$, and the superscripts of $p$ and $q$ label the species of fermions.
Since we choose the same energy for the massive fermions, $E-\omega$, both equalities hold at any time. 
Due to the orthogonality condition for the spinor spherical harmonics, we find 
\begin{align}
  \int d \cos \theta d \phi \, \Omega^\dag_{\kappa,m'} p_{\varkappa,m}(\mathbf{R}) & = P_{\varkappa} (ER) \delta_{\kappa,\varkappa} \delta_{m',m} \,, 
  \label{eq:orth_ssh1} \\
  \int d \cos \theta d \phi \, \Omega^\dag_{\kappa,m'} q_{\varkappa,m}(\mathbf{R}) & = Q_{\varkappa} (ER) \delta_{\kappa,-\varkappa} \delta_{m',m} \,.
  \label{eq:orth_ssh2}
\end{align}
Therefore, the matching conditions give independent relations for each $\varkappa$ and $m$, and we find the relations among the radial functions as follows.
\begin{align}
  P^{C_{+,m}}_\varkappa - P^{C_{-,m}}_\varkappa 
  & = \cos\beta \left( P^{D_{+,m}}_\varkappa - P^{D_{-,m}}_\varkappa \right)
  - \sin \beta \left( P^{F_{+,m}}_\varkappa - P^{F_{-,m}}_\varkappa \right) \,, \\
  P^{C_{+,m}}_{-\varkappa} + P^{C_{-,m}}_{-\varkappa} 
  & = \cos\beta \left( P^{D_{+,m}}_{-\varkappa} + P^{D_{-,m}}_{-\varkappa} \right)
  - \sin \beta \left( P^{F_{+,m}}_{-\varkappa} + P^{F_{-,m}}_{-\varkappa} \right) \,, \\
  Q^{C_{+,m}}_\varkappa - Q^{C_{-,m}}_\varkappa 
  & = \cos\beta \left( Q^{D_{+,m}}_\varkappa - Q^{D_{-,m}}_\varkappa \right)
  - \sin \beta \left( Q^{F_{+,m}}_\varkappa - Q^{F_{-,m}}_\varkappa \right) \,, \\
  Q^{C_{+,m}}_{-\varkappa} + Q^{C_{-,m}}_{-\varkappa} 
  & = \cos\beta \left( Q^{D_{+,m}}_{-\varkappa} + Q^{D_{-,m}}_{-\varkappa} \right)
  - \sin \beta \left( Q^{F_{+,m}}_{-\varkappa} + Q^{F_{-,m}}_{-\varkappa} \right) \,.
\end{align}

The wave functions inside the Q-ball corresponding to quarks are multiplied by the time-dependent phase factor as given by \cref{eq:wf_oscbkg}, while our energy assignment is given in \cref{eq:E_assign}.
The matching condition for the upper two-component of the quark wave function gives 
\begin{align}
  & e^{-iEt} i (- p^{A_-}_{\varkappa,-1/2} + p^{A_-}_{-\varkappa,-1/2} ) 
  + \sum_m \left\{  
    e^{-i(E-2\omega)t} i (p^{B_{+}}_{\varkappa,m} + p^{B_{+}}_{-\varkappa,m}) 
    + e^{-iEt} i (- p^{B_-}_{\varkappa,m} + p^{B_-}_{-\varkappa,m} ) 
  \right\} \nonumber \\
  & = e^{-i(E-\omega)t} \sum_m \left\{ \sin\beta 
    \left[ e^{i \omega t} (i p^{D_{+}}_{\varkappa,m} + i p^{D_{+}}_{-\varkappa,m} 
  + q^{D_{+}}_{\varkappa,m} + q^{D_{+}}_{-\varkappa,m})
  + e^{-i \omega t} (i p^{D_{+}}_{\varkappa,m} + i p^{D_{+}}_{-\varkappa,m} 
  - q^{D_{+}}_{\varkappa,m} - q^{D_{+}}_{-\varkappa,m}) \right] \right. \nonumber \\
  & + \sin\beta 
    \left[ e^{i \omega t} (- i p^{D_{-}}_{\varkappa,m} + i p^{D_{-}}_{-\varkappa,m} - q^{D_{-}}_{\varkappa,m} + q^{D_{-}}_{-\varkappa,m}) 
  + e^{-i \omega t} (- i p^{D_{-}}_{\varkappa,m} + i p^{D_{-}}_{-\varkappa,m} 
  + q^{D_{-}}_{\varkappa,m} - q^{D_{-}}_{-\varkappa,m}) \right] \nonumber \\
  & + \cos\beta 
    \left[ e^{i \omega t} (i p^{F_{+}}_{\varkappa,m} + i p^{F_{+}}_{-\varkappa,m} + q^{F_{+}}_{\varkappa,m} + q^{F_{+}}_{-\varkappa,m}) 
  + e^{-i \omega t} (i p^{F_{+}}_{\varkappa,m} + i p^{F_{+}}_{-\varkappa,m} - q^{F_{+}}_{\varkappa,m} - q^{F_{+}}_{-\varkappa,m}) \right] \nonumber \\
  & + \left. \cos\beta 
    \left[ e^{i \omega t} (- i p^{F_{-}}_{\varkappa,m} + i p^{F_{-}}_{-\varkappa,m} - q^{F_{-}}_{\varkappa,m} + q^{F_{-}}_{-\varkappa,m}) 
  + e^{-i \omega t} (- i p^{F_{-}}_{\varkappa,m} + i p^{F_{-}}_{-\varkappa,m} 
  + q^{F_{-}}_{\varkappa,m} - q^{F_{-}}_{-\varkappa,m}) \right] \right\} \,.
\end{align}
Meanwhile, the matching condition for the lower two-component of the quark wave function gives 
\begin{align}
  & e^{-iEt} (- q^{A_-}_{\varkappa,-1/2} + q^{A_-}_{-\varkappa,-1/2}) + \sum_m \left[  
    e^{-iEt} (- q^{B_-}_{\varkappa,m} + q^{B_-}_{-\varkappa,m})
    + e^{-i(E-2\omega)t} (q^{B_{+}}_{\varkappa,m} + q^{B_{+}}_{-\varkappa,m}) 
    \right] \nonumber \\
  & = \sum_m e^{-i(E-\omega)t} \left\{ 
    \sin\beta \left[ e^{i \omega t} (i p^{D_{+}}_{\varkappa,m} + i p^{D_{+}}_{-\varkappa,m} 
  + q^{D_{+}}_{\varkappa,m} + q^{D_{+}}_{-\varkappa,m})
  - e^{- i \omega t} (i p^{D_{+}}_{\varkappa,m} + i p^{D_{+}}_{-\varkappa,m} 
  - q^{D_{+}}_{\varkappa,m} - q^{D_{+}}_{-\varkappa,m})  \right] \right. \nonumber \\
  & + \sin\beta \left[ e^{i \omega t} (- i p^{D_{-}}_{\varkappa,m} + i p^{D_{-}}_{-\varkappa,m} - q^{D_{-}}_{\varkappa,m} + q^{D_{-}}_{-\varkappa,m}) 
  - e^{- i \omega t} (- i p^{D_{-}}_{\varkappa,m} + i p^{D_{-}}_{-\varkappa,m} 
  + q^{D_{-}}_{\varkappa,m} - q^{D_{-}}_{-\varkappa,m}) \right] \nonumber \\
  & + \cos\beta \left[ e^{i \omega t} (i p^{F_{+}}_{\varkappa,m} + i p^{F_{+}}_{-\varkappa,m} 
  + q^{F_{+}}_{\varkappa,m} + q^{F_{+}}_{-\varkappa,m}) 
  - e^{- i \omega t} (i p^{F_{+}}_{\varkappa,m} + i p^{F_{+}}_{-\varkappa,m} 
  - q^{F_{+}}_{\varkappa,m} - q^{F_{+}}_{-\varkappa,m}) \right]\nonumber \\
  & + \left. \cos\beta \left[ e^{i \omega t} (- i p^{F_{-}}_{\varkappa,m} + i p^{F_{-}}_{-\varkappa,m} 
  - q^{F_{-}}_{\varkappa,m} + q^{F_{-}}_{-\varkappa,m})
  - e^{- i \omega t} (- i p^{F_{-}}_{\varkappa,m} + i p^{F_{-}}_{-\varkappa,m} 
  + q^{F_{-}}_{\varkappa,m} - q^{F_{-}}_{-\varkappa,m})
  \right] \right\} \,.
\end{align}
Each matching condition gives two time-independent relations among functions $p$ and $q$, which are from the terms proportional to $e^{- i E t}$ and $e^{- i (E - 2 \omega) t}$.
Using the orthogonal relations of the spherical spinors (\ref{eq:orth_ssh1}) and (\ref{eq:orth_ssh2}), we obtain four equations from each time-independent relations. 
As a result, we have sixteen relations among $P_{\pm\varkappa}$ and $Q_{\pm\varkappa}$, 
\begin{align}
  - P_\varkappa^{A_-} - P_\varkappa^{B_{-1/2,-}} + P_\varkappa^{B_{-1/2,+}}
  & = 2 \sin\beta (P_\varkappa^{D_{-1/2,+}} - P_\varkappa^{D_{-1/2,-}} ) 
  + 2 \cos\beta (P_\varkappa^{F_{-1/2,+}} - P_\varkappa^{F_{-1/2,-}} ) \,, \\ 
  - P_\varkappa^{A_-} - P_\varkappa^{B_{-1/2,-}} - P_\varkappa^{B_{-1/2,+}}
  & = 2 i \sin\beta ( Q_{-\varkappa}^{D_{-1/2,+}} + Q_{-\varkappa}^{D_{-1/2,-}} )  
  + 2 i \cos\beta ( Q_{-\varkappa}^{F_{-1/2,+}} + Q_{-\varkappa}^{F_{-1/2,-}} ) \,, \\ 
  P_{-\varkappa}^{A_-} + P_{-\varkappa}^{B_{-1/2,-}} + P_{-\varkappa}^{B_{-1/2,+}}
  & = 2 \sin\beta ( P_{-\varkappa}^{D_{-1/2,+}} + P_{-\varkappa}^{D_{-1/2,-}} ) 
  + 2 \cos\beta ( P_{-\varkappa}^{F_{-1/2,+}} + P_{-\varkappa}^{F_{-1/2,-}} ) \,, \\ 
  P_{-\varkappa}^{A_-} + P_{-\varkappa}^{B_{-1/2,-}} - P_{-\varkappa}^{B_{-1/2,+}}
  & = 2 i \sin\beta ( Q_{\varkappa}^{D_{-1/2,+}} - Q_{\varkappa}^{D_{-1/2,-}} ) 
  + 2 i \cos\beta ( Q_{\varkappa}^{F_{-1/2,+}} - Q_{\varkappa}^{F_{-1/2,-}} ) \,, \\
  - P_\varkappa^{B_{1/2,-}} + P_\varkappa^{B_{1/2,+}}
  & = 2 \sin\beta ( P_\varkappa^{D_{1/2,+}} - P_\varkappa^{D_{1/2,-}} ) 
  + 2 \cos\beta ( P_\varkappa^{F_{1/2,+}} - P_\varkappa^{F_{1/2,-}} ) \,, \\ 
  P_\varkappa^{B_{-1/2,-}} + P_\varkappa^{B_{-1/2,+}} 
  & = - 2 i \sin\beta ( Q_{-\varkappa}^{D_{1/2,+}} + Q_{-\varkappa}^{D_{1/2,-}} )  
  - 2 i \cos\beta ( Q_{-\varkappa}^{F_{1/2,+}} + Q_{-\varkappa}^{F_{1/2,-}} ) \,, \\ 
  P_{-\varkappa}^{B_{1/2,-}} + P_{-\varkappa}^{B_{1/2,+}}
  & = 2 \sin\beta ( P_{-\varkappa}^{D_{1/2,+}} + P_{-\varkappa}^{D_{1/2,-}} ) 
  + 2 \cos\beta ( P_{-\varkappa}^{F_{1/2,+}} + P_{-\varkappa}^{F_{1/2,-}} ) \,, \\ 
  P_{-\varkappa}^{B_{1/2,-}} - P_{-\varkappa}^{B_{1/2,+}}
  & = 2 i \sin\beta ( Q_{\varkappa}^{D_{1/2,+}} - Q_{\varkappa}^{D_{1/2,-}} ) 
  + 2 \cos\beta ( Q_{\varkappa}^{F_{1/2,+}} - Q_{\varkappa}^{F_{1/2,-}} ) \,, 
\end{align}
and 
\begin{align}
  Q_{-\varkappa}^{A_{-,-1/2}} + Q_{-\varkappa}^{B_{-,-1/2}} + Q_{-\varkappa}^{B_{+,-1/2}}
  & = 2 \sin\beta (Q_{-\varkappa}^{D_{+,-1/2}} + Q_{-\varkappa}^{D_{-,-1/2}} )
  + 2 \cos\beta (Q_{-\varkappa}^{F_{+,-1/2}} + Q_{-\varkappa}^{F_{-,-1/2}})  \,, \\ 
  - Q_{-\varkappa}^{A_{-,-1/2}} - Q_{-\varkappa}^{B_{-,-1/2}} + Q_{-\varkappa}^{B_{+,-1/2}}
  & = 2 i \sin\beta (P_{\varkappa}^{D_{+,-1/2}} + P_{\varkappa}^{D_{-,-1/2}}) - 2 i \cos\beta (P_{\varkappa}^{F_{+,-1/2}} - P_{\varkappa}^{F_{-,-1/2}}) \,, \\
  - Q_{\varkappa}^{A_{-,-1/2}} - Q_{\varkappa}^{B_{-,-1/2}} + Q_{\varkappa}^{B_{+,-1/2}} 
  & = 2 \sin\beta (Q_{\varkappa}^{D_{+,-1/2}} - Q_{\varkappa}^{D_{-,-1/2}} )
  + 2 \cos\beta (Q_{\varkappa}^{F_{+,-1/2}} - Q_{\varkappa}^{F_{-,-1/2}})  \,, \\ 
  Q_{\varkappa}^{A_{-,-1/2}} + Q_{\varkappa}^{B_{-,-1/2}} + Q_{\varkappa}^{B_{+,-1/2}} 
  & = 2 i \sin\beta (P_{-\varkappa}^{D_{+,-1/2}} + P_{-\varkappa}^{D_{-,-1/2}}) 
  + 2 i \cos\beta (P_{-\varkappa}^{F_{+,-1/2}} + P_{-\varkappa}^{F_{-,-1/2}}) \,, \\
  Q_{- \varkappa}^{B_{-,1/2}} + Q_{- \varkappa}^{B_{+,1/2}} 
  & = 2 \sin\beta (Q_{-\varkappa}^{D_{+,1/2}} + Q_{-\varkappa}^{D_{-,1/2}}) 
  + 2 \cos\beta (Q_{-\varkappa}^{F_{+,1/2}} + Q_{-\varkappa}^{F_{-,1/2}}) \,, \\
  Q_{- \varkappa}^{B_{-,1/2}} - Q_{- \varkappa}^{B_{+,1/2}} 
  & = - 2 i \sin\beta (P_\varkappa^{D_{+,1/2}} - P_\varkappa^{D_{-,1/2}}) 
  - 2 i \cos\beta (P_\varkappa^{F_{+,1/2}} - P_\varkappa^{F_{-,1/2}}) \,, \\ 
  - Q_{\varkappa}^{B_{-,1/2}} + Q_{\varkappa}^{B_{+,1/2}} 
  & = 2 \sin\beta (Q_{\varkappa}^{D_{+,1/2}} - Q_{\varkappa}^{D_{-,1/2}}) 
  + 2 \cos\beta (Q_{\varkappa}^{F_{+,1/2}} - Q_{\varkappa}^{F_{-,1/2}})  \,, \\
  - Q_{\varkappa}^{B_{-,1/2}} - Q_{\varkappa}^{B_{+,1/2}} 
  & = - 2 i \sin\beta (P_{-\varkappa}^{D_{+,1/2}} + P_{-\varkappa}^{D_{-,1/2}}) 
  - 2 i \cos\beta (P_{-\varkappa}^{F_{+,1/2}} + P_{-\varkappa}^{F_{-,1/2}}) \,.
\end{align}
The equations with $m = 1/2$ consist only of the radial functions for the reflected quark and the Majorana fermions inside the Q-ball since we assume the incoming quark has the left-handed chirality ($m = -1/2$) and due to the angular momentum conservation.
Thus, the coefficients of these $P_{\pm\varkappa}$ and $Q_{\pm\varkappa}$ vanish, and we drop the subscripts regarding $m$ in the following and in the text.

In general, the overall coefficients of $P_\varkappa$ ($Q_\varkappa$) and $P_{- \varkappa}$ ($Q_{- \varkappa}$) for each wave function are independent.
As shown in \cref{eq:plain_expansion}, we use the linear combination for an incoming state with a definite chirality, and the coefficients for $P^{A_\pm}_{\pm\varkappa}$ ($Q^{A_\pm}_{\pm\varkappa}$) are given by
\begin{align}
  A_\varkappa = \pm i^{\varkappa-1} \sqrt{\varkappa} A_{\pm} \,, \quad 
  A_{-\varkappa} = i^{\varkappa-1} \sqrt{\varkappa} A_{\pm} \,. 
\end{align}
Here, $+$ ($-$) corresponds to the right (left) chirality. 
When the energy of the incoming quark is large, $E R \gg \varkappa$, and the other fermion masses are large, the relation between $A_{\pm\varkappa}$ and $B_{\pm\varkappa}$ leads to  
\begin{align}
  B_\varkappa = \mp (-1)^{-\varkappa} i e^{-2i E R} i^{\varkappa-1} \sqrt{\varkappa} A_{\pm}\,, \quad 
  B_{-\varkappa} = (-1)^{-\varkappa} i e^{-2i E R} i^{\varkappa-1} \sqrt{\varkappa} A_{\pm} \,.
\end{align}
The relative sign of $B_\varkappa$ and $B_{-\varkappa}$ is opposite to that of $A_\varkappa$ and $A_{-\varkappa}$. 
In other words, when a left-handed quark comes into the Q-ball, the right-handed anti-quark is reflected. 
Meanwhile, when the energy of the incoming quark is sufficiently small such that $E R  \ll \varkappa$, the relation between the amplitudes of the incoming quark and the outgoing quark is given by $A_\kappa = B_\kappa$.
Thus, the relative sign of  $B_\varkappa$ and $B_{-\varkappa}$ is same as that of $A_\varkappa$ and $A_{-\varkappa}$.
When a left-handed quark comes into the Q-ball, the left-handed quark is reflected. 

We find the solution to the matching conditions for the generic energy $E$ (while $M_{(\pm)} R \gg \varkappa$) as follows:
\begin{align}
  B_{+} & = \sqrt{\frac{E}{E-2\omega}} \frac{h_{\varkappa-1}^{(1)} (E R) h_\varkappa^{(2)} (E R) - h_\varkappa^{(1)} (E R) h_{\varkappa-1}^{(2)} (ER)}{h_{\varkappa-1}^{(1)} (E R) h_{\varkappa-1}^{(1)} [(E-2\omega)R] - h_\varkappa^{(1)} (E R) h_\varkappa^{(1)} [(E-2\omega)R]} A_{-} \,, \\
  B_{-} & = - \frac{h_{\varkappa-1}^{(2)} (E R) h_{\varkappa-1}^{(1)} [(E-2\omega)R] - h_\varkappa^{(2)} (E R) h_\varkappa^{(1)} [(E-2\omega)R]}{h_{\varkappa-1}^{(1)} (E R)h_{\varkappa-1}^{(1)} [(E-2\omega)R] - h_\varkappa^{(1)} (E R) h_\varkappa^{(1)} [(E-2\omega)R]} A_{-} \,, \\
  C_{+} & = \frac{1}{4} (1+ i) \sqrt{E M_\lambda} R \tan \beta e^{M_\lambda R} \left\{ h_{\varkappa-1}^{(1)} [(E-2\omega) R] + i h_{\varkappa}^{(1)} [(E-2\omega) R] \right\} \nonumber \\ 
  & \qquad \times \frac{h_{\varkappa-1}^{(1)} (E R) h_{\varkappa}^{(2)} (E R) - h_\varkappa^{(1)} (E R) h_{\varkappa-1}^{(2)} (E R)}{h_{\varkappa-1}^{(1)} (E R) h_{\varkappa-1}^{(1)} [(E-2\omega)R] - h_\varkappa^{(1)} (E R) h_\varkappa^{(1)} [(E-2\omega)R]} A_{-} \,, \\
  C_{-} & = - \frac{1}{4} (1+ i) \sqrt{E M_\lambda} R \tan \beta e^{M_\lambda R} \left\{ h_{\varkappa}^{(1)} [(E-2\omega) R] + i h_{\varkappa-1}^{(1)} [(E-2\omega) R] \right\} \nonumber \\ 
  & \qquad \times \frac{h_{\varkappa-1}^{(1)} (E R) h_{\varkappa}^{(2)} (E R) - h_\varkappa^{(1)} (E R) h_{\varkappa-1}^{(2)} (E R)}{h_{\varkappa-1}^{(1)} (E R) h_{\varkappa-1}^{(1)} [(E-2\omega)R] - h_\varkappa^{(1)} (E R) h_\varkappa^{(1)} [(E-2\omega)R]} A_{-} \,,
\end{align}
and 
\begin{align}
  D_{\pm} & = 0 \,, \\
  F_{+} & = \frac{2}{\cos\beta} \sqrt{\frac{M_-}{M_\lambda}} e^{-i E R} e^{-(M_\lambda + M_- )R} C_{+} \,, \\
  F_{-} & = \frac{2}{\cos\beta} \sqrt{\frac{M_-}{M_\lambda}} e^{-i E R} e^{-(M_\lambda + M_- )R} C_{-} \,. 
\end{align}
See \cref{eq:Blimit_1,eq:Blimit_2,eq:Blimit_3} in the text for the coefficients for reflected quarks $B_{\pm}$ in some limits. 

\section{\texorpdfstring{$S$}{S}-matrix for the Q-ball Scattering \label{app:QM_Qball}}

We use the two-component spinors to describe the Majorana fermions inside the Q-ball and the Weyl fermions outside the Q-ball in (3+1)--dimensional spacetime, which correspond to the fermions with two degrees of freedom. 
In \cref{sec:infinite}, we consider the oblique-incident scattering of the (colorless) quark on the infinitely-large Q-ball wall. 
The incoming quark may be regarded as the (1+1)--dimensional fermion for the normal-incident scattering for $\theta = 0$: the fermions with two degrees of freedom in the (1+1)--dimension are the Dirac fermion.

Now, we discuss the correspondence between the fermions in terms of the (3+1)--dimension and in terms of the (1+1)--dimension.
Let us assume that the fermions move along the $z$ axis. 
We first consider the correspondence inside the Q-ball: the fermions are described by the Majorana fermions in terms of the (3+1)--dimension, while are described by the Dirac fermions in terms of the (1+1)--dimension. 
The Majorana fermions do not have any internal charges, but the Dirac fermions in terms of the (1+1)--dimension should have a kind of ``charges''. 
The Majorana fermions have the up- and down-spin components, which get the opposite phase factors under the rotation around the $z$-axis. 
Meanwhile, since the fermions in the (1+1)--dimension do not feel the rotation around the $z$-axis, it is regarded as an internal symmetry. 
The Majorana fermion in the (3+1)--dimension is interpreted as the Dirac fermion in the (1+1)--dimension, which is composed of two degrees of freedom with the opposite ``charges'' arising from the spin degrees of freedom in the (3+1)--dimension.

We may consider the $S$-matrix structure of the one-dimensional scattering process. 
The $S$-matrix element is defined as the expansion of ``in'' state in terms of ``out'' state.
\begin{align}
  | a \rangle_\mathrm{in} 
  = \sum_\beta S_{b a} | b \rangle_\mathrm{out} \,.
\end{align}
Here, $a \,, b$ denote the label of degrees of freedom.

We consider the one-dimensional scattering with $\omega = 0$. 
The ``in'' state is labelled by mass inside the Q-ball, helicity, and the moving direction.
Since a massless fermion just passes through the Q-ball, its $S$-matrix elements are trivial. 
Moreover, the $S$-matrix is block-diagonalized with respect to the mass inside the Q-ball: 
\begin{align}
  S' = 
  \begin{pmatrix}
    \mathbf{1}_4 & 0 \\
    0 & S_4
  \end{pmatrix} \,.
\end{align}
We define the ``in'' state as a direct product of the mass eigenstates and the states labeled by the helicity and the moving direction.
We choose the four-component basis consisting of the ``in'' states as 
\begin{align}
  \begin{pmatrix}
    R + \\
    L - \\
    R - \\
    L + \\
  \end{pmatrix}_\mathrm{in} \,.
\end{align}
Here, $L \,, R$ denotes the moving direction, and $\pm$ denotes the helicity $\pm 1/2$. 
In \cref{sec:infinite}, we find that the right-moving quark is reflected as the left-moving anti-quark on the Q-ball boundary for $\theta = 0$. 
The coefficient of the reflected anti-quark is given by $B = i e^{-2iER} A$, and hence ${}_\mathrm{out}\langle L +| R - \rangle_\mathrm{in} = i e^{-2iER} $ and other $S$-matrix elements associated with $| R - \rangle_\mathrm{in}$ vanish.

We find the other matrix elements using the (3+1)-dimensional $CP$ and $T$ [(1+1)-dimensional $P$ and $CT$] transformations.
Under $CP$ and time-reversal transformations, the ``in'' states transform as,
\begin{align}
  CP
  \begin{pmatrix}
    R + \\
    L - \\
    R - \\
    L + \\
  \end{pmatrix}_\mathrm{in} = i 
  \begin{pmatrix}
    L - \\
    R + \\
    L + \\
    R - \\
  \end{pmatrix}_\mathrm{in} \,, \qquad 
  T
  \begin{pmatrix}
    R + \\
    L - \\
    R - \\
    L + \\
  \end{pmatrix}_\mathrm{in} 
  = 
  \begin{pmatrix}
    L + \\
    R - \\
    - L - \\
    - R + \\
  \end{pmatrix}_\mathrm{out}
  \,.
\end{align}
Here, we take into account the intrinsic parity of the Majorana fermion to be $\eta = + i$ and the relative sign under the time-reversal (originating from $i \sigma_2$). 
Only four entries of $S_4$ are non-zero, and all of them are equal:
\begin{align}
  S_4 = i e^{-2iER}
  \begin{pmatrix}
    0 & 1 & 0 & 0 \\
    1 & 0 & 0 & 0 \\
    0 & 0 & 0 & 1 \\
    0 & 0 & 1 & 0 \\
  \end{pmatrix} \,.
\end{align}
This $S$-matrix is given in the mass basis inside Q-ball.

\section{Non-relativistic Scattering of Nucleons \label{app:NRpot}}

From the discussions about scattering of colorless quarks on the Q-ball (with $\omega = 0$), one might naively expect that some of nucleons incident on the Q-ball is reflected as anti-nucleons. 
In particular, once the delta baryon, $\Delta^{++}$ or $\Delta^{-}$, comes into Q-ball, it is expected that the delta baryon is totally reflected as its anti-baryon since they are spin-$3/2$ states and all spins of quarks inside are aligned. 
To begin with, we consider the one-dimensional non-relativistic scattering of nucleons on the (spherically symmetric) Q-ball in order to examine this behavior.
The Schr\"odinger equation describes the non-relativistic scattering:
\begin{align}
  \left[- \frac{\nabla^2}{2 m_N} + V_\mathrm{NR} \right] \varphi(x) = E \varphi(x) \,.
\end{align}
Here, $\varphi$ denotes the two-component wave function for nucleon and anti-nucleon:  
\begin{align}
  \varphi = 
  \begin{pmatrix}
    \varphi_N \\
    \varphi_{\overline{N}}
  \end{pmatrix} \,.
\end{align}
We assume that the center of the Q-ball locates at $x = 0$ and that its radius is $R$.
The potential $V_\mathrm{NR}$ for this system is zero outside Q-ball ($|x|>R$). 
Meanwhile, along our discussion using the colorless toy model, the Q-ball has a uniform distribution and may affect the wave function for nucleons as a position-independent potential. 
This potential is parameterized by two parameters inside Q-ball ($|x|<R$):%
\footnote{
  In general, we can also include the boundary terms proportional to the delta function. 
  The following discussion does not significantly change once we include them.
}
\begin{align}
  V_\mathrm{NR} = V_0 \theta(R - |x|)
  \begin{pmatrix}
    1 & \beta \\
    \beta^\ast & 1 
  \end{pmatrix} 
  \,.
\end{align}
We assume $CP$ invariance of the potential, and thus the diagonal components of the potential are same. 
The potential matrix is hermitian, and $\beta$ can take a complex value.
This complex phase can be absorbed by the phase rotation of wave function, and thus we simply assume the positive $\beta$ in this study.

Now, we consider the situation where only a nucleon is coming from the left-side of the Q-ball.
Outside the Q-ball, the wave functions are given by linear combinations of the free solutions: 
\begin{align}
  \varphi_N(x) & = e^{ikx} + \overline R e^{-ikx} \,, \qquad (x < - R) \,, &
  \varphi_{\overline{N}}(x) & = \widetilde R e^{-ikx} \,, \qquad (x < - R) \,, \nonumber \\
  & = \overline T e^{ikx} \,, \qquad (x > R) \,, &
  & = \widetilde T e^{ikx} \,, \qquad (x > R) \,, 
\end{align}
where $\overline R$ ($\widetilde R$) denotes the reflection amplitude of a nucleon (an anti-nucleon), while $\overline T$ ($\widetilde T$) denotes the transmission amplitude.
Meanwhile, the eigenfunctions inside the Q-ball are given by the linear combinations of the damping and growing solutions: 
\begin{align}
  \varphi_\pm(x) = A_\pm e^{-\kappa_\pm x} + B_\pm e^{\kappa_\pm x} \,, 
\end{align}
where $\kappa_\pm^2 = - k^2 + 2 m_N V_0(1\pm\beta)$. 
We assume $\kappa_\pm > 0$ in order that there is no propagating mode in the Q-ball, since we are interested in a solution with a nucleon being totally reflected as an anti-nucleon.
Matching the wave functions and their derivatives with respect to $x$ at $x = R$ and $-R$, one finds the solutions to the matching conditions in the large $\kappa_\pm R$ limit as follows. 
\begin{align}
  \overline R & = - \frac12 \left( \frac{i k + \kappa_+}{- i k + \kappa_+} + \frac{i k + \kappa_-}{- i k + \kappa_-} \right) e^{-2ik R} \,, \quad
  \widetilde R = - \frac12 \left( \frac{i k + \kappa_+}{- i k + \kappa_+} - \frac{i k + \kappa_-}{- i k + \kappa_-} \right) e^{-2ik R} \,, \\
  \overline T & = 2 i k e^{- 2 i k R} \left[ \frac{e^{-2\kappa_+ R} \kappa_+}{(k+i \kappa_+)^2} + \frac{e^{-2\kappa_- R} \kappa_-}{(k+i \kappa_-)^2} \right] \,, \quad
  \widetilde T = 2 i k e^{- 2 i k R} \left[ \frac{e^{-2\kappa_+ R} \kappa_+}{(k+i \kappa_+)^2} - \frac{e^{-2\kappa_- R} \kappa_-}{(k+i \kappa_-)^2} \right] \,, \\
  A_\pm & = \pm \sqrt2 e^{- \kappa_\pm R} e^{-i k R} \frac{k}{k+i\kappa_\pm} \,, \quad 
  B_\pm = \mp \sqrt2 e^{- 3 \kappa_\pm R} e^{-i k R} \frac{k (k-i\kappa_\pm)}{(k+i\kappa_\pm)^2} \,.
\end{align}

If the nucleon incident on the Q-ball is completely reflected as the anti-nucleon at the boundary, the reflection amplitude for the nucleon vanishes (namely $\overline R = 0$) and thus $k^2 + \kappa_+ \kappa_- = 0$.
However, the left-hand side, $k^2 + \kappa_+ \kappa_-$, is always positive, and hence the reflection amplitude for the nucleon never vanishes in this model. 
Meanwhile, in the range of $\kappa_- < k < \kappa_+$ (keeping $\kappa_- R$ large), the reflection amplitude for the anti-nucleon is comparable with that for the nucleon.
This is realized for a large range of $k$ when $\beta \simeq 1$.
To further extend the range of $k$, we also consider the case where there is a propagating mode inside the Q-ball.
The solution in the large $\kappa_+ R$ limit is approximately given by
\begin{align}
  \overline R + \widetilde R & \simeq - e^{-2ik R} \,, \quad 
  \overline R  - \widetilde R \simeq\frac{(k^2 - k_-^2) (e^{4ik_- R}-1) e^{-2ik R}}{ (e^{4ik_- R}-1) k^2 - 2 (e^{4ik_- R}+1) k k_- + (e^{4ik_- R}-1) k_-^2} \,, \\
  \overline T + \widetilde T & \simeq - \frac{4 i k e^{- 2 i k R} e^{-2 \kappa_+ R}}{\kappa_+ } \,, \quad 
  \overline T - \widetilde T \simeq - \frac{4 i k k_- e^{- 2 i (k - k_-) R}}{(k - k_-)^2 e^{4 i k_- R} - (k + k_-)^2 } \,, \\
  A_+ & = - \sqrt2 i e^{- \kappa_+ R} e^{-i k R} \frac{k}{\kappa_+} \,, \quad 
  A_- = - \sqrt2 \frac{e^{-i (k-3k_-) R} k (k - k_-)}{(k - k_-)^2 e^{4 i k_- R} - (k + k_-)^2 } \,, \\
  B_+ & = - \sqrt2 i e^{- 3 \kappa_+ R} e^{-i k R} \frac{k}{\kappa_+} \,, \quad 
  B_- = - \sqrt2 \frac{e^{-i (k-k_-) R} k (k + k_-)}{(k - k_-)^2 e^{4 i k_- R} - (k + k_-)^2 } \,, 
\end{align}
where $k_-^2 = k^2 + 2 m_N V_0(\beta-1) > 0$.
If $k_- \gg k$, the reflection amplitudes for the anti-nucleon is suppressed.
On the other hand, when $k_- \simeq k$, the reflection amplitudes for the nucleon and for the anti-nucleon are almost close to each other, $|\overline R| \simeq |\widetilde R| \simeq 1/2$, and the transmitted amplitudes for them also have the same absolute value, $|\overline T| \simeq |\widetilde T| \simeq 1/2$.
In particular, $\beta = 1$ is an interesting choice where $k_- = k$ and the amplitude $|\widetilde R|$ is comparable with the others for any $k$.

Furthermore, we examine how the above observation in one-dimensional scattering is realized in three-dimensional scattering.
We also note that unlike quark (inside a nucleon) scattering, one-dimensional analysis is not justified for nucleon scattering since $k R$ can be smaller than unity, namely, Q-ball cannot be regarded as a wall for a low-velocity nucleon.
Under the spherical potential, the radial Schr\"odinger equation is given by 
\begin{align}
  \left[- \frac{1}{2 m_N r^2} \frac{d}{dr} \left( r^2 \frac{d}{dr} \right) + \frac{\ell(\ell+1)}{2 m_N r^2} + V_\mathrm{NR} (r) \right] \varphi(r) = E \varphi(r) \,, 
\end{align}
with the constant spherical potential inside Q-ball 
\begin{align}
  V_\mathrm{NR} (r) = V_0 \theta(R - r)
  \begin{pmatrix}
    1 & \beta \\
    \beta^\ast & 1 
  \end{pmatrix} \,.
\end{align}
Here, we again take $\beta$ to be positive. 
The wave functions outside the Q-ball are given by linear combinations of free solutions as follows:
\begin{align}
  \varphi_N(r) & = A h_\ell^{(2)} (kr) + \overline R h_\ell^{(1)} (kr) \,, \\
  \varphi_{\overline N}(r) & = \widetilde R h_\ell^{(1)} (kr) \,,
\end{align}
Here, $h_\ell^{(1,2)} (x)$ is the spherical Hankel functions. 

Let us start with the case where there is no propagating mode inside the Q-ball.
The wave functions inside the Q-ball are given by the modified spherical Hankel functions, which is regular at the origin. 
\begin{align}
  \varphi_\pm(r) & = A_\pm i_\ell (\kappa_\pm r) \,, 
\end{align}
with the parameters $\kappa_\pm$ which are the same as before: $\kappa_\pm^2 = - k^2 + 2 m_N V_0(1\pm\beta)$.
By matching wave functions and their derivative at $r = R$, we find the coefficients as follows. 
\begin{align}
  \overline R + \widetilde R = - \frac{k i_\ell (\kappa_+ R) h_\ell^{(2)'} (k R) - \kappa_+ i_\ell' (\kappa_+ R) h_\ell^{(2)} (k R)}{k i_\ell (\kappa_+ R) h_\ell^{(1)'} (k R)  - \kappa_+ i_\ell' (\kappa_+ R) h_\ell^{(1)} (k R)} A \,, \\
    \overline R - \widetilde R = - \frac{k i_\ell (\kappa_- R) h_\ell^{(2)'} (k R) - \kappa_- i_\ell' (\kappa_- R) h_\ell^{(2)} (k R)}{k i_\ell (\kappa_- R) h_\ell^{(1)'} (k R)  - \kappa_- i_\ell' (\kappa_- R) h_\ell^{(1)} (k R)} A \,, 
\end{align}
and 
\begin{align}
  A_\pm = - \frac{1}{\sqrt{2}} \frac{h_\ell^{(1)} (k R) h_\ell^{(2)'} (k R) - h_\ell^{(1)'} (k R) h_\ell^{(2)} (k R)}{k i_\ell (\kappa_\pm R) h_\ell^{(1)'} (k R)  - \kappa_\pm i_\ell' (\kappa_\pm R) h_\ell^{(1)} (k R)} A \,. 
\end{align}
At large $\kappa_\pm R$, we find the approximate forms of the solutions as follows.
\begin{align}
  \overline R + \widetilde R \simeq - \frac{h_\ell^{(2)} (k R) - \frac{k R}{\kappa_+ R} h_\ell^{(2)'} (k R)}{h_\ell^{(1)} (k R) - \frac{k R}{\kappa_+ R}h_\ell^{(1)'} (k R) } A \,, \\
  \overline R - \widetilde R \simeq - \frac{h_\ell^{(2)} (k R) - \frac{k R}{\kappa_- R} h_\ell^{(2)'} (k R)}{h_\ell^{(1)} (k R) - \frac{k R}{\kappa_- R}h_\ell^{(1)'} (k R) } A \,.
\end{align}
The reflection amplitude for the anti-nucleon, $\widetilde R / A$ is suppressed by $k/\kappa_-$ (for $kR \gg 1$) or $ 1/ (\kappa_- R)$ (for $kR \ll 1$) compared to elastic amplitude $\overline R / A - 1$.
Therefore, conversion (partial-wave) cross section
\begin{align}
{\tilde \sigma}_{\ell} = \frac{\pi}{k^{2}} (2 \ell + 1) \left|  \frac{\widetilde R}{A}\right|^{2} \,,
\end{align}
is suppressed by $(k/\kappa_-)^{2}$ or $1/ (\kappa_- R)^{2}$ compared to elastic (partial-wave) cross section
\begin{align}
{\overline \sigma}_{\ell} = \frac{\pi}{k^{2}} (2 \ell + 1) \left| \frac{\overline R}{A} - 1\right|^{2} \,
\end{align}
as in the one-dimensional case.
Therefore, we again consider the case where $\beta \simeq 1$ and also there is a propagating mode inside the Q-ball. 
The wave function $\varphi_-(r)$ is regular at the origin, and thus is given by $\varphi_-(r) = A_- j_\ell (k_- r)$. 
In this case, the solution is modified as follows.
\begin{align}
  \overline R - \widetilde R & = - \frac{k j_\ell (k_- R) h_\ell^{(2)'} (k R) - k_- j_\ell' (k_- R) h_\ell^{(2)} (k R)}{k j_\ell (k_- R) h_\ell^{(1)'} (k R)  - k_- j_\ell' (k_- R) h_\ell^{(1)} (k R)} A \,, \\
  A_- & = - \frac{1}{\sqrt{2}} \frac{h_\ell^{(1)} (k R) h_\ell^{(2)'} (k R) - h_\ell^{(1)'} (k R) h_\ell^{(2)} (k R)}{k j_\ell (k_- R) h_\ell^{(1)'} (k R)  - k_- j_\ell' (k_- R) h_\ell^{(1)} (k R)} A \,, 
  \label{eq:Rdiff}
\end{align}
where $k_-^2 = k^2 + 2 m_N V_0(\beta-1)$. 

Lastly, we show the approximate forms of the reflection amplitudes in $k R \gg 1$ and $k R \ll 1$ limits with keeping $k/\kappa_+$ small. 
The reflection amplitudes in the high-energy limit, $k R \gg 1$, are approximated as follows.
\begin{align}
  \overline R + \widetilde R \simeq \left( 1 + \frac{2 i k R}{\kappa_+ R} \right) e^{- 2 i k R + i \ell \pi} A \,, \quad
  \overline R - \widetilde R \simeq \frac{k_{-} j_\ell' (k_- R) + i k j_\ell (k_- R)}{k_{-} j_\ell' (k_- R) - i k j_\ell (k_- R)} e^{- 2 i k R + i \ell \pi} A  \,.
\end{align}
If $k_{-} \gg k$, the conversion amplitude is suppressed by $k/k_{-}$ compared to the elastic amplitude.
On the other hand, when $k_- \simeq k$,
\begin{align}
  \overline R \simeq \frac{e^{- 2 i k R + i \ell \pi}+1}{2} A \,, \quad
  \widetilde R \simeq \frac{e^{- 2 i k R + i \ell \pi}-1}{2} A \,,
\end{align}
and thus conversion and elastic cross sections are equal and given by a quarter of hard-sphere scattering cross section.
The reflection amplitudes in the low-energy limit, $k R \ll 1$, with keeping the leading order of $\kappa_+ R$ corrections are expanded as follows.
\begin{align}
  \overline R + \widetilde R & \simeq A - \frac{2 i (k R)^{2 \ell + 1}}{(2 \ell+1)!} \left( 1 - \frac{2 \ell + 1}{\kappa_+ R} \right)A \,, \\
  \overline R - \widetilde R & \simeq A - \frac{2 i (k R)^{2 \ell + 1}}{(2 \ell+1)!} \frac{k_- R j_\ell' (k_- R) - \ell j_\ell (k_- R)}{k_- R j_\ell' (k_- R) + (\ell+1) j_\ell (k_- R)} A\,.
\end{align}
If $k_{-} R \gg 1$, the conversion amplitude is suppressed by $1/(k_{-} R)$ compared to the elastic amplitude.
On the other hand, when $k_- R \ll 1$,
\begin{align}
  \overline R \simeq  A - \frac{i (k R)^{2 \ell + 1}}{(2 \ell+1)!} A \,, \quad
  \widetilde R \simeq - \frac{i (k R)^{2 \ell + 1}}{(2 \ell+1)!} A \,.
\end{align}
and thus conversion and elastic cross sections are equal and given by a quarter of hard-sphere scattering cross section.
We note that $\beta = 1$ is an interesting choice, where conversion and elastic cross sections are comparable for any $k$ as in the one dimensional case.

\bibliographystyle{utphys}
\bibliography{ref}
\end{document}